\documentclass[10pt,journal]{IEEEtran}

\usepackage{enumitem}
\usepackage{balance}
\usepackage{booktabs} 
\usepackage[caption=false]{subfig}
\usepackage{graphicx}
\usepackage{subfig}
\usepackage{url}
\usepackage{tikz}
\usepackage{adjustbox}
\usepackage{booktabs}
\usepackage{rotating}
\usepackage{multirow}
\usepackage{lineno}
\usepackage[hidelinks]{hyperref}
\usepackage{array,ragged2e}
\usepackage{inputenc}
\newcolumntype{P}[1]{>{\RaggedRight\arraybackslash}p{#1}}
\usepackage[T1]{fontenc}
\usepackage{color, colortbl}
\definecolor{grey}{gray}{0.9}
\usepackage[misc,geometry]{ifsym}

\usepackage{dirtytalk}
\usepackage[normalem]{ulem}
\usepackage{amsmath}
\usepackage{blindtext}
\usepackage{tabularx}
\usepackage{tabulary}
\usepackage{wrapfig}
\usepackage{amssymb}
\usepackage{multirow}
\usepackage{stix}
\usepackage{float}
\usepackage{svg}
\usepackage{cite}
\newcommand{\orcid}[1]{\href{https://orcid.org/#1}{\includesvg[width=10pt]{orcid}}}
\usepackage{fancyhdr}
\begin{document}

\title{A Survey on Security and Privacy Issues of UAVs}

\author{Yassine~Mekdad,~\IEEEmembership{Student~Member,~IEEE,}
        Ahmet~Aris,
        Leonardo~Babun,~\IEEEmembership{Member,~IEEE,}       Abdeslam~EL~Fergougui,
        Mauro~Conti,~\IEEEmembership{Senior~Member,~IEEE,}
       Riccardo~Lazzeretti,~\IEEEmembership{Senior~Member,~IEEE,}
        and~A.~Selcuk~Uluagac
\IEEEcompsocitemizethanks{
\IEEEcompsocthanksitem Y. Mekdad and A. El Fergougui are with the Laboratory of Computer Networks and Systems, Moulay Ismail University of Meknes, 11201, Zitoune, Meknes, Morocco (e-mail: y.mekdad@edu.umi.ac.ma; a.elfergougui@umi.ac.ma).
\IEEEcompsocthanksitem A. Aris, L. Babun, and  A. Selcuk Uluagac are with the Cyber-Physical Systems Security Lab, Department of Electrical and Computer Engineering, 
Florida International University, Miami, FL 33174 USA (e-mail: aaris@fiu.edu; lbabu002@fiu.edu; suluagac@fiu.edu).
\IEEEcompsocthanksitem M. Conti is with the Department of Mathematics, University of Padua, 35121, Padua, Italy (e-mail: conti@math.unipd.it).
\IEEEcompsocthanksitem Riccardo Lazzeretti is with the Department of Computer, Control, and Management Engineering "Antonio Ruberti", Sapienza University of Rome, 00185 Rome, Italy (e-mail: lazzeretti@diag.uniroma1.it).\protect\\
}

}


\maketitle

\begin{abstract}
In the 21st century, the industry of drones, also known as Unmanned Aerial Vehicles (UAVs), has witnessed a rapid increase with its large number of airspace users. The tremendous benefits of this technology in civilian applications such as hostage rescue and parcel delivery will integrate smart cities in the future. Nowadays, the affordability of commercial drones expands its usage at a large scale. However, the development of drone technology is associated with vulnerabilities and threats due to the lack of efficient security implementations. Moreover, the complexity of UAVs in software and hardware triggers potential security and privacy issues. Thus, posing significant challenges for the industry, academia, and governments.

In this paper, we extensively survey the security and privacy issues of UAVs by providing a systematic classification at four levels: Hardware-level, Software-level, Communication-level, and Sensor-level. In particular, for each level, we thoroughly investigate (1) common vulnerabilities affecting UAVs for potential attacks from malicious actors, (2) existing threats that are jeopardizing the civilian application of UAVs, (3) active and passive attacks performed by the adversaries to compromise the security and privacy of UAVs, (4) possible countermeasures and mitigation techniques to protect UAVs from such malicious activities. In addition, we summarize the takeaways that highlight lessons learned about UAVs' security and privacy issues. Finally, we conclude our survey by presenting the critical pitfalls and suggesting promising future research directions for security and privacy of UAVs.

\end{abstract}

\begin{IEEEkeywords}
Unmanned Aerial Vehicles, UAVs, security, privacy, drones.
\end{IEEEkeywords}

\section{Introduction}

\IEEEPARstart{I}{n} the past decades, the Unmanned Aerial Vehicles (UAVs) global market has meaningfully increased and gained more attention from governments and commercial industries due to its wide civilian and military applications such as traffic monitoring, search-and-rescue operations, surveillance, and biochemical sensing~\cite{Hayat2016,Gupta2016,HosseinMotlagh2016Low-AltitudePerspectives}. Currently, there is a socio-technical debate about the use of UAVs for passenger transportation, so-called "air taxis" that will replace commercial helicopters because of their electric Vertical Takeoff and Landing (eVTOL) capabilities~\cite{Kellermann2020DronesReview}. A recent report shows that the commercial drone market revenue forecast will reach 129.33 billion dollars by 2025~\cite{Market}. According to the Federal Aviation Administration (FAA), the size of the commercial drone market could triple by 2023~\cite{AviationAdministrationFAATables}. Thus, the introduction of UAVs into the civilian market will increase the demand for their commercial use in different sectors. Nowadays, with the rise of drone technology, the industrial players have their interest investing in UAVs~\cite{LiuRiseIssues}. Therefore, the UAVs will represent an essential part of our technological society as their civilian popularity is significantly increasing.

Although the worldwide development of the drone business model and the benefits offered by commercial UAVs, a considerable number of drones incidents are reported every week~\cite{DroneIncidents}. Thus, highlighting the need to counter the UAV threats efficiently. To that end, one line of argument suggests detecting and identifying the UAV threats at their early phases~\cite{Sedjelmaci2016}. This approach would provide the operator a reasonable amount of time to deploy the required tools to neutralize such threats. It is of utmost importance to consider the malicious use of such technology and its potential threats for civilian users. Its exponential growth triggers different vulnerabilities to their cyber and physical components~\cite{Guo2020}. The security and privacy aspects of their deployment into the national airspace have become a significant concern for governments, as the UAVs threat landscape becomes wide. In addition, most of the existing commercial UAVs are not equipped with security mechanisms such as intrusion detection systems (IDS). Therefore, they present perfect targets for adversaries. 

Recently, the world has witnessed a series of successful cyber attacks on UAVs~\cite{Yahuza2021a}. Performing real-world cyber attacks against civilian UAVs has become a matter of national security. Upon integrating UAVs in the national airspace, their security issues have created a substantial discussion among governments and agencies in the public and private sectors. From a security point of view, the variety of existing cyber attacks demonstrate that UAVs are vulnerable at different levels. Indeed, malicious actors benefit from the ubiquity of drone usage in civilian applications. They exploit different vulnerabilities across commercial drones creating an active threat to the safety of people. Furthermore, drones manufacturers lack considering security and privacy concerns in the early phases of their production.

It is worth mentioning that the active use of civilian UAVs in many applications can pose new security and privacy challenges~\cite{Yaacoub2020}. With this in mind, existing countermeasures to detect compromised drones and secure drone systems are weak. To that end, cyber attacks against UAVs are feasible due to the lack of implementing appropriate security measures that guarantee the classical CIA triad (Confidentiality, Integrity, and Availability)~\cite{Watkins2018}. Hence, we need to investigate Unmanned Aerial Vehicles from a security and privacy perspective. On the other hand, the integration of UAVs in the national airspace can also violate public users' privacy and sensitive facilities such as chemical industries and nuclear power plants. Indeed, most UAVs are equipped with onboard camera capabilities, which might potentially disclose sensitive details of human activities~\cite{Li2019b}.

In general, we consider UAVs as complex aerial vehicles. A flying UAV operates under a set of onboard sensors (e.g., GPS, accelerometer, etc.) that provide sensor readings to the Flight Controller, which sends data through a communication channel to the operator. According to the received data, the operator sends the control signal to the Flight Controller. In this scenario, four fundamental components of the UAV system need to correlate and operate to maintain the desired state. Namely, the sensors, the hardware, the software, and the communication link. Moreover, the potential failure of any components might result in grounding and crashing the UAV system. Motivated by this vision and from an adversarial perspective, we consider the abovementioned elements as critical attack points of the UAV system. Hence, we aim to investigate the security and privacy issues of UAVs according to these components that are organized into four levels: the \textit{Sensor-level}, the \textit{Hardware-level}, the \textit{Software-level}, and the \textit{Communication-level}.

\subsection{Contributions}
In this paper, we aim to provide a comprehensive survey targeting the security and privacy issues of Unmanned Aerial Vehicles and their related concepts. We summarize our main contributions as follows:
\begin{itemize}
    \item We shed light on the background of UAVs, emphasizing the main components characterizing the UAV system such as the hardware and software architecture, the communication principles, and the sensing technology;
    \item We provide the first comprehensive categorization of the security issues of UAVs into four different levels: the \textit{Sensor-level}, the \textit{Hardware-level}, the \textit{Software-level}, and the \textit{Communication-level}. For each level, we investigate common vulnerabilities, threats, attacks, and existing countermeasures. We believe that this categorization can provide a reference for future researchers to start investigating the UAV security;
    \item We systematically consider how commercial drones can affect people's privacy by discussing the primary privacy invasion attacks and possible countermeasures;
    \item Finally, we discuss the lessons learned, pitfalls and promising directions for future research in the field of security and privacy of UAVs.
\end{itemize}

\subsection{Roadmap}
The remainder of the article is structured as follows. Firstly, we provide an overview of related work in Section~\ref{sec:related-work}. 
Section~\ref{sec:background} provides background on UAVs describing their general architecture, communication principles, and security requirements. In Section~\ref{sec:security-issues}, we discuss the main security issues targeting UAVs. In particular, we classify these issues into four different levels: \textit{Sensor-level}, the \textit{Hardware-level}, the \textit{Software-level}, and the \textit{Communication-level}. For each level, we list the vulnerabilities and threats. Then, we discuss the potential attacks and existing countermeasures. In Section~\ref{sec:privacy-issues}, we focus on the privacy issues of commercial UAVs, including existing defense mechanisms against privacy-invasion attacks. Section~\ref{sec:pitfalls} discusses the lessons learned, pitfalls and future research directions. Finally, Section~\ref{sec:conclusion} concludes the survey. 

\section{Related Work}\label{sec:related-work}

Unmanned Aerial Vehicles are considered as a new emerging type of "flying IoT" devices~\cite{DronesSociety}. They incorporate several applications. For example, drones can provide immediate assistance for patients, such as delivering blood and medical supplies. However, security and privacy challenges might occur when integrating UAVs in modern healthcare systems~\cite{Iqtidar2021ASystems}. With the introduction of synchronized IT components in the Enterprise Architecture (EA) domain, commercial UAVs are extensively used for business development (e.g., safe inspection of critical infrastructures, aerial data collection, etc.). In contrast, the security implications on the use of UAVs within companies and organizations need to be properly considered~\cite{RondonSurveyPerspective2}. In the past decade, the evolution of UAV technology has 
faced security and privacy issues. In this context, prior works have been published to cover different aspects of UAVs' security and privacy issues.
\begin{table*}[h]
\centering
\caption{Comparison of our survey and existing surveys on Security and privacy issues of UAVs}
\label{Survey_Comparison}
\begin{tabular}{|l|l|c|c|c|c|c|c|c|c|c|c|c|c|c|c|c|c|c|c|}
\hline
\multirow{3}{*}{\textbf{Year}} & \multirow{3}{*}{\textbf{Work}}                                                        & \multicolumn{16}{c|}{\textbf{Security issues}}                                                                                                                                                                                                                                                            & \multicolumn{2}{c|}{\multirow{2}{*}{\textbf{Privacy issues}}} \\ \cline{3-18}
                               &                                                                                            & \multicolumn{4}{c|}{\textbf{Software-level}}                              & \multicolumn{4}{c|}{\textbf{Hardware-level}}                              & \multicolumn{4}{c|}{\textbf{Communication-level}}                         & \multicolumn{4}{c|}{\textbf{Sensor-level}}                            & \multicolumn{2}{c|}{}            \\ \cline{3-20} 
                               &                                                                                            & V                & T                & A                & C                & V                & T                & A                & C                & V                & T                & A                & C                & V              & T              & A                & C                & A                                & C
                               \\ \hline
              2016              &  Hayat et al. \cite{Hayat2016}     &     $\square$     &     $\square$    &  $\square$     &   $\square$      &   $\square$      &  $\square$       & $\square$ & $\square$  & $\squarelrblack$ & $\squarelrblack$ &   $\square$     & $\square$ &    $\square$   &  $\square$     &      $\square$   &   $\square$      &    $\square$    &    $\square$
                                 \\ \hline
                         2017   & Altawy et al. \cite{Altawy2017}                                           & $\squarelrblack$ & $\squarelrblack$ & $\squarelrblack$ & $\squarelrblack$ & $\squarelrblack$ & $\squarelrblack$ & $\squarelrblack$ & $\squarelrblack$ & $\square$        & $\blacksquare$   & $\blacksquare$   & $\blacksquare$   & $\square$      & $\square$      & $\squarelrblack$ & $\squarelrblack$ & $\squarelrblack$                 & $\squarelrblack$           
                               
                                 \\ \hline
                          2017     & Krishna el al. \cite{Krishna2017}                                         & $\squarelrblack$ & $\square$        & $\squarelrblack$ & $\square$        & $\squarelrblack$ & $\square$        & $\square$        & $\square$        & $\blacksquare$   & $\squarelrblack$ & $\squarelrblack$ & $\squarelrblack$ & $\square$      & $\square$      & $\square$        & $\square$        & $\square$                        & $\square$                   
                               
                                 \\ \hline
                       2017          &  Maxa et al. \cite{Maxa2017b}    &         $\square$     &     $\square$    &  $\square$     &   $\square$      &   $\square$      &  $\square$       & $\square$ & $\square$  & $\squarelrblack$ & $\squarelrblack$ &   $\squarelrblack$     & $\squarelrblack$ &    $\square$   &  $\square$     &      $\square$   &   $\square$      &    $\square$    &    $\square$
                                 \\ \hline
                      2018           &   Choudhary et al. \cite{Choudhary2018a}    &        $\square$     &     $\square$    &  $\squarelrblack$    &   $\square$      &   $\square$      &  $\square$       & $\squarelrblack$ & $\square$  & $\squarelrblack$ & $\squarelrblack$ &   $\blacksquare$     & $\square$ &    $\square$   &  $\square$     &      $\square$   &   $\square$      &    $\squarelrblack$    &    $\square$
                                 \\ \hline
                        2018         &   Lin et al. \cite{Lin2018}    &          $\square$     &     $\square$    &  $\square$     &   $\square$      &   $\square$      &  $\square$       & $\square$ & $\square$  & $\square$ & $\square$ &   $\square$     & $\squarelrblack$ &    $\square$   &  $\square$     &      $\square$   &   $\square$      &    $\squarelrblack$    &    $\squarelrblack$
                                 \\ \hline
                        2019  & Shakhatreh et al. \cite{Shakhatreh2019}                                   & $\square$        & $\square$        & $\square$        & $\square$        & $\square$        & $\square$        & $\square$        & $\square$        & $\square$        & $\square$        & $\blacksquare$   & $\blacksquare$   & $\square$      & $\square$      & $\squarelrblack$ & $\squarelrblack$ & $\square$                        & $\square$                  
                               
                                 \\ \hline
                           2019     &  Nassi et al. \cite{Nassi2019}    &         $\square$     &     $\square$    &  $\squarelrblack$     &   $\square$      &   $\square$      &  $\square$       & $\squarelrblack$ & $\squarelrblack$  & $\square$ & $\square$ &   $\squarelrblack$    & $\squarelrblack$ &    $\square$   &  $\square$     &      $\squarelrblack$   &   $\squarelrblack$      &    $\squarelrblack$    &    $\squarelrblack$
                                 \\ \hline
                             2019  & Fotouhi el al. \cite{Fotouhi2019a}                                        & $\square$        & $\square$        & $\square$        & $\square$        & $\square$        & $\square$        & $\squarelrblack$ & $\square$        & $\squarelrblack$ & $\squarelrblack$ & $\squarelrblack$ & $\squarelrblack$ & $\square$      & $\square$      & $\square$        & $\square$        & $\square$                        & $\square$  
                               
                                 \\ \hline
                              2019  & Chriki et al.  \cite{Chriki2019FANET:Issues}    &      $\square$     &     $\square$    &  $\square$     &   $\square$      &   $\square$      &  $\square$       & $\square$ & $\square$  & $\square$ & $\square$ &   $\squarelrblack$     & $\squarelrblack$ &    $\square$   &  $\square$     &      $\square$   &   $\square$      &    $\square$    &    $\square$
                                 \\ \hline
                           2020   & Yaacoub el al. \cite{Yaacoub2020}    & $\squarelrblack$ & $\squarelrblack$ & $\squarelrblack$ & $\squarelrblack$ & $\squarelrblack$ & $\squarelrblack$ & $\squarelrblack$ & $\squarelrblack$ & $\squarelrblack$ & $\square$        & $\blacksquare$   & $\blacksquare$   & $\square$      & $\square$      & $\square$        & $\square$        & $\squarelrblack$                 & $\squarelrblack$   
                               
                                 \\ \hline
                            2020   & Boccadoro et al. \cite{Boccadoro2020}                                     & $\square$        & $\square$        & $\square$        & $\square$        & $\square$        & $\square$        & $\squarelrblack$ & $\squarelrblack$ & $\squarelrblack$ & $\squarelrblack$ & $\square$        & $\squarelrblack$ & $\square$      & $\square$      & $\square$        & $\square$        & $\squarelrblack$                 & $\squarelrblack$  
                               
                                 \\ \hline
                          2020   & Wang et al. \cite{Wang2020}                                               & $\square$        & $\square$        & $\square$        & $\square$        & $\square$        & $\square$        & $\square$        & $\square$        & $\square$        & $\square$        & $\squarelrblack$ & $\square$        & $\square$      & $\square$      & $\squarelrblack$ & $\square$        & $\squarelrblack$                 & $\square$       
                               
                                 \\ \hline
                            2020    & Hentati et al. \cite{Hentati2020}      &      $\square$     &     $\square$    &  $\square$     &   $\square$      &   $\square$      &  $\square$       & $\square$ & $\square$  & $\square$ & $\square$ &   $\squarelrblack$     & $\squarelrblack$ &    $\square$   &  $\square$     &    $\squarelrblack$   &   $\squarelrblack$      &    $\square$    &    $\square$
                                 \\ \hline
                            2020    & Zhi el al. \cite{Zhi2020}                                                 & $\square$        & $\square$        & $\squarelrblack$ & $\square$        & $\square$        & $\square$        & $\square$        & $\square$        & $\square$        & $\square$        & $\squarelrblack$ & $\square$        & $\square$      & $\square$      & $\squarelrblack$ & $\square$        & $\squarelrblack$                 & $\square$      
                               
                                 \\ \hline
                            2020 & Sharma el al. \cite{Sharma2020}                                           & $\square$        & $\square$        & $\square$        & $\square$        & $\square$        & $\square$        & $\square$        & $\square$        & $\squarelrblack$ & $\squarelrblack$ & $\squarelrblack$ & $\squarelrblack$ & $\square$      & $\square$      & $\square$        & $\square$        & $\square$                        & $\square$                 
                               
                                 \\ \hline
                            2020   &  Noor et al. \cite{Noor2020a}      &        $\square$     &     $\square$    &  $\square$     &   $\square$      &   $\square$      &  $\square$       & $\square$ & $\square$  & $\squarelrblack$ & $\squarelrblack$ &   $\squarelrblack$     & $\square$ &    $\square$   &  $\square$     &      $\square$   &   $\square$      &    $\square$    &    $\square$
                                 \\ \hline
                              2020 & Mishra et al. \cite{Mishra2020}       &   $\square$     &     $\square$    &  $\square$     &   $\square$      &   $\square$      &  $\square$       & $\square$ & $\square$  & $\squarelrblack$ & $\squarelrblack$ &   $\square$     & $\square$ &    $\square$   &  $\square$     &      $\square$   &   $\square$      &    $\square$    &    $\square$
                                 \\ \hline
                           2020 & Syed et al. \cite{Syed2020a}      &       $\square$     &     $\square$    &  $\square$     &   $\squarelrblack$      &   $\square$      &  $\square$       & $\square$ & $\squarelrblack$ & $\square$ & $\square$ &   $\square$     & $\squarelrblack$ &    $\square$   &  $\square$     &      $\square$   &  $\squarelrblack$      &    $\square$    &    $\squarelrblack$
                                 \\ \hline
                             2021 & Yahuza et al. \cite{Yahuza2021}       &        $\square$     &     $\squarelrblack$    &  $\squarelrblack$     &   $\squarelrblack$      &   $\square$      &  $\squarelrblack$       & $\squarelrblack$ & $\squarelrblack$  & $\square$ & $\squarelrblack$ &   $\blacksquare$     & $\squarelrblack$ &    $\square$   &  $\squarelrblack$     &      $\square$   &   $\square$      &   $\squarelrblack$    &    $\squarelrblack$
                                 \\ \hline
                               2021 & Nassi et al. \cite{Nassi2021SoK:Drones}        &        $\square$     &     $\square$    &  $\squarelrblack$     &   $\squarelrblack$      &   $\square$      &  $\square$       & $\squarelrblack$ & $\squarelrblack$  & $\squarelrblack$ & $\squarelrblack$ &   $\squarelrblack$     & $\squarelrblack$ &    $\square$   &  $\square$     &      $\squarelrblack$   &   $\squarelrblack$      &    $\squarelrblack$    &   $\squarelrblack$
                                 \\ \hline
                              2021 & Shafique et al. \cite{Shafique2021a}      &      $\squarelrblack$     &     $\square$    &  $\square$     &   $\squarelrblack$      &  $\squarelrblack$      &  $\square$       & $\square$ & $\squarelrblack$  & $\squarelrblack$ & $\square$ &  $\squarelrblack$     & $\squarelrblack$ &    $\squarelrblack$   &  $\square$     &      $\square$   &   $\squarelrblack$      &    $\square$    &    $\squarelrblack$
                                 \\ \hline
                             2021 & Hassija et al. \cite{Hassija2021}       &    $\square$     &     $\square$    &  $\square$     &   $\square$      &   $\square$      &  $\square$       & $\square$ & $\square$  & $\squarelrblack$ & $\squarelrblack$ &   $\squarelrblack$    & $\squarelrblack$ &   $\square$   &  $\square$     &      $\square$   &   $\square$      &    $\squarelrblack$    &    $\squarelrblack$
                              \\ \hline
                              2021 & This work                                                                                 & $\blacksquare$   & $\blacksquare$   & $\blacksquare$   & $\blacksquare$   & $\blacksquare$   & $\blacksquare$   & $\blacksquare$   & $\blacksquare$   & $\blacksquare$   & $\blacksquare$   & $\blacksquare$   & $\blacksquare$   & $\blacksquare$ & $\blacksquare$ & $\blacksquare$   & $\blacksquare$   & $\blacksquare$                   & $\blacksquare$             \\ \hline
\end{tabular}
\\$\blacksquare$ = Survey the category, $\square$ = Does not survey the category, $\squarelrblack$ = Partially survey the category 
\\ V = Vulnerabilities, T = Threats, A = Attacks, C = Countermeasures
\end{table*}

\noindent \textbf{Security and Privacy Challenges of UAVs.} 
Wang et al.~\cite{Wang2020} discussed the security and privacy challenges of UAV networks from a cyber-physical system (CPS) perspective. The authors considered the significant components of UAVs that are vulnerable to several cyber attacks either from the cyber or the physical domain. A similar work presented the security challenges of UAV's communication networks and proposed their essential security requirements~\cite{Hentati2020}. Shakhatreh et al.~\cite{Shakhatreh2019} reviewed UAV's civil applications and their major key challenges. Krishna et al.~\cite{Krishna2017} conducted a review on cybersecurity vulnerabilities of UAVs. The authors proposed a taxonomy to classify different types of UAVs cyber attacks. Recently, a work by Shafique et al.~\cite{Shafique2021a} surveyed the security protocols and their vulnerabilities in UAVs. Syed et al.~\cite{Syed2020a} surveyed the emerging technologies used in the literature to overcome the security and privacy challenges in UAVs. Their work primarily covers the application of Blockchain, Machine Learning (ML), and watermarking technologies. 

\noindent \textbf{Security and Privacy Issues of Commercial UAVs.} 
In~\cite{Altawy2017}, the authors surveyed the security, privacy, and safety aspects of commercial drones. In particular, they identified the major vulnerabilities, cyber and physical threats, as well as potential attacks that can result in crashing the drone during a flight mission. Similarly, in~\cite{Yaacoub2020}, the authors investigated the emerging cyber attacks and challenges facing commercial drones. In~\cite{Nassi2019}, the authors reviewed the current threats and malicious use of drones in civilian applications. In their recent work, Nassi et al.~\cite{Nassi2021SoK:Drones} carried out a systematic literature review of security and privacy issues of commercial drones. In~\cite{Zhi2020}, the researchers analyzed the potential threats of wireless communications in commercial UAVs such as Wi-Fi-based UAVs communications. Further, they highlighted the privacy disclosure caused by UAVs through aerial photos.

\noindent \textbf{Security and Privacy Issues of UAV Communications.}
Fotouhi et al.~\cite{Fotouhi2019a} surveyed the important security issues of UAV-assisted cellular communications. Mishra et al.~\cite{Mishra2020} pointed that the integration of UAVs to cellular networks such as 5G 
triggers security challenges that need to be thoroughly investigated by the research community. Hayat et al.~\cite{Hayat2016} addressed the safety, security, and privacy issues of UAV networks from a communication viewpoint. Then, provided the general communication requirements of UAV networks for a safe, secure, and privacy-preserving deployment of UAVs. The authors in~\cite{Sharma2020} provided a comprehensive review of the latest UAV communication technologies and the need to secure the collected and transmitted data to the Ground Control Station (GCS). Hassija et al.~\cite{Hassija2021} presented a 
survey covering the major security issues in UAV communications and their potential vulnerabilities. 

\noindent \textbf{Security and Privacy Issues of UAV networks.}
Boccadoro et al.~\cite{Boccadoro2020} provided a 
survey on the Internet of Drones (IoD). They discussed the security and privacy issues of the drone-2-drone communications and their existing solutions. They also considered the security aspects in specific application scenarios involved in the IoD architecture, such as public safety and smart farming. In another work, Noor et al.~\cite{Noor2020a} considered the security and privacy challenges associated with the design of UAV networks. One of the main challenges is the communication among multiple UAVs in an Ad hoc fashion. This type of communication is known as Flying Ad hoc Network (FANET). FANETs security issues are also surveyed by Chriki et al.~\cite{Chriki2019FANET:Issues}. The authors discussed the need to develop robust security schemes before deploying FANET networks in realistic scenarios.
Additionally, Maxa et al.~\cite{Maxa2017b} surveyed 
the main security challenges of UAV routing protocols. Additionally, the work proposed by Sharma et al.~\cite{Sharma2020} outlined the security mechanisms for communication and networking technologies of UAVs. In this context, the authors discussed the underlying security vulnerabilities and threats of UAVs communication protocols.\\ 
\noindent \textbf{Differences from existing surveys.} 
Differently from prior works, our work aims to 
extensively survey the security and privacy issues of UAVs by categorizing them into different levels. 
Most existing surveys and tutorials in this line of research categorize UAVs' security and privacy issues in terms of attack vectors or according to the fundamental principles of information security. However, such categorization cannot fully explain the vulnerabilities, threats, attacks, and countermeasures of UAVs. Moreover, prior works consider analyzing specific components of the UAV system, such as communications and networking. Instead, our survey is focused on the security and privacy aspects of the complete drone system, covering the end-to-end components, including sensors, hardware, software, and communication. 
In our work, we survey the security and privacy issues of commercial UAVs. In particular, we dissect from a security perspective the vulnerabilities, threats, attacks, and existing countermeasures of commercial UAVs into four different levels: \textit{(i) Sensor-level}, \textit{(ii) Hardware-level}, \textit{(iii) Software-level}, and \textit{(iv) Communication-level}. These levels are the most important levels of the functionality of a UAV system. Moreover, we discuss the attacks targeting the privacy aspect of UAVs and their existing mitigation techniques. Throughout our survey, we offer readers a good understanding and visibility of UAVs' most current security and privacy issues at each level. To demonstrate the differences between existing surveys and our work, we perform a comparison of our survey and existing surveys in the literature as shown in Table~\ref{Survey_Comparison}.

\section{Background}\label{sec:background}
In this section, we provide 
background information for Unmanned Aerial Vehicles.  In our survey, our focus is only on commercial drones. Military drones are out of the scope of our work. In addition, for the rest of our paper, we use drones and Unmanned Aerial Vehicles interchangeably. 
In this section, we start by systematically introducing the hardware and software architecture of UAVs. Then, we highlight existing UAV communication capabilities and protocols. Afterward, we present the onboard sensing elements of UAVs that are part of the payload. Finally, we list the security and privacy requirements of the UAVs for mission-driven civilian applications. 

\subsection{General Architecture of UAVs}

The development of UAV technology has created various types of drones with different shapes and weights. To the best of our knowledge, there is no existing standard to classify UAVs. 
A UAV system generally consists of the \textit{Unmanned Aircraft}, the \textit{Ground Control Station} (GCS), and the \textit{Communication Link} (CL). The \textit{Unmanned Aircraft}, also known as UAV, constitutes the core of an Unmanned Aerial Vehicle system~\cite{Altawy2017}, and is monitored by the operator either through the GCS or using a \textit{Remote Controller} (RC).

\begin{figure}[!t]
\begin{center}
\includegraphics[width=0.43\textwidth]{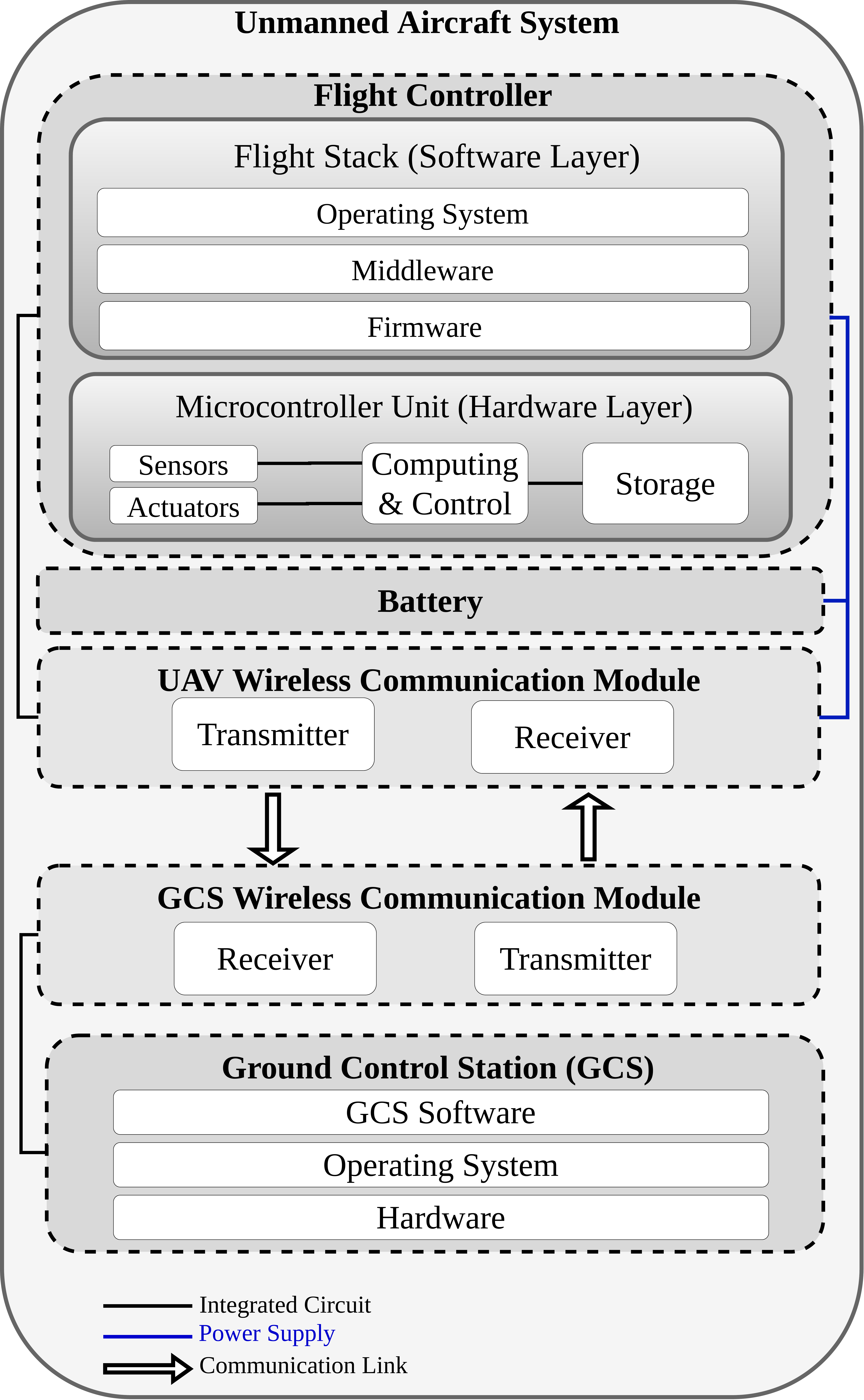}
\caption{General architecture of an Unmanned Aerial Vehicle.}
\label{Architecture}
\end{center}
\end{figure}

\textbf{Hardware Architecture.} The inner hardware architecture of an \textit{Unmanned Aircraft} device includes: a \textit{Flight Controller} (FC), \textit{rechargeable batteries}, \textit{actuators}, a set of \textit{sensors} such as GPS and accelerometer, and a \textit{wireless communication module}. A high-level architecture of an Unmanned Aerial Vehicle is depicted in Figure~\ref{Architecture}.   
\begin{itemize}
    \item  \textit{The Flight Controller:} It serves as the central processing unit of the UAV that interfaces between the software and the onboard devices. It is a microcontroller board equipped with a computing and control unit and storage (e.g., Raspberry Pi~\cite{TeachPi}, BeagleBoard~\cite{BeagleBoard.orgMaking}, etc.). 
    \item  \textit{The rechargeable batteries:} Lithium polymer-based batteries that provide the power supply for the whole UAV. 
    \item  \textit{The actuators:} They consist of the brushless motors and the propellers. Moreover, they produce the appropriate actuation needed for the UAV during the flight mission, thus ensuring high stability. 
    \item  \textit{The sensors:} They are crucial parts of the UAV. They enable sensing functionalities by providing physical measurements of the surrounding environment, such as height, speed, and geospatial references. These measurements are translated into data that the Flight Controller processes and  transmitted to the operator. 
    \item  \textit{The wireless communication module:} It is directly connected to the circuit board of the Flight Controller and includes a transmitter and a receiver. It is designed to send and receive signals from other devices such as the Remote Controller, the Ground Control Station, and nearby unmanned aircrafts. 
\end{itemize}

 \textit{The Ground Control Station (GCS)}  is a fundamental component of any UAV system. It allows controlling and monitoring the UAV remotely during the flight mission using wireless communication. The GCS hardware is a ground-based computer processing unit used to control and administer the flight mission~\cite{patent}. It is equipped with a wireless data link module that: (1) generates and transmits control commands to the UAVs, and (2) receives real-time data from UAVs.   

\textbf{Software Architecture.} The software architeture of the \textit{Unmanned Aircraft} operates in a layered system. The integration between these layers constitutes the \textit{flight stack}, and consists of three main layers: \textit{the Firmware}, \textit{the Middleware}, and \textit{the Operating System}. Examples of open source flight stack are: Arducopter~\cite{CopterDocumentation}, Crazyflie 2.1~\cite{crazy}, and KKMultiCopter~\cite{KKMulticopterlazyzero.de}. The firmware and the middleware are subject to real-time constraints. 
\begin{itemize}
    \item \textit{The Firmware;} It is the lower layer of the flight stack and provides instructions from machine code to the Flight Controller's processor. 
    \item \textit{The Middleware:} It constitutes the layer responsible for proper control of the flight by managing the communication between the services such as guidance, navigation, and telecommunication. Thus, operating the UAV system as a distributed embedded system.
    \item \textit{The Operating System:} It is the highest layer of the flight stack and most of the time labeled as a Real-Time Operating System (RTOS). A Real-Time Operating System handles real-time data processing and enables the autopilot software to manage different processes such as flight operations, video recording, and path planning. 
\end{itemize}

According to the recent FAA regulations, and with the integration of UAVs into the national airspace; all UAVs are required to have a Remote ID (or a System ID), which can be defined as the ability of a flying drone to provide its identification and location information to third parties such as law enforcement, and federal agencies~\cite{UASOverview}.

The \textit{Ground Control Station} software is also known as a mission planner. It includes a human-machine interface that displays the flight parameters and typically runs on laptops, tablets, or any devices in the field.

\textbf{Communication Link.} The communication link represents the wireless communication between the GCS and the UAV. It enables data transmission during the flight mission. However, due to the weather conditions and limited power supply, transmission frequencies and flight range may pose several challenges. We identify two types of communication streams: data communication and control communication. In data communication, the UAV sends data signals such as telemetry and status information to the GCS. While in control communication, the GCS sends commands and control signals 
to the UAV~\cite{Petricca2011}. In what follows, we highlight the UAV communication principles. 

\subsection{Communication Principles}
UAV communications can take place between a UAV and another end point, which can be referred to as UAV-2-X communication. In this subsection, 
we first explain the UAV-2-X communications. Afterwards, we explain the UAV communication architectures, networks of UAVs, as well as their routing protocols. Following that, we shed light on the well-known communication protocols.


\subsubsection{UAV-2-X Communication Types}

During a flight mission, a UAV communicates with several entities. As depicted in Figure~\ref{UAVComsChannel}, we categorize four endpoints of UAV-2-X communications:

\begin{figure}[!t]
\centering
\includegraphics[width=0.45\textwidth]{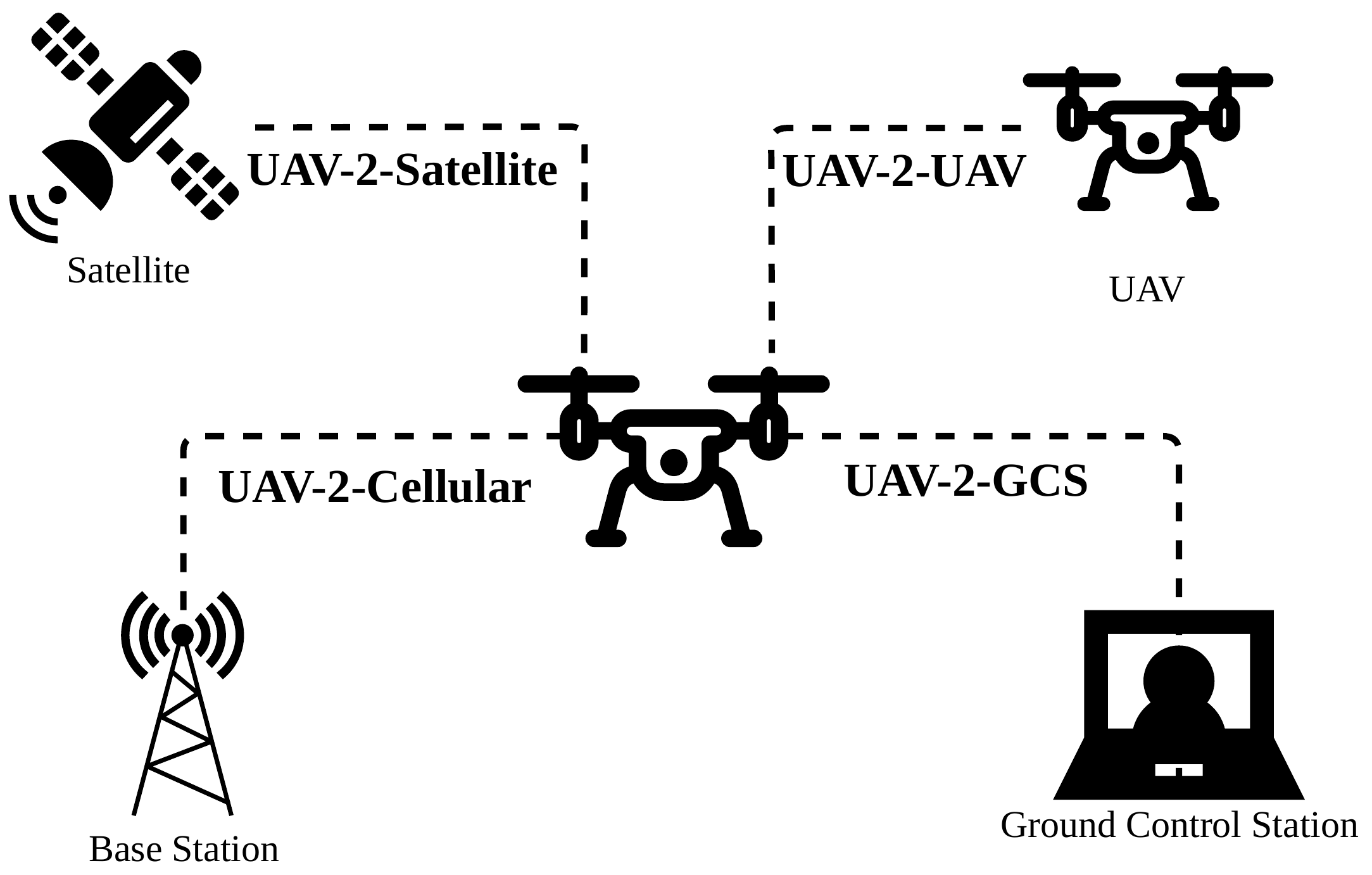}
\caption{UAV-2-X communication types.}
\label{UAVComsChannel}
\end{figure}

\begin{enumerate}[label=(\roman*)]

\item \textit{UAV-2-GCS communication}: It is the fundamental type of communication for UAVs. The GCS exchanges data with UAVs through uplinks and downlinks, enabling monitoring traffic and controlling the flight mission. We consider three classes of transmitted traffic in UAV-2-GCS communications: The control traffic, the coordination traffic, and the sensing traffic~\cite{Andre2014}. The control traffic encompasses controlling and monitoring commands. In particular, mission-specific commands and the real-time status of UAVs (e.g., telemetry data, battery level, etc.). The coordination traffic handles the collaboration between multiple UAVs during the flight mission and tasks performed independently from the GCS, such as collision avoidance processes. The sensing traffic encloses onboard sensor readings that are transmitted to the GCS. We mention that all different types of traffic in the UAV-2-GCS communications are based on wireless technologies with limited range such as Bluetooth or Wi-Fi 802.11, and most of the time not secure~\cite{Hooper2016SecuringAttacks}; thus making them vulnerable to passive and active attacks.
    
\item \textit{UAV-2-Satellite communication}: In the Beyond Line-of-Sight (BLOS) missions, the operator needs to locate UAV's position for safe navigation. Therefore, UAVs can establish a satellite communication link to gather their real-time GPS location, then transmit it back to the GCS through the satellite. Furthermore, satellite communications are useful at long distances without fixed infrastructure and provide reliable communication with high transmission bandwidth. Moreover, we can leverage commercial satellite communications to control UAVs~\cite{Valavanis2015}. However, they are energy-consuming and expensive in terms of maintenance costs and they can introduce high latency issues. 

\item \textit{UAV-2-Cellular communication}: At high altitude whether in urban or rural environments, UAVs guarantee a wide coverage area and incorporate cellular networks with the coexistence of ground users to provide a reliable wireless communication~\cite{Fotouhi2019a}. In this integration, the UAVs operate either as aerial User Equipments (UEs) or as aerial Base Stations (BSs)~\cite{Hentati2020}. When they act as User Equipments, also known as \textit{cellular-connected UAVs}, they establish a UAV-2-Cellular communication with the terrestrial base station, and the ground pilot can directly control UAVs through cellular networks. Differently, UAVs as aerial Base Stations are complementary to ground base stations. They provide reliable and cost-effective wireless cellular networks to cover areas where ground base stations are not reachable. Although given the advantages of using UAVs in cellular networks in both scenarios, their real-world deployments face several challenges such as limited performance and energy-efficiency~\cite{Mozaffari2019}.  

\item \textit{UAV-2-UAV communication}: Referred as Air-to-Air communications, and takes place during flight missions that require multiple UAVs. In such scenarios, UAVs collaborate and coordinate over wireless technologies with low-power consumption (e.g., Bluetooth, Zigbee, etc.) to exchange information directly or through multi-hop wireless links. In this case, a single UAV operates within a network of UAVs to share data and accomplish the desired flight mission. However, UAV-2-UAV communications have a very low throughput and transmission bandwidth. 
\end{enumerate}

\subsubsection{UAV-2-X Communication Architecture}
UAV-2-X communications operate under a layered architecture and include the physical \& MAC layer, the network layer, and the transport layer. Unfortunately, implementing security solutions for these layers is challenging due to UAV's characteristics, such as battery life, insufficiency of resources, real-time computation, and autonomous control. This problem triggers various vulnerabilities at the communication level.

\textit{\textbf{Physical \& MAC Layer.}} The physical \& MAC layer defines the communication between the UAV and the transmission medium. In the Physical \& MAC layer of UAV-2-X communications, the UAVs utilize different wireless communication technologies such as Wi-Fi, Zigbee, and Bluetooth. 

\begin{figure}[!t]
\begin{center}
\includegraphics[width=0.47\textwidth]{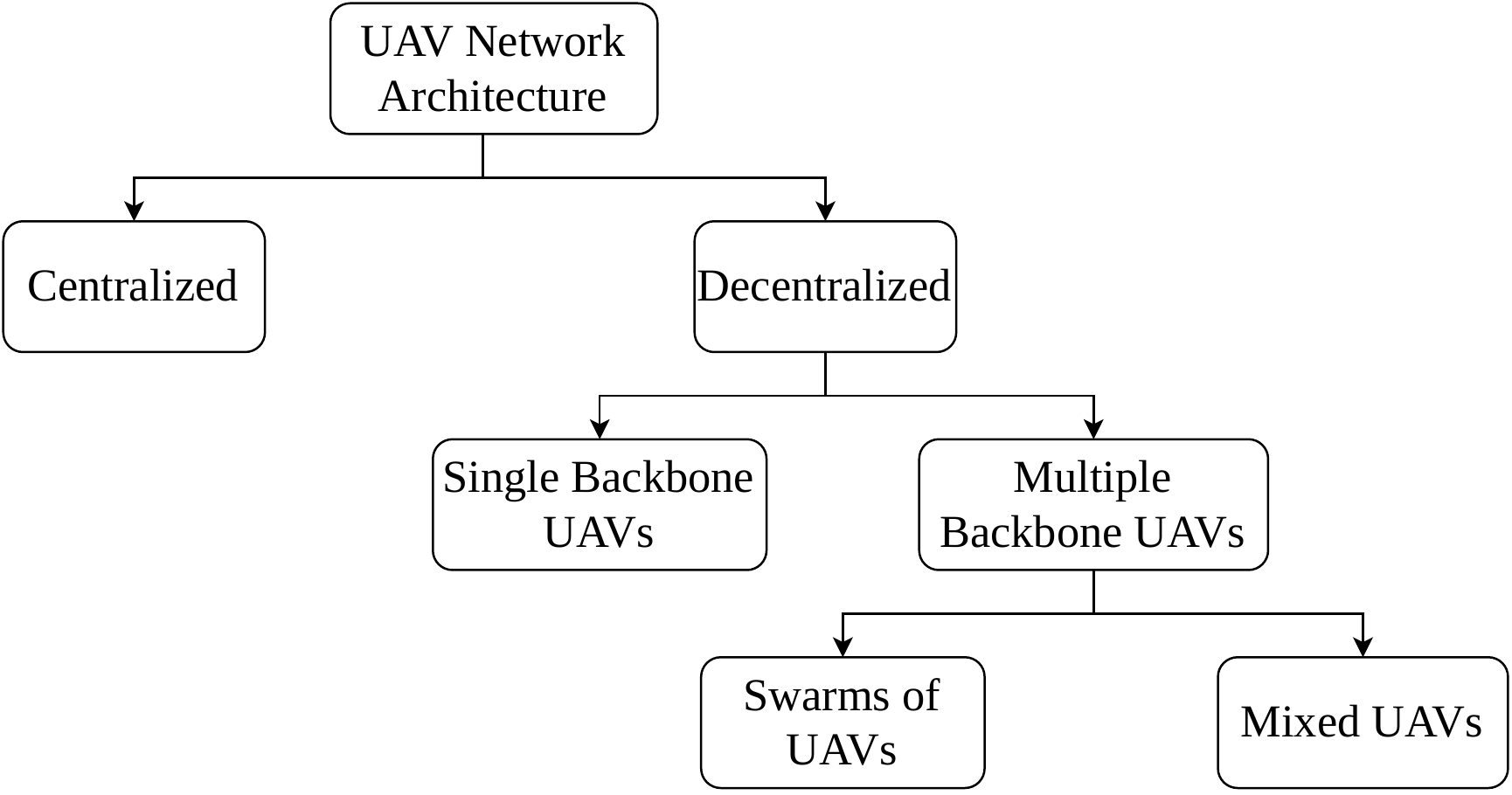}
\caption{UAV communication network architectures.}
\label{UAVNetwork}
\end{center}
\end{figure}

\textit{\textbf{Network Layer.}}
In multi-UAV systems, UAV communication networks are aerial and they are notably different from the mobile ad hoc, and vehicular ad hoc networks in terms of node mobility and topology change~\cite{Gupta2016}. The unique properties and challenges of these networks create a new category of ad hoc networks, namely flying ad hoc networks (FANETs)~\cite{Bekmezci2013}. In Multi-UAV operations, the features and the nature of FANETs make them vulnerable to various cyber attacks~\cite{Chriki2019FANET:Issues}. Indeed, challenging issues arise in multi-UAVs systems due to their very low node density, topology change, and architectural design~\cite{Gupta2016}. As shown in Figure~\ref{UAVNetwork}, we distinguish two broad categories of UAV communication network architectures: centralized architecture and decentralized architecture~\cite{Li2013}. 

In the centralized architecture, the UAVs transmit to and receive data and control commands from a single GCS that serves as a central station. The centralized architecture is applicable in small and straightforward missions. An example of this type of communication is in crowd surveillance applications in urban areas~\cite{Chriki2019UAV-GCSApplications}. In such a network architecture, any UAV-2-UAV communication must go through the GCS. This routing results in a delay in data transmission. Therefore, the centralized architecture is not suitable for long-distance communications, especially for resource-constrained UAVs.

In contrast, the decentralized architecture enables UAV-2-UAV communications without routing information to the GCS. We consider two sub-types of decentralized UAV network architectures: single backbone UAVs and multiple backbone UAVs. For both scenarios, a single UAV or multiple UAVs operate as a gateway node and transmit exchanged data to the GCS either directly or through another networking infrastructure such as cellular-based or satellite-based systems. In a single backbone UAV ad hoc network, UAVs form a connection group, and only one backbone UAV serves as a gateway between the GCS and the other UAVs. However, the single backbone UAV architecture may not be practical for flight missions that require a significant number of UAVs. In this case, we rely on two types of multiple backbone UAV architecture. In the first type, multiple groups of UAVs in a collective behavior form a swarm, such that each group consists of a single backbone UAV architecture. The latter one consists of grouping all single backbone UAVs of all groups. Each group can transmit data to the other group without being routed through the GCS, and only one backbone UAV exchanges data with the GCS.       
We note that all the abovementioned UAV network architectures have their strengths and limitations regarding communication needs, autonomy, and scalability. Therefore, the appropriate type of architecture to deploy depends on the flight mission requirements. For example, in Search And Rescue (SAR) missions where time is crucial, the decentralized architecture is more efficient than centralized ones due to the collaboration and coordination between multiple UAVs.

In multi-UAVs networks, routing protocols are essential to provide a reliable end-to-end data transmission between UAV nodes~\cite{Arafat2019a}. Several routing protocols have been proposed in the literature with different classifications~\cite{Oubbati2019,Arafat2019a,ShumeyeLakew2020}. One approach classifies these protocols either on the network architecture or data forwarding~\cite{Arafat2019}. Another approach suggests classifying UAV routing protocols according to their design constraints such as dynamic topology, energy consumption, scalability, security, and allocated bandwidth~\cite{Oubbati2019}. However, given UAV's unique characteristics, all these protocols cannot fulfill UAV's security requirements. 

\textit{\textbf{Transport Layer.}} The transport layer provides reliable data transfer between end-to-end components. Two well known examples of UAV communication protocols at the transport layer include the MAVLink protocol and the UranusLink protocol.

\textit{MAVLink Protocol.} The Micro Air Vehicle Link (MAVLink) protocol is a lightweight point-to-point networking protocol primarily used in UAV-2-GCS communications to exchange control and telemetry data~\cite{Koubaa2019}. It uses bidirectional communication between UAVs and GCS over wireless channels for real-time applications. Its transmissions can be performed through different wireless mediums such as Wi-Fi and Bluetooth, with sub-GHz frequencies. MAVLink protocol comes in two versions: v1.0 and v2.0. \textit{MAVLink v2.0} is currently the recommended one. It is a backward-compatible and improved version compared to \textit{MAVLink v1.0}. \textit{MAVLink v2.0} protocol header contains new features and adds new fields to the existing structure of MAVLink messages, such as message extensions and packet-signing. Commercial UAVs extensively use \textit{MAVLink v2.0} since it provides reliable communication and packet-signing. 
However, only a few studies addressed security implementations of the MAVLink communication protocol. Therefore, MAVLink protocol is prone to several attacks such as flooding and packet injection~\cite{Kwon2018}.

\textit{UranusLink Protocol.} UranusLink is a packet-oriented protocol for wireless UAV-2-GCS communications~\cite{Kriz2015}. Its design satisfies radio communication requirements such as data throughput and low latency, making it useful for aerospace and robotic applications. UranusLink operates in a half-duplex mode under 2.4 GHz frequency and with a maximal throughput of 250 kbps. It is suitable for UAVs with small overhead. Although UranusLink employs an integrity protection scheme, it does not encrypt message payloads that can result in replay attacks~\cite{Khan2020}.  

\subsection{Sensing Technology}


UAVs possess a wide range of sensors to accomplish their flight missions. These sensors represent critical components for the functionality of the UAV system, and they are designed to measure physical quantities of the surrounding environment such as altitude, speed, and GPS location. The outputs of these quantities are then directly transferred to the Flight Controller to decide the appropriate actuation/action. In Table~\ref{Sensors}, we present the well-known sensors of most of the commercial UAVs. We mention that for each type of UAV application, there exists a corresponding set of onboard sensors. It is worth mentioning that the Flight Controller cannot distinguish between legitimate or malicious sensor inputs, even with the robust design of UAV sensors.

\begin{table}[h]
\centering
\caption{Sensors of Unmanned Aerial Vehicles \label{Sensors}}
\begin{tabular}{|p{1.55cm}|p{6.4cm}|}
\hline
\rowcolor{grey}
\textbf{Components} & \textbf{Functionality}\\ \hline
GPS & Many UAVs use Global Positioning System in outdoor applications to determine geospatial references from the satellite within its range.\\ \hline
3D \newline Accelerometers & Three accelerometer sensors used to provide the non-gravitational acceleration of UAVs for each axis X, Y, and Z. They rely on the piezoelectric effect and handle the hover capability of UAVs.\\ \hline
3D gyroscopes & 3D Gyroscopes can measure or maintain orientation and angular velocity in pitch, roll, and yaw. They are essential for navigation and provides orientation stability of UAVs. Moreover, they collaborate with 3D accelerometers to handle rotational and linear movements.\\ \hline
Magnetometers & Magnetometers provide additional geographical direction of UAVs using the magnetic field. However, these sensors might be defective when placed together with motors and electrical devices.\\ \hline
Infrared \newline cameras & Also known as thermographic cameras, they provide detailed images using infrared energy of objects even in the darkness. Mainly used in military UAV applications. This type of a camera could potentially spy on people in challenging environments (e.g., forest, private houses). \\ \hline
Gas sensors & Gas sensors can detect different gasses such as toxic or explosive gasses and measure their concentrations. They have many industrial and military applications.\\ \hline
Radiation sensors & Very useful in nuclear industries. UAVs can be equipped with radiation sensors to determine radiation levels and provide gamma radiation readings for large areas.\\ \hline
Cameras & Crucial device of UAVs. A wide range of cameras for UAVs exists with different types and sizes. With many civilian and military applications, they can capture images and record videos. Moreover, they help the pilot to navigate in indoor missions. However, the zoom function of these cameras triggers privacy challenges.\\ \hline
Microphones & Practical for search and rescue operations or spying missions, microphones can record audio and gather information remotely. However, using microphones can violate personal privacy.\\ \hline
Biosensors & Biosensors are electrochemical sensing technologies mainly used to detect airborne biological hazards.\\ \hline
Pressure \newline sensors & Pressure sensors aim to detect the atmospheric pressure and convert it into altitude. They provide UAVs altitude stabilization.\\ \hline
LiDAR \newline sensors & Light detection, and ranging sensors provide a high-resolution map with laser light. They have several applications such as archeology, agriculture, and landscaping.\\ \hline
\end{tabular}
\end{table}

\subsection{Security and Privacy Requirements}

The wide use of UAVs in civilian applications raise a large amount of 
vulnerabilities~\cite{Krishna2017}. To that end, different features are essential to protect UAVs from disclosure, disruption, modification, and destruction~\cite{Altawy2017}. To guarantee the these properties, we identify the following major security and privacy requirements needed to establish a secure UAV flight mission.
\begin{itemize}
    \item \textbf{Confidentiality.} It is crucial to protect private information and data exchange between UAVs and the GCS from unauthorized access, as it could be a source of sensitive information leakage of the flight mission such as telemetry data and control commands. To prevent the adversary from obtaining such information, we need to consider implementing robust cryptographic solutions.
    \item \textbf{Integrity.} Preserving data integrity is of utmost importance. It is a requirement for the success of a flight mission and it prevents adversaries from forging the network traffic. Compromising the integrity could change the behavior of the UAV system and lead to a mission failure. Hence, any communication has to be protected and verified. We can guarantee this requirement through authenticated encryption algorithms~\cite{Altawy2017}.
    \item \textbf{Availability.} UAVs must be operational without intentional or unintentional interruptions. All the resources needed for a flight mission have to be available for authorized users. Moreover, it is required from the UAV system to resist classical Denial of Service (DoS) attacks, which are compromising its availability. Such attacks can be mitigated using IDS~\cite{Choudhary2018}. 
    \item \textbf{Authenticity.} The authentication process is a fundamental step towards establishing secure communication between different components of the UAV system. It allows verifying the authenticity and identity of UAVs participating in the flight mission. We ensure the trustworthiness of each UAV through authentication, and only authenticated UAVs can participate in the flight mission. Moreover, the authentication protects the UAV network from adversaries that are spoofing the legitimate nodes.
    \item \textbf{Non-Repudiation.} The users cannot deny their actions (e.g., transmitting or receiving data) within UAV networks. Otherwise, we may deal with accountability issues in case of a mission failure. This property prevents the denial of the user's operations. Furthermore, the UAV system has to develop proper mechanisms ensuring non-repudiation, such as the digital signature of the exchanged messages. 
    \item \textbf{Authorization.} Data exchange in the UAV system has to be shared only with authorized users. We note that unauthorized users are not allowed to perform any action in the UAV network. Besides, the UAV system has to specify what resources an authorized user can access. Granting access to such resources has to be monitored through access control policies.
    \item \textbf{Non-disclosure.} In addition to the abovementioned security requirements, we consider the non-disclosure property in the privacy requirements for UAV systems. Indeed, sensitive information exchanged between the GCS and the UAV, such as captured images and video footage, should not be disclosed to a third party~\cite{Koubaa2019}.  
\end{itemize}

\section{Security Issues of UAVs}\label{sec:security-issues}
\begin{figure}[h]
\begin{center}
\includegraphics[width=0.47\textwidth]{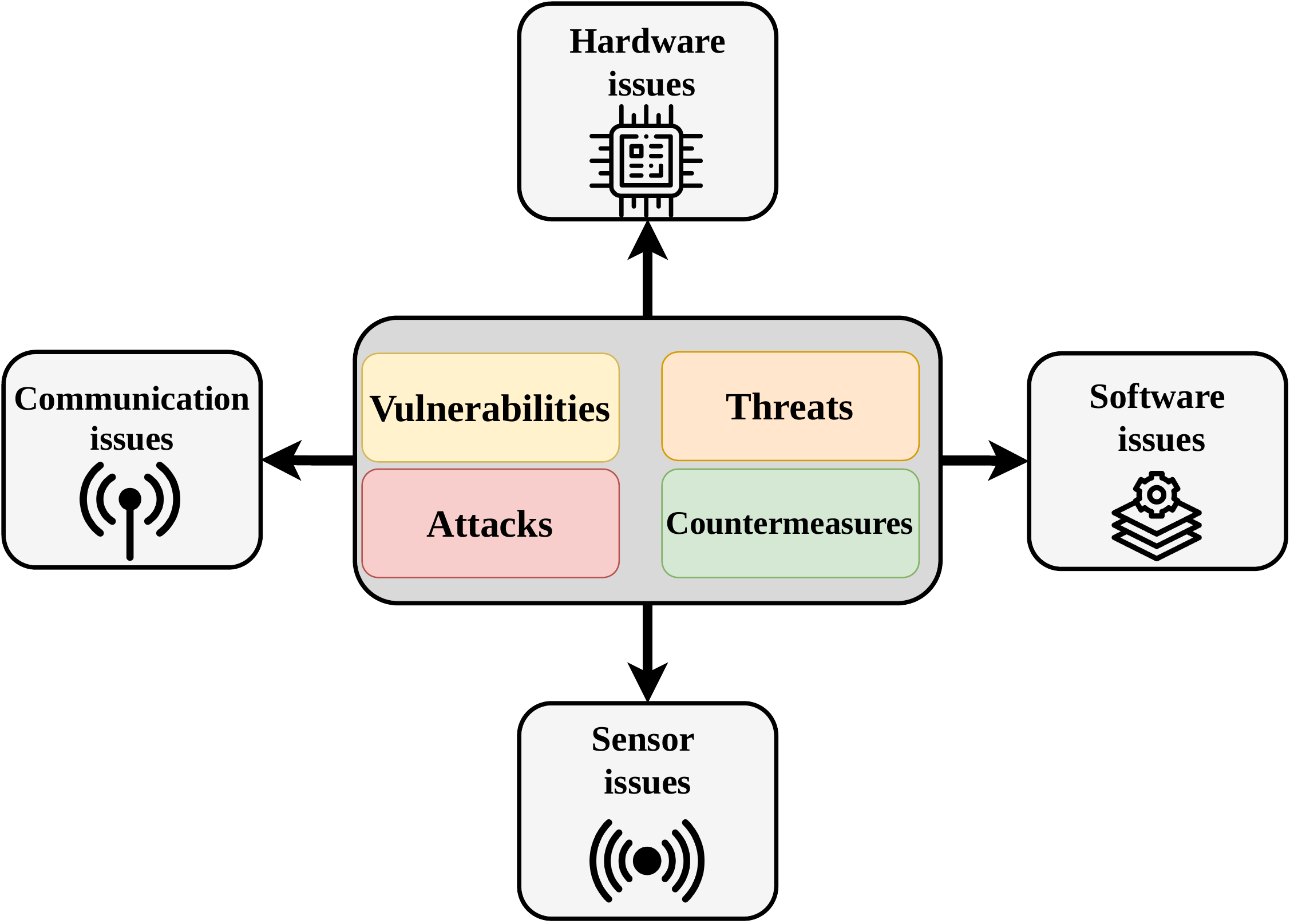}
\caption{Taxonomy of UAV security issues.}
\end{center}
\label{securityIssues}
\end{figure}

Security issues associated with UAVs in the national airspace greatly increase the likelihood of performing passive and active attacks. In this section, we categorize the security issues of UAVs into four different levels: \textit{Sensor-level}, \textit{Hardware-level}, \textit{Software-level}, and \textit{Communication-level}. As shown in Figure~\ref{securityIssues}, we provide a detailed overview about the threats and vulnerabilities targeting UAVs for each level. Then, we review the attacks and their existing countermeasures.

\subsection{Sensor-Level Issues} 

UAVs rely on sensors to gather data about the surrounding environment. These data are sensitive and need to be protected from malicious actors. Under adversarial conditions, compromising UAV sensors might cause the UAV system to fail. In what follows, we provide different sensor-level vulnerabilities, threats, and potential attacks against UAVs. Afterward, we highlight existing countermeasures against  sensor-based attacks on UAVs.    

\subsubsection{Sensor Vulnerabilities and Threats}
UAVs are extremely sensor-driven devices. They are equipped with a variety of sensors such as cameras, GPS, and accelerometers. Therefore, they rely on sensor readings to operate efficiently. However, these sensors handle sensitive information and could be used by a malicious operator to compromise the flight mission. For example, civil GPS signals are unencrypted and unauthenticated. Therefore, an adversary can exploit this vulnerability by simulating a GPS signal to delude the operator. From an attacker's perspective, exploiting the onboard sensors' real-time data may cause the UAV system to malfunction. This exploit could happen because the Flight Controller does not evaluate the authenticity of sensor readings. The introduction of sensor vulnerabilities into the UAV system can also be performed through malicious software. Due to the practicality of sensory-channel attacks in real-world scenarios, this class of vulnerabilities exposes a new attack vector for the adversary to fully control commercial UAVs~\cite{Uluagac2014SensoryCall, Sikder2021AApplications}.

\subsubsection{Sensor-based Attacks}
Sensor-based attacks include GPS data jamming, false sensor data injection, and sensory-channel attacks.

\textit{GPS data jamming.} During a flight mission, the onboard GPS receiver gathers its GPS location from the satellite and sends it to the GCS. GPS data jamming attack occurs when the adversary blocks the navigation feed of the GPS signals, forcing the UAV into a disoriented mode~\cite{AruSaputro2020ImplementationTest}. Performing such attacks results in losing control of the UAV, and therefore possible hijacking of the drones.

\textit{False sensor data injection.} Injecting false sensor data readings in the Flight Controller can compromise external sensors such as electro-optical and infrared sensors~\cite{mo2010}. This attack leads to thwart the UAV's stabilization. An attacker can inject false sensor data into UAVs by accessing the onboard Flight Controller system or by altering the sensor readings through system calls. Otherwise, he can directly transmit fake signals to the sensors, and therefore compromising the flying UAV. A well-known example of false sensor data injection attacks is GPS spoofing. Since GPS signal broadcasts are most of the time unencrypted and unauthenticated, the attacker performs a spoofing attack on the GPS by faking the generated signal, which can eventually alter the UAV's GPS receiver~\cite{Wesson2013HackingDrones}. Consequently, the attacker gains control over the UAV. In~\cite{Kerns2014UnmannedSpoofing,Seo2015EffectSignal}, the authors demonstrate a GPS spoofing attack on UAVs. The GPS spoofing attack forces the drone to reply to fake signals, consequently affecting its navigation system.

\begin{table*}[h]
\centering
\caption{Summary of sensor-level security issues, existing countermeasures and their limitations \label{SumSens}}
\begin{tabular}{l|p{6cm}|p{6cm}}
\hline
\textbf{Sensor-based attacks/threats} & \textbf{Countermeasures} & \textbf{Limitations} \\ \hline
\rowcolor{grey}
Sensory channel attacks \cite{Uluagac2014SensoryCall} &  
-Physical isolation for acoustic sensory channels to shield the sound noise \cite{Roth2009SimulationGyroscopes}.

-Building robust optical flow algorithms for optical flow sensors \cite{Fischler1981RandomCartography}.
 & -A large number of sensory channels to consider. \\ \hline

GPS data spoofing \cite{Kerns2014UnmannedSpoofing, Seo2015EffectSignal} & 

-Implementing anti-GPS-spoofing methods into the Flight Controller \cite{Feng2020, Vaeshosaz2019, Feng2019AnUnit}.

-The use of collaborative data attestation approach that verifies the correctness of GPS coordinates \cite{Abera2019DIAT:Systems}.

-The adoption of authenticated schemes for GPS signals.

-Detection of unusual signal power changes that indicate the beginning of a spoofing attack.
 & 
-Authenticated GPS signals require additional changes in the infrastructure of the satellite. 

\\ \hline

\rowcolor{grey}
GPS data jamming \cite{AruSaputro2020ImplementationTest} & 

-Enabling the autonomous navigation without GPS signal \cite{Wu2013AutonomousSensors}.

-The use of additional sensors for alternative navigation \cite{Wu2013AutonomousSensorsb}.

-Adopting machine learning-based IDS to detect sensor-based attacks \cite{Whelan2020} \cite{Arthur2019DetectingIDS}.  & 

-Limited energy and computation costs for realistic implementations. \\ \hline

False sensor data injection  \cite{mo2010} & 

-Modeling UAV's physical properties \cite{Choi2018DetectingApproach}. 

-Securing sensor readings in the presence of physical invariants \cite{Quinonez2020SAVIOR:Invariants}.

-Cross-verification of data by gathering sensor readings from an alternative set of sensors.
 & -Adopting the existing solutions to other types of on-board sensors is still unknown.\\ \hline

\rowcolor{grey}
MEMS gyroscopes attacks \cite{Son2015} & -Physical isolation for acoustic sensory channels to shield the sound noise \cite{Roth2009SimulationGyroscopes}. & 
-The physical isolation could increase the temperature and cause a malfunctioning of the UAVs.\\ \hline

Optical flow camera sensor attack \cite{Davidson2016} & -Building robust optical flow algorithms for optical flow sensors \cite{Fischler1981RandomCartography}. & -Practical limits of the optical flow estimation due to its inherent noisy nature. \\ \hline

\end{tabular}

\end{table*}

\textit{Sensory-channel attacks.} UAVs use a set of sensors in which their sensory channels (e.g., infrared, acoustic, light, etc.) serve as a vector for attacks. In~\cite{Son2015}, the authors demonstrate that UAVs equipped with Micro-Electro-Mechanical Systems (MEMS) gyroscopes can fail using intentional sound noise. The study shows that MEMS gyroscopes resonate at audible frequencies. Another study has shown that optical flow camera sensors which are used to stabilize UAVs can be compromised by influencing the surrounding environment~\cite{Davidson2016}.

\subsubsection{Countermeasures for Sensor-based Attacks}
To mitigate GPS jamming attacks, the authors in~\cite{Wu2013AutonomousSensors} provided a solution enabling autonomous navigation when the Flight Controller does not receive GPS signals. Other approaches rely on ML-based IDS to detect known and unknown sensor-based attacks~\cite{Whelan2020}~\cite{Arthur2019DetectingIDS}. These solutions collect training datasets from onboard components of the UAVs, such as flight logs and sensors readings. However, real-world implementation is challenging due to the limited energy and computation resources of the UAVs. In another work, Wu et al.~\cite{Wu2013AutonomousSensorsb} proposed the use of additional sensors as an alternative navigation solution when GPS signals are not available. The authors used a monocular camera visual sensor combined with an Inertial Measurement Unit (IMU) sensor to enable the autonomous flight of UAVs in a loss-of-GPS scenario. 

To prevent injecting falsified Flight Controller sensors data, we can cross-verify the data by gathering readings from an alternative set of sensors. Another solution to detect external sensor attacks is by modeling UAV's physical properties through a control invariant approach~\cite{Choi2018DetectingApproach}. The control invariant approach checks the consistency of the UAV's physical state with its expected state, which is identified by its control model. Similarly, in~\cite{Quinonez2020SAVIOR:Invariants}, the authors presented an architecture to secure sensor readings in the presence of physical invariants. Physical invariants of the UAVs are unique features that can be modeled to predict sensor measurements according to their behavior. These features consist of nonlinear differential equations that model UAVs' speed, angles, position, and angular speed. The study shows that the use of well-known physical invariants provides learning of their parameters and enables the detection of sensor-based stealthy attacks.

Preventing the adversary from performing a GPS spoofing attack could be achieved by detecting unusual signal power changes, which indicates the beginning of a spoofing attack. In Multi-UAVs scenarios, the authors in~\cite{Abera2019DIAT:Systems} proposed a collaborative data attestation approach that verifies the correctness of shared information such as GPS coordinates, thus the detection of GPS spoofing attacks. Another countermeasure against GPS spoofing attacks is by adopting GPS signal authentication schemes with classical cryptographic approaches. However, the implementation of such solutions requires additional changes in the infrastructure of the satellite~\cite{Altawy2017}. We also note that some anti-GPS-spoofing methods are suitable to be implemented into the Flight Controller, enabling an efficient hijacking detection solution~\cite{Feng2020, Vaeshosaz2019, Feng2019AnUnit}. 
 
A set of countermeasures have been proposed in the literature to mitigate each type of sensory-channel attack. Acoustic sensory channels are protected by the physical isolation that can shield the sound noise~\cite{Son2015}. Optical flow sensors rely on optical flow algorithms, which are utilized to measure visual motion. Building robust optical flow algorithms such as the RANSAC algorithm~\cite{Fischler1981RandomCartography} constitutes a defense-in-depth mechanism for spoofing optical flow sensors. 

The attacker's capabilities to compromise UAV sensors are outlined in Table~\ref{SumSens}. The reported sensor-based attacks aim to compromise the sensory channel, GPS signals, and also inject false sensor data. The solutions proposed in the existing literature are specific for each type of sensor. For example, implementing anti-GPS-spoofing methods or using a collaborative data attestation approach to verify the correctness of GPS coordinates helps prevent GPS data spoofing. The cross-verification of data by gathering sensor readings from different sensors protects the UAVs from gathering false sensor data. However, we also need to consider that the proposed countermeasures for sensor-based attacks have primary shortcomings. For instance, realistic implementations to prevent GPS jamming attacks will increase the computation costs. Moreover, given many sensory channels, providing a set of alternative sensors for each sensory channel is not efficient.


\subsection{Hardware-Level Issues}
The adversaries consider UAVs as a potential means to conduct physical attacks in the national airspace. The hardware components of a UAV system consist of the onboard Flight Controller (FC) and the Ground Control Station (GCS). Both hardware devices are subject to security issues that can potentially lead to cyber or physical attacks. We devote this subsection to present the vulnerabilities that an adversary can exploit to compromise the Hardware-level of a UAV system. Then, we provide existing defense mechanisms to mitigate hardware-based attacks. 

\subsubsection{Hardware Vulnerabilities and Threats} Hardware-level vulnerabilities and threats include hardware trojans, physical UAV collision, hardware failures, and flying skills issues.

\textit{Hardware trojans.} Hardware trojans involve the modifications of the electronic hardware (e.g., tampering with the hardware circuit, resizing the logic gate, etc.)~\cite{Vosatka2017IntroductionTrojans}. In particular, hardware trojans target the Flight Controller, making the UAV system vulnerable to several attacks. The hardware trojans are maliciously embedded by a non-trusted third party in the semiconductor supply chain of the Flight Controller~\cite{Rahman2020IntrusionVehicles}. The adversary leverages these modifications to compromise the functionalities and security features of the FC's Integrated Circuit (IC) (e.g., decreasing the rotation speed of the propellers, leaking the cryptographic keys of the Flight Controller, etc.). An example of a trojan was found in the Actel ProASIC chip of the Boeing 787 jet~\cite{GilCasals2013GenericDetection}. The backdoor allowed the attacker to monitor the avionics system and control the aircraft, therefore jeopardizing the safety of onboard passengers.

\textit{Physical UAV collision.} During a flight mission that requires the cooperation and collaboration between multiple UAVs, physical collisions could happen, resulting in crashing the drones. To prevent such collisions in the civilian airspace, the UAVs rely heavily on Collision Avoidance Systems (CAS)~\cite{Collision}. However, these systems do not encompass built-in security features and cannot satisfy the collision avoidance threat caused by malicious actors~\cite{Hannah2020}.

\textit{Hardware failures.} UAVs can go through malfunctioning of their hardware components, such as battery life or motor issues. These technical failures constitute a threat to the flight mission and could lead to an unsafe landing of the UAVs in an unexpected location~\cite{Alwateer2019DroneProcessing}. In this case, if the UAVs store unencrypted data, the adversary can disclose sensitive mission-related information and violate the flight mission's confidentiality.  

\textit{Flying skills issues.} These issues occur when human operators remotely control non-autonomous or semi-autonomous UAVs, especially those that are very sensitive under wind disturbance due to their complex dynamics and size~\cite{Lee2020}. They require flying skills such as remote control of the speed, height, and orientation of the UAV. In such scenarios, the operator's lack of these technical skills might crash the drone and cause an operational failure. Consequently, the UAVs can be easily exposed to physical theft.  

\subsubsection{Hardware-based Attacks}
Hardware-based attacks include hijacking, supply chain attacks, battery attacks, and radio frequency module attacks.

\textit{Hijacking.} Due to the nature of UAVs, they are visible at a low altitude, making them the perfect targets for hijacking. The adversary hijacks a flying drone either directly or remotely through malicious software. The straightforward technique to disable and hijack UAVs is by using the anti-drone rifles~\cite{hodgkins}. They are usually in possession by law enforcement to protect malicious UAVs hovering in restricted flight areas. Nevertheless, the attacker can also use the same rifle to ground the drones and hijack them.

\textit{Supply chain attacks.} With the drone industry's growth, the adversaries have a wider window compromising the UAVs through supply chain attacks. This type of attack consists of exploiting the vulnerabilities in the supply chain process of an organization by targeting the less-secure and sensitive components such as the propellers, airframes, and actuators. Consequently, the end product that is delivered to the customer is already compromised. A practical supply chain attack against UAVs is demonstrated by Belikovetsky et al.~\cite{BelikovetskyDr0wned-Cyber-PhysicalManufacturing}. The researchers carried out a physical supply chain attack for UAVs with Additive Manufacturing (AM). The attack consists of sabotaging a given UAV by remotely manipulating the design files of the propellers. The adversary reduces the 3D printed propeller's fatigue life and creates delayed damage during a flight mission. This study shows that sabotage attack detection for additive manufacturing systems remains a challenging research problem.

\textit{Battery attacks.} Prevalent UAVs are powered with Lithium-Ion rechargeable batteries. These batteries are supported by the Battery Management System (BMS) to provide reliable energy to different components of the UAV system. However, an adversary can exhaust the battery's energy by performing potential battery depletion attacks~\cite{Desnitsky2021}, which results in a malfunctioning of the UAV system, and consequently compromising the availability, integrity, and confidentiality of the batteries~\cite{lopez2017}. The attacker compromises the availability of UAV batteries by physically tampering or swapping legitimate batteries with faulty ones to fail the UAV system. Another possible attack may occur when the adversary generates a deep discharging of the batteries. This type of attack could happen by compromising other components of UAVs, such as spoofing the sensors or injecting malicious software, leading to exhausting the UAV batteries~\cite{lopez2017}. Attacking the integrity of UAV batteries includes modifying real battery information to the operator through the UAV-2-GCS data transmission. Furthermore, the confidentiality of UAV batteries can be compromised by leaking sensitive battery-related data such as the State-of-Charge (SoC), which represents the ratio of available charge to the UAV battery capacity.

\textit{Radio Frequency modules attacks.} Radio Frequency modules (RF) are used to transmit and receive radio signals from two different devices. In the context of UAVs, an operator might use a typical remote controller or the GCS to send control signals to the flying drones. In this case, the adversary can jam the control signals and disable the UAV-2-GCS communication, resulting in the drones' lost-link state. In~\cite{Rodday2016HackingDrone}, the authors demonstrated a replay attack on the XBee 868LP protocol, a low power radio frequency module used for UAV-2-GCS communications. In this attack, the adversary alters the UAV-2-GCS communication using a third XBee chip. In particular, the attacker compromises the security of the communication channel of the main XBee by combining existing features of the chip to access the address of the XBee communication channel.

\subsubsection{Countermeasures for Hardware-based Attacks}
Given physical vulnerabilities and threats of UAVs, physical protection approaches should be considered and enhanced to address those threats. To guarantee a trojan-free drone, possible mitigation of hardware trojans consists of building ML-based IDSs to detect such hardware attacks~\cite{Rahman2020IntrusionVehicles}. Detecting the presence of tampered data or commands using IDS solutions is achieved by: (1) learning the model based on the average data generated by the Pulse Width Modulation (PWM) signals. These signals are commonly used in the IC of UAVs. (2) training the model with malicious data. These data are generated by compromising the firmware or injecting hardware trojans. Thus, affecting the PWM signals. Another mitigation technique consists of performing a fine-grained circuit analysis to enable the detection of hardware trojans~\cite{Nigh2021AdaTrust:Construction}.

Securing both the GCS and UAVs from illegal access using authenticated encryption, and keeping them malware-free will significantly prevent malicious actors from taking over and hijacking the flying UAV. Further, changing the flight paths could prevent the adversary from identifying the flight pattern, thus making the target more difficult for physical theft. In~\cite{McNeely2016DetectionStatistics}, the authors proposed a hijacking detection method for UAVs based on a statistical analysis of standard flight patterns. The simulation of different hijacking scenarios shows the effectiveness of their detection algorithm. However, their algorithm fails when simulation parameters such as control instability are changed, which motivates further testing and improvement of the quality of the simulation data.      

Supply chain attacks can be mitigated by managing the supply chain's security during the manufacturing process to avoid using compromised UAV components~\cite{Williams2008SupplyAgenda,Altawy2017}. Besides, tamper-proofing solutions (e.g., tamper-proof microprocessors, Anti-tamper software, etc.) will disable unauthorized physical or logical modifications that could sabotage the authenticity of the UAV's critical components. 

Existing countermeasures to mitigate battery depletion attacks include the use of safety circuits in the Battery Management System that ensures physical battery protection for UAVs~\cite{lopez2017}. Moreover, a pre-flight diagnosis of the UAV batteries would be an equitable procedure to guarantee a safe flight mission. Another solution could also detect depletion attacks during the flight mission, which consists of monitoring the battery discharging process in real-time~\cite{Desnitsky2021}. However, if the UAV-2-GCS data transmission is unauthenticated, the adversary may counterfeit the transmission and display an incorrect battery level to the operator. Therefore, we need to adopt cryptographic solutions to secure the UAV-2-GCS data transmission. Further, we can leverage the use of ML techniques to detect UAV battery depletion attacks automatically. This can be achieved using the features extracted from simulated battery depletion attacks~\cite{Desnitsky2021}. 

To mitigate the radio frequency modules attacks, the manufacturer can adopt the onboard encryption of the chip. However, this solution remains limited because it decreases the bandwidth and increases the latency of the chip. In this case, the authors in~\cite{Rodday2016HackingDrone} suggested possible outsourcing of the encryption to a second separate chip. Although this remediation guarantees the confidentiality of the data sent over the radio channel, it would not prevent the adversary from executing remote commands since they are sent directly to the chip. Therefore, the adversary can perform a DoS attack by setting random values to destination addresses. Another approach considers encrypting the Radio Control (RC) link. In~\cite{Podhradsky2017ImprovingLink}, the author implemented an encrypted RC link based on Galois Embedded Crypto (GEC) library~\cite{GitHubSMACCM}, which is compatible with resource-constrained devices. The proposed design enables secure communication between the UAV and the RC transmitter. 

\begin{table*}[!ht]
\centering
\caption{Summary of hardware-level security issues, existing countermeasures and their limitations \label{SumHard}}
\begin{tabular}{l|p{6cm}|p{6cm}}
\hline
\textbf{Hardware-based attacks/threats} & \textbf{Countermeasures} & \textbf{Limitations} \\ \hline
\rowcolor{grey}
Hardware trojans \cite{Vosatka2017IntroductionTrojans} & -Building ML-based IDSs to detect hardware trojans~\cite{Rahman2020IntrusionVehicles}. \newline
-Performing a fine-grained circuit analysis \cite{Nigh2021AdaTrust:Construction}.
& -Hardware obfuscation techniques can bypass the existing detection methods. \\ \hline

Physical collisions \cite{He2021AnAttack} & -The development of Collision Avoidance Systems~\cite{Collision}. & -Collision Avoidance Systems do not implement security features. \\ \hline

\rowcolor{grey}
Hardware failures \cite{Alwateer2019DroneProcessing} & -Adopting encryption techniques on the flying UAVs prevent the adversary from capturing the stored data in the case of hardware failures~\cite{Shafique2021a}. & -Data encryption might prevent forensics analysts from recovering evidence about the hardware failures. \\ \hline

Hijacking \cite{hodgkins} & -Secure the GCS and UAVs from unauthorized access using authenticated encryption \cite{Pu2020}.

-Consistent change of the flight path to avoid the adversary from identifying the flight pattern \cite{McNeely2016DetectionStatistics}. & -The use of counter-drone technology from malicious users to hijack legitimate UAVs.  \\ \hline

\rowcolor{grey}
Supply chain attacks \cite{BelikovetskyDr0wned-Cyber-PhysicalManufacturing} & -Managing the supply chain's security during the manufacturing process~\cite{Williams2008SupplyAgenda}. 

-Adopting tamper-protected devices \cite{Paul2008TamperDevices}. & -Internal attacks during the manufacturing process. \\ \hline

Battery depletion attacks \cite{Desnitsky2021} & -The use of safety circuits in the Battery Management System \cite{lopez2017}.

-Pre-flight diagnosis of the UAV batteries.

-Monitoring the real-time battery discharging process \cite{Desnitsky2021}.& -For unauthenticated communications, the adversary can display incorrect battery levels to the operator. \\ \hline

Attacks on Radio Frequency Modules \cite{Rodday2016HackingDrone}&

-Encryption of the  Radio Control (RC) link \cite{Podhradsky2017ImprovingLink}.

-Onboard encryption of the Flight Controller.

& 
-Onboard encryption decreases the bandwidth and increases the latency of the chip.

\\ \hline

\end{tabular}
\end{table*}

Hardware-level security issues, their countermeasures, and limitations are summarized in Table~\ref{SumHard}. As outlined in Table~\ref{SumHard}, the existing attacks against UAVs on the Hardware-level include the supply chain attacks, the battery depletion attacks, the use of hijacking techniques, and attacks on Radio Frequency Modules. The security measures proposed by the research community consist of developing defense mechanisms at the Hardware-level. For instance, managing the supply chain's security during the manufacturing process, performing a fine-grained circuit analysis, and using safety circuits in the Battery Management System. Although the existing countermeasures aim to protect UAVs from hardware-based attacks, there are still limitations that need to be considered. For example, the hardware obfuscation techniques can hinder the fine-grained circuit analysis; the onboard encryption on the Radio Frequency Modules decreases the bandwidth and increases the latency of the chip. Furthermore, the development of Collision Avoidance Systems does not consider security implementations. 


 \subsection{Software-Level Issues}
Having discussed the Hardware-level issues, we introduce the Software-level issues by presenting the vulnerabilities, threats, and attacks targeting the software layer of UAVs. Afterward, we provide existing defense mechanisms to protect against such attacks. 

\subsubsection{Software Vulnerabilities and Threats}
Software-level vulnerabilities and threats on UAVs consists of malicious software and zero-day vulnerabilities.

\textit{Malicious software.} The Ground Control Station and the Flight Controller are prone to malicious software. The threats posed by UAV malware can lead to the loss of sensitive data and control of the operated UAV system. The accessibility of an attacker to the UAV's flight stack could potentially lead him to shutdown the UAV system, which results in a denial-of-service and consequently disrupts the flight mission. Embedding such malware into UAVs can significantly compromise their security and privacy. For instance, Maldrone is a virus infecting the Flight Controller, enabling the attacker to control the UAV~\cite{PaganiniPierluigi2015AAffairs}. It behaves as a proxy for the drone's Flight Controller and sensor communications, thus making the compromised drone land at any chosen location. SkyJack is a hijacking malware that can be implanted on a malicious drone~\cite{skyjack}. It can wirelessly take over other legitimate drones through the Wi-Fi de-authentication attack and compromise the whole system. Snoopy is a spyware that can be equipped on a drone with the ability to steal personal information from public users~\cite{ThisPhone}. It uses impersonation techniques to trick the users into joining a fake Wi-Fi network. Afterward, Snoopy tracks its users and harvests their personal information. Recently, there has been an emerging type of malware that consists of encrypting a user's data or locking the system, and holding it encrypted or locked until the user pays a ransom for the adversary. This type of malware is known as ransomware~\cite{Oz2021ASolutions}. To the best of our knowledge, ransomware attacks have not targeted UAVs yet. However, it is essential to consider that future ransomware might target UAVs, given their popularity and civilian applications.

\textit{Zero-day vulnerabilities.} Unknown vulnerabilities may exist in the UAV's flight stack or the GCS software (e.g., buffer overflow, DoS, etc.). These vulnerabilities are unknown to the UAV's manufacturers and can present critical threats to the operators. The adversaries can continuously exploit zero-day vulnerabilities until the UAV's manufacturers release appropriate patches. However, the operators need to update their UAV systems for every patch released.  

\subsubsection{Software-based Attacks}
Software-based attacks to UAVs include operating system attacks, tampering captured videos, and system ID spoofing.

\textit{Operating systems attacks.} Potential attacks against civilian or military missions could happen through the Flight Controller's system software. As a result, the compromised system software will lead to the loss of the UAVs and their payloads. Parcel-copters of the Prime Air service developed by Amazon is an example of the civilian applications that can be subject to operating system attacks~\cite{amazon}. Attacking the delivery system can potentially bring down the delivery package for the recipient and consequently crash the drone. Attacking UAV operating systems consists of remotely injecting malicious software to UAVs such as Maldrone~\cite{PaganiniPierluigi2015AAffairs}, then hijacking the drone by taking control of the system. To that end, the adversary can extract the FC's cryptographic key and steal the stored unencrypted data.   

\begin{table*}[!h]
\centering
\caption{Summary of Software-level security issues, existing countermeasures and their limitations \label{SumSoft}}
\begin{tabular}{l|p{6cm}|p{6cm}}
\hline
\textbf{Software-based attacks/threats} & \textbf{Countermeasures} & \textbf{Limitations} \\ \hline
\rowcolor{grey}
Malicious software \cite{PaganiniPierluigi2015AAffairs,skyjack,ThisPhone}& -Firewall implementations. 

-The use of antivirus and IDS solutions. & -Real-time detection of malware increases the computation costs. \\ \hline

Zero days vulnerabilities \cite{Hooper2016SecuringAttacks} & -Periodic system update.  & -Some manufacturers can release the patches weeks after the zero-day disclosures. \\ \hline

\rowcolor{grey}
Operating systems attacks \cite{PaganiniPierluigi2015AAffairs} & 
-Adopting the authorization mechanisms for UAV system resources.

-Software-based attestation approaches \cite{Seshadri2004SWATT:Devices} \cite{Dushku2020SARA:Systems}.
& -In a multi-UAVs network, managing authorizations for a swarm of UAVs is challenging. \\ \hline

Tampering captured videos \cite{mo2010} & -Firewall implementations.

-Software-based attestation approaches \cite{Seshadri2004SWATT:Devices} \cite{Dushku2020SARA:Systems}.
& -Even with proper security measures, a legitimate user who joins the UAV network can still tamper the captured videos. 
\\ \hline

\rowcolor{grey}
System ID spoofing \cite{Choudhary2018a} & -Periodic system update. 

-Firewall implementations. & -The use of social engineering techniques can reveal the System ID of UAVs since their manufacturers provide them. \\ \hline

\end{tabular}

\end{table*}

\textit{Tampering captured videos.} To guarantee safe navigation and avoid collisions during a flight mission, the operating system uses system calls that enable capturing the videos from the onboard camera~\cite{mo2010}. However, a knowledgeable adversary with the system parameters can intercept the issued system calls to hijack UAVs. The adversary might also combine the tampering attack with a GPS spoofing attack to control the flying drone. Unlike the operating system attacks, the adversary's primary goal is to compromise the navigation's safety and produce collisions. 

\textit{System ID spoofing.} According to the FAA's regulations~\cite{UASOverview}, UAVs should provide their System ID and location to third parties such as federal agencies and law enforcement when required. However, since most existing UAVs do not implement encryption mechanisms, the attacker can impersonate a third-party and execute an \textit{identity spoofing attack} to compromise the communication link and get the System ID of a UAV~\cite{Choudhary2018a}.

\subsubsection{Countermeasures for Software-based Attacks}
A regular operating system update can prevent compromising the UAVs and their payloads. In addition, firewall implementations on the GCS can block sending malicious traffic to the UAVs. Also, software-based solutions such as antivirus and IDSs can monitor the network traffic and secure UAVs against malicious activities. However, implementing onboard IDS is challenging due to the computation and energy constraints. Further, enabling the authorization mechanisms for UAV system resources can help to protect malicious code from execution. A promising solution against software-based attacks is the use of software-based attestation approaches. They ensure the integrity of software running on the flight stack~\cite{Seshadri2004SWATT:Devices,Dushku2020SARA:Systems}. Remote attestation solutions are low-cost and they provide a strong legitimacy of the software stack. 

At the Software-level, the adversary leverages malicious software and zero-days to infect the flight stack. Moreover, the adversary can tamper the captured videos to mislead the operator. These software-based attacks are mitigated by adopting antivirus and IDS solutions. Furthermore, the operator should keep his operating system updated and should implement software-based attestation solutions to verify the legitimacy of the code running on the operating system. However, It is worth mentioning that the provided defense mechanisms against software-based attacks cannot fully protect the flight stack from malicious activities. The patching process can take several weeks for disclosed zero-day vulnerabilities. Thus, making the UAVs vulnerable to adversaries. Furthermore, using IDS solutions or firewall implementations on the GCS can increase the computation costs and cause latency issues. Table~\ref{SumSoft} summarizes the the software security issues of UAVs, their existing countermeasures, and limitations.

\subsection{Communication-Level Issues} 
Communication is the critical component of the UAV system for flight control and data transmission. Majority of the UAVs use wireless communication for data and command exchange with the GCS. In this subsection, we provide the communication-level vulnerabilities, threats, and attacks against UAVs that compromise confidentiality, integrity, authenticity, and availability. 

\subsubsection{Communication Vulnerabilities and Threats}
Communication-level vulnerabilities and threats can be categorized based on the layers of communication as follows.

\textit{Physical \& MAC Layer Vulnerabilities and Threats.} The complexity of the UAV-2-GCS wireless communication network opens potential vulnerabilities. For example, in~\cite{Hooper2016SecuringAttacks} the authors demonstrated three different attacks affecting commercial Wi-Fi-based UAVs. These attacks are (1) Buffer overflow attack, (2) DoS attack, and (3) ARP cache poisoning attack. Their experimental results reveal massive security issues in UAV-2-GCS wireless communications. Choosing the correct type of wireless communication technology depends on the specification of the mission requirements (e.g., transmission range, operating frequency, category, etc.). However, this choice does not guarantee the flight's success since we have to consider the security issues of each type of wireless communication technology. Therefore, the fundamental question that remains unanswered is which type of wireless communication technology achieves a high level of security for UAVs for each application domain. 

\begin{figure*}[ht]
\centering
\subfloat[Centralized architecture.]{\includegraphics[width=0.41\textwidth]{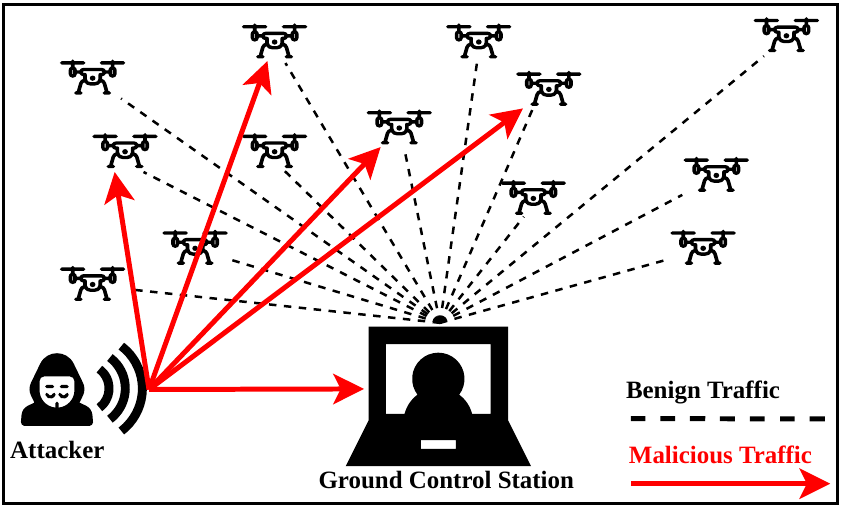}\label{CentralizedThreat}}
\hfill
\subfloat[Single backbone in a decentralized architecture.]{\includegraphics[width=0.4\textwidth]{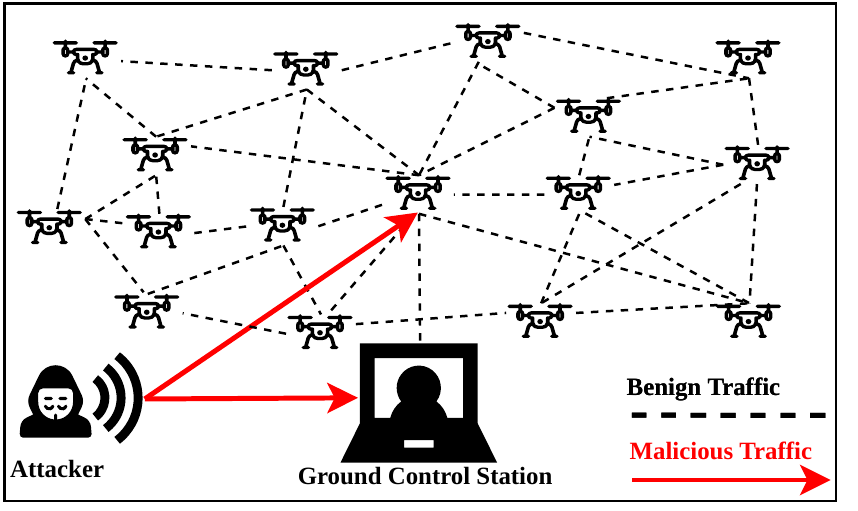}\label{SingleThreat}}
\hfill
\subfloat[Swarm of UAVs in a decentralized architecture.]{\includegraphics[width=0.42\textwidth]{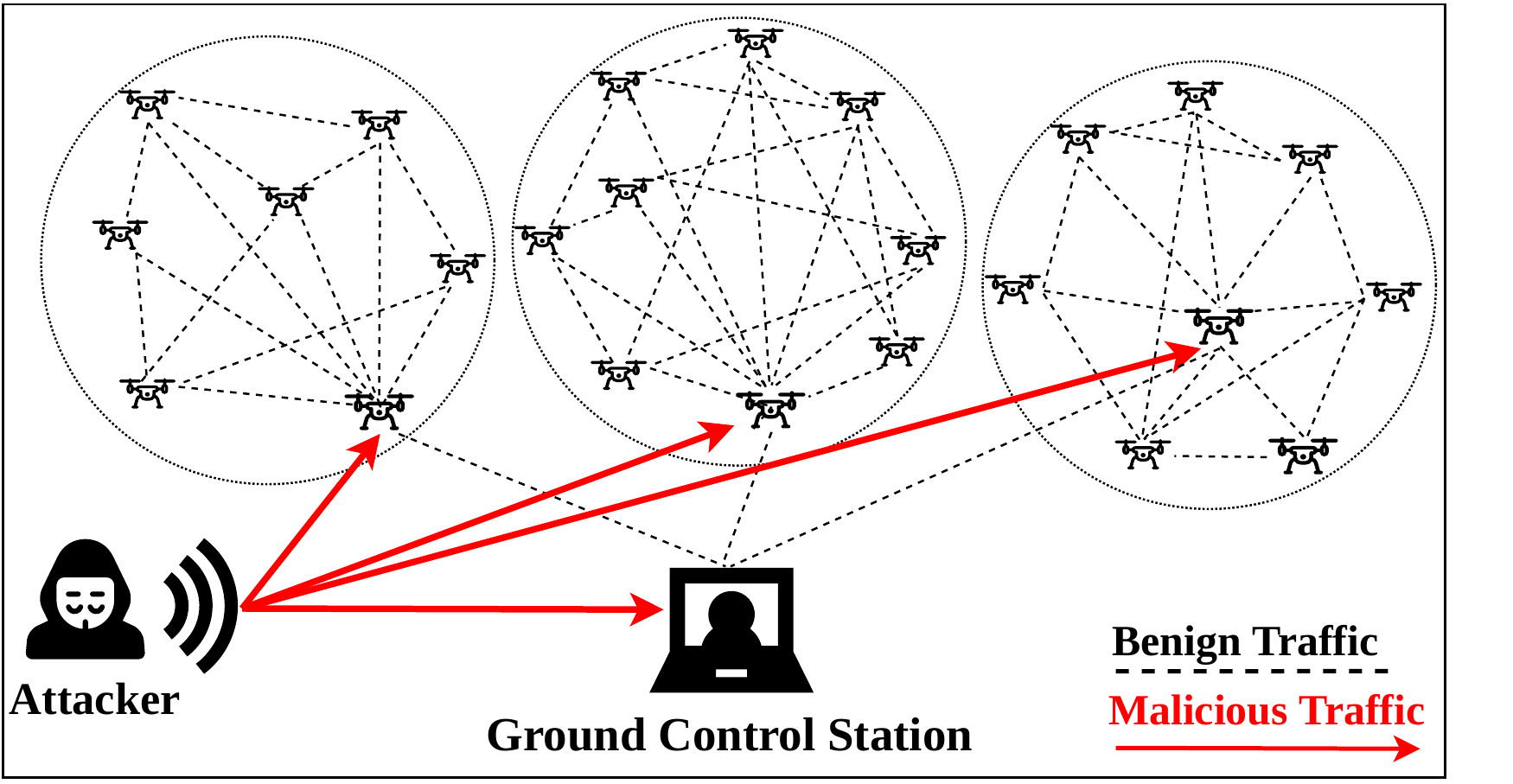}\label{SwarmThreat}}
\hfill
\subfloat[Mixed UAVs in a decentralized architecture.]{\includegraphics[width=0.4\textwidth]{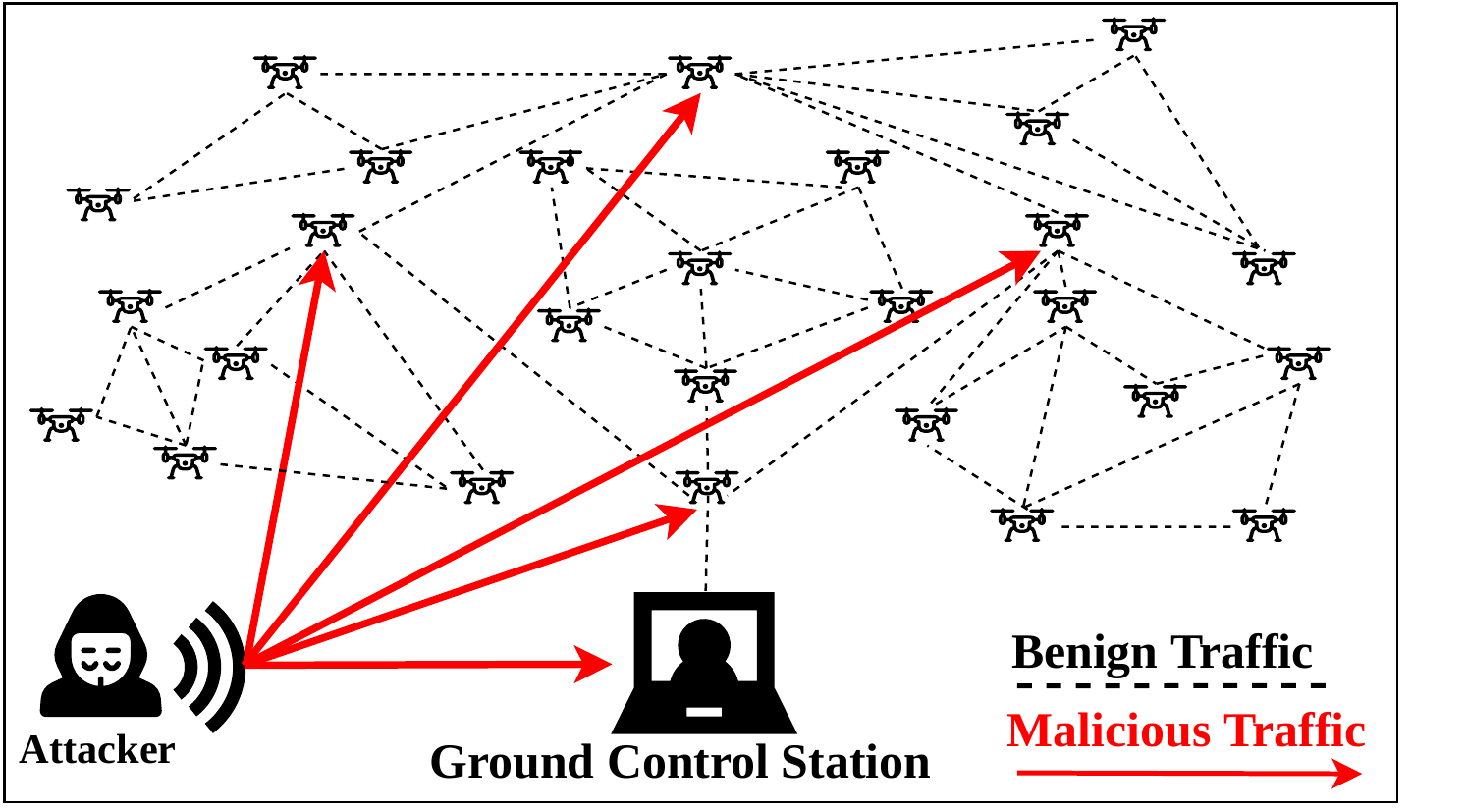}\label{MixedThreat}}

\caption{Threats for UAV communication networks in different architectures.}
\end{figure*}   

\textit{Network Layer Vulnerabilities and Threats.} The UAV network operates in an ad hoc fashion, commonly called FANETs. These networks have a dynamic topology and they present critical threats. A prior work presented general security threats of drone-assisted public safety networks~\cite{He2017}. It shows that the increase in UAV network's complexity results in more vulnerabilities to attacks. These attacks target mainly sensor inputs and communication modules. UAV communication threats such as intercepting or blocking the communication link between the Flight Controller and the GCS might cause a potential DoS attack. Furthermore, given FANETs unique characteristics that include the latency and computational power to route data, there is a need to build cryptographic algorithms for FANETs that takes these characteristics into consideration~\cite{Chriki2019FANET:Issues}. An attacker can disrupt the UAV network by sending malicious traffic directly through the GCS or indirectly through the UAVs. Whether in a centralized or decentralized architecture, the GCS is constantly threatened by adversaries. In both architectures, the GCS represents a single point of failure, and the security of the whole UAV network depends on the security of the GCS. However, even though the security mechanisms are implemented for the GCS, the attacker can still interrupt the flight mission by compromising the flying UAVs. It should be emphasized that in some scenarios, the flight mission can still be considered successful even though if one or multiple UAVs are compromised. In this case, and depending on the civilian application, the operator requires a minimum amount of legitimate (uncompromised) UAVs to accomplish the mission. 

In a centralized architecture, as illustrated in Figure~\ref{CentralizedThreat}, the adversary needs to target and send malicious traffic to a specific number of UAVs, such that the minimum number of legitimate UAVs required for the flight mission to succeed cannot be satisfied. Consequently, the adversary causes the flight mission to fail by disrupting the entire UAV network.  

Alternatively, for a decentralized UAV network architecture, the adversary needs only to compromise a particular UAV or UAVs to cause the flight mission to fail. In fact, for a single backbone UAV network, as depicted in Figure~\ref{SingleThreat}, the adversary needs to send malicious traffic only to the backbone UAV since it serves as a gateway between the other UAVs and the GCS. When the attacker compromises the single backbone UAV, the group of UAVs or the whole network is disrupted. Therefore, the single backbone UAV constitutes the weakest link in the UAV network. For a single backbone UAV architecture, the UAV network's security depends on the security of the GCS, the single backbone UAV, and their communication link. However, if the UAVs are similar in terms of shape, size, and color, it is challenging for the adversary to determine the backbone UAV. In multiple backbone UAVs architectures, the GCS and the backbone UAVs of each swarm are particularly critical for the success of the flight mission. However, even if a backbone UAV is compromised, the flight mission could be completed. From Figure~\ref{SwarmThreat}, we notice that the adversary needs to compromise four backbone UAVs or the GCS to disrupt the entire UAV network. Moving forward to more advanced UAV network architectures in mixed UAVs, Figure~\ref{MixedThreat} shows that securing the network of backbone UAVs is as crucial as securing the whole network. It is worth mentioning that the threats increase at the same level as the network complexity and the number of UAVs increase. In Table~\ref{Attack}, we summarize the different attack points described for each UAV network architecture that, if compromised by the adversary, the flight mission will fail.

UAV routing protocols are vulnerable due to the inherent characteristics posed by UAV networks such as dynamic topology, limited resources, and lack of encryption in their wireless links~\cite{Maxa2017b}. In this context, the adversary leverages these constraints to perform different routing attacks in the network layer. The adversary can disclose critical information in UAV networks which do not implement security mechanisms. With eavesdropping techniques, the adversary can leak routing information, topology information, and UAV positions~\cite{Maxa2017b}. Furthermore, without authentication and integrity considerations, UAVs routing protocols are prone to additional attacks such as DoS attacks or route-cache poisoning attack~\cite{Wu2007ANetworks}, where the adversary inserts incorrect routing information into the caches of legitimate UAVs.  
\begin{table}[!h]
\centering
\caption{Attack points of different UAV network architectures \label{Attack}}
\begin{tabular}{|p{3cm}|p{5cm}|}
\hline
\rowcolor{grey}
\textbf{Network Architecture} & \textbf{Attack Points}\\ \hline
Centralized Architecture & -The Ground Control Station. 

-Specific number of UAVs.\\ \hline
Single Backbone UAV & -The Ground Control Station.

-The backbone UAV. 

-The GCS-2-UAV communication link.\\ \hline
Swarm of UAVs & -The Ground Control Station.

-The backbone UAVs of each swarm. 

-The GCS-2-UAV communication link of backbone UAVs.

-The network of backbone UAVs.\\ \hline

Mixed UAVs & -The Ground Control Station.

-The backbone UAVs.

-The network of backbone UAVs.

-The GCS-2-UAV communication link of backbone UAV.\\ \hline

\end{tabular}
\end{table}

\textit{Transport Layer Vulnerabilities and Threats.} UAV communication protocols suffer from vulnerabilities leading to various attacks if not properly secured. Despite their communication features, they must ensure basic security requirements such as confidentiality, integrity, availability, and authenticity. Recent studies show that MAVLink protocol, one of the most well-known UAV communication protocols, is vulnerable to ICMP flooding and packet injection attacks~\cite{Kwon2018}. In addition, another transport layer protocol, the UranusLink, checks only the integrity of the message. Consequently, the adversary gathers the exchanged packets and discloses its content~\cite{Khan2020}.

\subsubsection{Communication-based Attacks}
In what follows, we present the common attacks exploiting UAV communications on the physical \& MAC layer, the network layer, and the transport layer.

\textit{Attacks on the Physical \& MAC Layer.} Given the significant difference between aerial networks and traditional wireless networks, there is a need to choose the most suitable wireless technology for UAVs~\cite{Hayat2016}. In this context, we categorize for each wireless communication technology in the physical \& MAC layer, its unique features, and specific security issues. Although it is possible to find surveys on the security of each wireless communication technology on its own, we briefly list the major characteristics and security issues of wireless communication technologies in UAV systems in Table~\ref{Wireless}. 

\begin{table*}[h]
\centering
\caption{Security Issues of the Wireless Communication Technologies used in physical \& MAC layer of UAV communications \label{Wireless}}
\begin{tabular}{P{3.5cm}|p{1cm}|p{1.2cm}|p{1.1cm}|p{8.8cm}}
\hline
\textbf{Communication Technology} & \textbf{Category} & \textbf{Frequency} & \textbf{Range} & \textbf{Security Issues} \\ \hline
\rowcolor{grey}
Wi-Fi & WLAN & 2.4-5 Ghz & 20-120 m & -Commercial Wi-Fi-based UAVs are vulnerable to basic security attacks, such as Wi-Fi de-authentication attack~\cite{Hooper2016SecuringAttacks}.

-Unencrypted Wi-Fi networks allow the adversary to perform spoofing or jamming attacks~\cite{Pleban2014Hacking}.

-Popular attacks against the IEEE 802.11 standard exist in the literature (e.g., flooding attacks, Key Retrieving Attacks, ARP injection attacks, etc.) ~\cite{Kolias2016IntrusionDataset}.\\ \hline

Bluetooth & WPAN & 2.4 GHz & 10-200 m & -The sequence extraction of Frequency Hopping Spread Spectrum (FHSS)-type controllers using a Software Defined Radio (SDR) enables the likelihood of performing Bluetooth sniffing~\cite{Shin2016SecurityController}.

-A family of different vulnerabilities of Bluetooth communication known as BRAKTOOTH can be applied in UAVs scenarios~\cite{Garbelini2021BRAKTOOTH:Manager}. \\ \hline
\rowcolor{grey}
Zigbee & WPAN & 2.4 GHz & 10-100 m & -Threat analysis of autonomous UAVs shows that multiple vulnerabilities allow the adversary to locate the Zigbee transmitter, perform DoS and replay attacks~\cite{Olawumi2003ThreeLearned}. 

-KillerBee is an example of an open-source exploitation framework designed to perform reconnaissance and exploit Zigbee vulnerabilities~\cite{WrightKillerBee:World}. \\ \hline

Long Range (Lora) & LPWAN & 868 MHz 

915 MHz & 05-15 km & -The LoRa Alliance does not consider security implementations and lacks security controls on the network servers \cite{Gemalto2017LoRaWANPROVIDERS}. 

-End-to-end security between the application server and the end device is not covered. Therefore, no protection of data transmission exists in terms of confidentiality and integrity \cite{You2018AnSystem}.

-Prone to various security attacks: jamming attacks, replay attacks, and wormhole attacks \cite{Chacko2018SecurityLPWAN}.
\\ \hline
\rowcolor{grey}
Sigfox & LPWAN & 868 MHz

902 MHz & 03-30 km & -Lack of data confidentiality and authentication~\cite{Chacko2018SecurityLPWAN}.

-Sigfox does not support encryption~\cite{Neji2019CommunicationAgriculture}.  \\ \hline

Narrowband-Internet of Things (NB-IoT) & LPWAN & 200 KHz & 10-35 km &
-Several layerwise passive and active attacks exist: Malicious code injection, Man-in-the-Middle attack, and jamming attack~\cite{Chacko2018SecurityLPWAN,Kumar2020NB-IoTSurvey}.
 \\ \hline
\rowcolor{grey}
Worldwide Interoperability for Microwave Access (WiMAX) & WMAN & 2.3-5.8 GHz & 01-48 km & -Considering UAVs as a collection of mobile nodes communicating within a WiMAX network, when compromised, they create a byzantine failure and disrupt the whole network \cite{Saranya2017ANetworks}.

-DoS attacks can target different resources: storage and processing resources (e.g., memory, storage, CPU), energy resources (e.g., battery), and bandwidth~\cite{Han2009PotentialNetworks}.  \\ \hline

Cellular Technology (GPRS, EDGE, UMTS/WCDMA, UMTS/HSPA, LTE, LTE Advanced - 4G, 5G) & WWAN & Sub-6 Ghz & World wide &

-Prone to jamming, spoofing, eavesdropping, hijacking, and DoS attacks~\cite{Fotouhi2019a}. \\ \hline
\end{tabular}

\end{table*}

\textit{Attacks on the Network Layer.} The attacks on the network layer of UAV communications include eavesdropping, DoS, man-in-the-middle, forgery, replay, and other attacks on the FANETs. 

\textit{a) Eavesdropping attacks.} An attacker can perform an eavesdropping attack through the UAV-2-GCS communication link by gathering data such as live video feeds, sensor readings, and GPS data sent by the UAVs to the GCS. Since most UAVs avoid encrypting the wireless communication for the sake of improving communication performance~\cite{mo2010}, the attacker can eavesdrop on exchanged information, including telemetry feeds and GCS commands. Therefore, he can violate confidentiality of the communication and the data by gathering sensitive information such as sensor readings and GPS data.

\textit{b) DoS attacks.} An adversary can compromise a UAV system by launching a DoS attack. In this case, the attacker can flood the flying UAV's network card with random traffic by sending multiple requests, causing an overload of its resources and disrupting its availability. The impact of performing such attacks on UAVs can result in a substantial increase in the network latency and a decrease in the quality of video streaming applications for the user~\cite{Vasconcelos2016}. Another way to perform a DoS attack is by sending large packets to the GCS within a specific range to disable the control signal. Once the signal is disabled, the drone goes into a lost link-state, which results in a malfunctioning of the data link. Consequently, the operator can no longer send or receive data signals to the Flight Controller, which results in disrupting the communication link and losing control of the UAV. In~\cite{Muzzi2015}, the authors simulated a Distributed DoS (DDoS) attack on UAVs using botnets. The DDoS attack was simulated by flooding the network traffic using User Datagram Protocol (UDP) packets. This type of simulation demonstrates the possibility of performing real-world DDoS attacks on UAVs. Besides, performing de-authentication attacks can also disable the operator from controlling the UAV. The de-authentication attack is a DoS attack that consists of sending de-authentication packets to the UAVs to disrupt the UAV-2-GCS communication. As a result, the adversary blocks the UAV-2-GCS communication, and eventually, the UAVs are disconnected from the network. An example of such attacks is demonstrated by Skyjack~\cite{skyjack}.     

\textit{c) Man-in-the-Middle attacks.} In this one of the most well known attack~\cite{Conti2016AAttacks}, the adversary controls the UAV-2-GCS wireless channel and alters the benign packets with malicious content~\cite{Rodday2016ExploringVehicles}. Thus, the adversary can act as a bridge between the UAV and the GCS, and compromise the bidirectional UAV-2-GCS communication. A video replay attack is an example of a Man-in-the-Middle attack, where the adversary fools the operator by transmitting malicious live feed data. VideoJak~\cite{VideoJak:Calls} is an example of such attacks.  

\textit{d) Forgery attacks.} The adversary can compromise UAVs communication integrity by transmitting a forged request to unauthenticated UAVs~\cite{He2017}. In this attack, the adversary generates the malicious request by impersonating a legitimate request and disrupts the UAV-2-GCS communication.   

\textit{e) Replay attacks.} In UAV networks, the adversary can perform an eavesdropping attack to intercept several requests, then replay valid data to the UAVs. In this case, the UAVs might receive repeated data, and if no replay protection is implemented, the UAVs cannot distinguish the legitimate requests from the malicious ones~\cite{He2017}.   

\textit{f) Attacks on FANETs routing.} Different passive and active attacks can occur in Mobile Ad hoc Networks (MANETs) routing protocols which consist of injecting malicious nodes, controlling the network traffic, or disrupting the routing functionality~\cite{Maxa2017b}. Most existing attacks targeting routing protocols on MANETs are transferable to routing protocols on FANETs since FANETs is a subcategory of MANETs. To illustrate these attacks, we classify them on three categories based on their routing functionality~\cite{Maxa2015}: \textit{i) the route discovery attacks:} they target the traffic control and include the blackhole~\cite{Tseng2011ANetworks}, sleep deprivation~\cite{Pirretti2006TheDefense}, sybil~\cite{Douceur2002TheAttack}, and wormhole~\cite{HuWormholeNetworks} attacks. \textit{ ii) The route maintenance attacks:} they aim to corrupt the routing control packets. Examples of such attacks are flooding~\cite{routing2007} and Byzantine~\cite{Awerbuch2004MitigatingNetworks} attacks. \textit{iii) The data forwarding attacks:} they include the type of attacks that impact the payload traffic, such as real-time video traffic~\cite{Hu2003RushingProtocols}.

\textit{Attacks on the Transport Layer.} Attacks on the transport layer of UAV communication can be grouped based on the UAV transport layer protocols. 

\textit{a) UranusLink Protocol Attacks.} To the best of our knowledge, there is no existing attack against the UranusLink protocol. According to the design and implementation of UranusLink for real-world applications~\cite{Kriz2015}, UranusLink provides only integrity protection via the checksum field in the messages. However, an adversary with the ability to capture the exchanged packets can benefit from this vulnerability and disclose mission-related information~\cite{Khan2020}.

\textit{b) MAVLink Protocol Attacks.} 
Authors in~\cite{Koubaa2019} classify MAVLink attacks into four classes depending on how data is compromised: interception, modification, interruption, and fabrication attacks. Since the MAVLink protocol does not provide authentication and encryption, the adversary can capture communication traffic through eavesdropping and thus collect exchanged data between the GCS and the UAVs. Moreover, he can perform system ID spoofing attacks. Authors in~\cite{Highnam2016AnArchitecture} presented a realistic scenario of compromising different UAVs operating under MAVLink protocol. The considered specimen attack scenario demonstrates an attacker's ability to perform a stealthy attack by capturing a flight mission's system-ID and spoofing MAVLink packets.

\subsubsection{Countermeasures for Communication-based Attacks}
Different security approaches have been proposed in the literature to ensure confidentiality, authentication, availability, and data integrity in UAV communications. In what follows, we present existing countermeasures against UAV communication-based attacks at the physical \& MAC layer, network layer, and  transport layer.

\textit{Countermeasures for the Physical \& MAC Layer Attacks.} Securing the physical properties of the communication channel (e.g., transmission medium, physical topology, etc.) is one of the mitigations against the physical \& MAC layer attacks of UAVs. Given the wide use of UAVs across different wireless communication technologies, it is important to consider that securing wireless communications at the physical \& MAC layer is challenging due to the characteristics of each communication technology (e.g., category, frequency, range, etc.). 
In addition, encryption algorithms such as AES can be employed at the physical \& MAC layer communications. Moreover, artificial noise techniques that transmit generated noise to illegitimate users can also be used~\cite{Zhang2017SecuringOverview}. In addition to these, one of the best practice for secure communication in this layer is to keep the device firmware and and the related software up to date using the released security patches. We would like to note that, the attacks and the countermeasures for the wireless communication technologies outlined in  Table~\ref{Wireless} are vast and it is possible to find a survey on the attacks and mitigations for each communication technology in the list. For this reason, we do not go into much details with the countermeasures against the attacks on these well-known and widely used communication technologies in this survey.

\textit{Countermeasures for the Network Layer Attacks.} To mitigate eavesdropping attacks on UAV networks, the operator can adopt authenticated encryption~\cite{Bellare2000AuthenticatedParadigm}. It consists of protecting the UAV-2-GCS communications by ensuring the confidentiality and authenticity of the exchanged data. In~\cite{Zhang2017SecuringOptimization}, the authors proposed an anti-eavesdropping power control algorithm in UAV communication systems. Power control algorithms present an efficient approach for building a UAV network topology that ensures the Quality of Service (QoS), and they are also used to prevent eavesdropping attacks. In the presence of an eavesdropper, the algorithm proposed by Zhang et al.~\cite{Zhang2017SecuringOptimization} demonstrates that by optimizing the trajectory and transmit power control between the UAV and the GCS, we maximize the secrecy rate (the difference between the rate of the UAV-2-GCS communication channel and the maximum rate of the eavesdropper~\cite{Barros2006SecrecyChannels}). Moreover, adopting a continuous authentication against eavesdropping attacks can identify a pilot's unique profile during the flight mission~\cite{Nassi2021SoK:Drones}. Another solution aims to use fingerprinting techniques to authenticate UAVs~\cite{Alladi2020SecAuthUAV:Communication}.  

The use of cryptographic primitives such as public-key cryptography guarantees the integrity and confidentiality of UAV communications. In~\cite{He2017SecureNetwork}, the authors proposed a secure communication scheme for UAVs network using hierarchical identity-based broadcast encryption (HIBBE) technique. The proposed approach guarantees message confidentiality and authentication through identity-based signcryption. Their performance analysis results showed that the proposed scheme is resistant to DoS attacks. Another work presented a secure communication protocol based on an efficient certificateless Signcryption Tag KeyEncapsulation mechanism (eCLSC-TKEM)~\cite{WonAObjects}.
Furthermore, the protocol is energy-efficient and meets security and efficiency requirements for UAV communications. To secure commercial WiFi-based UAVs, the authors in~\cite{Hooper2016SecuringAttacks} presented a comprehensive multi-layer security framework. Their proposed framework is efficient against basic security attacks such as ARP cache poisoning attacks and DoS attacks. In~\cite{Shoufan2015}, the authors presented a lightweight FPGA hardware solution to secure UAV-2-GCS communication of commercial Wi-Fi-based UAVs. It contains a cryptographic engine responsible for encrypting the communication control data. Thus, ensuring confidentiality and authentication. However, enabling cryptography-based approaches will require additional computation in both GCS and UAVs and increase energy consumption. Hence, these solutions may reduce the performance of the UAV-2-GCS communication.

IDSs aim to detect malicious intrusion activities such as DoS attacks. They can be deployed on the flying UAV or in the GCS. We distinguish three intrusion detection approaches~\cite{Choudhary2018}: \textit{i) Rule-based intrusion detection}, where specific rules for UAVs are applied in which rules follow the expected behavior of the UAV system~\cite{Choudhary2018IntrusionSurvey}. \textit{ii) Signature-based intrusion detection}, which relies on attack signatures~\cite{ADS2016}, and  \textit{iii) Anomaly-based detection} that detects known and unknown attacks based on learning or filtering mechanisms. However, these three approaches mentioned above cannot fully detect UAV intrusions. For example, the signature-based detection approach is weak against attacks that frequently change their patterns, which results in changing their signature. Additionally, the anomaly-based approach may suffer from false positives and false negatives. A recent work uses a hybrid detection approach that combines two or more approaches for accurate detection of unknown attacks~\cite{Condomines2019NetworkValidation}. Other intrusion detection solutions rely on using packet analysis techniques to ensure data integrity and network availability in UAVs~\cite{Sedjelmaci2016}.

In the literature, different security solutions have been proposed to secure MANETs routing protocols from malicious actors~\cite{Maxa2017b,Oubbati2019}. These approaches can also be used in FANETs and include cryptographic schemes such as message authentication, digital signatures, and hashing. Hence, enabling the confidentiality and integrity of the UAV network. We distinguish the use of \textit{secure-based routing protocols} for FANETs to guarantee the routing process and reliability in the presence of malicious nodes. This category includes the use of security mechanisms in the routing protocols~\cite{Oubbati2019}. Examples of secure-based routing protocols for UAVs networks are: SUANET (Secure UAV Ad hoc NETwork)~\cite{Maxa2015a}, PASER (Position-Aware, Secure, and Efficient mesh Routing)~\cite{Sbeiti2016}, SUAP (Secure UAV Ad hoc routing Protocol)~\cite{Maxa2015}, AODV-SEC (Ad hoc On-demand Distance Vector-Secure)~\cite{Aggarwal2012}, and SRPU (Secure Routing Protocol for UAVs)~\cite{Maxa2016}. Each of these protocols uses a specific strategy to satisfy the security and privacy of the routing path. For instance, the SUANET protocol uses a key management strategy between UAVs to enable confidentiality and authentication services~\cite{Maxa2015a}. In contrast, PASER protocol utilizes cryptographic functions to secure the routing packets in the UAV network~\cite{Sbeiti2016}. SUAP routing protocol prevents the flooding attack~\cite{Maxa2015}. AODV-SEC routing protocol ensures a secure route discovery process~\cite{Aggarwal2012}. However, the implementation of \textit{secure-based routing protocols} in realistic scenarios is challenging due to their complexity and high density.

\begin{table*}[h]
\centering
\caption{Summary of UAV Network and Transport Layer security issues, existing countermeasures and their limitations \label{SumComm}}
\begin{tabular}{p{1cm}|p{4cm}|p{6.2cm}|p{5cm}}
\hline
\textbf{Layer} & \textbf{Attacks/threats} & \textbf{Countermeasures} & \textbf{Limitations} \\ \hline
\rowcolor{grey} Network \newline Layer &
Eavesdropping attacks \cite{mo2010}& 

-The use of anti-eavesdropping power control algorithm in UAV communications \cite{Zhang2017SecuringOptimization}.

-Adopting authenticated encryption~\cite{Bellare2000AuthenticatedParadigm}.
& -Cryptography-based approaches require additional computation and might increase energy consumption. \\ \hline

Network \newline Layer &
DoS attacks~\cite{Vasconcelos2016,Muzzi2015}&

-Building IDS solutions~\cite{Choudhary2018IntrusionSurvey,ADS2016}.
& 
-Impact on the performance of the GCS-2-UAV communication.

-The signature-based IDS fails against attacks that change their patterns.

-The anomaly-based IDS can suffer from false positives and false negatives. 
\\ \hline

\rowcolor{grey}Network \newline Layer &
Man-in-the-Middle attacks \cite{Rodday2016ExploringVehicles}& -Encrypting the communication control data \cite{Shoufan2015}.

-Implementing fingerprinting techniques to authenticate UAVs~\cite{Alladi2020SecAuthUAV:Communication}.
 & -Latency issues for time-critical UAVs applications. \\ \hline

Network \newline Layer &
Forgery attacks \cite{He2017} & -Enabling a multi-layer security framework~\cite{Hooper2016SecuringAttacks}. & -The complexity of the network increases in multi-UAVs scenarios. \\ \hline

\rowcolor{grey}Network \newline Layer &
Replay attacks~\cite{He2017} & -Establishing a secure communication scheme (e.g., identity-based encryption)~\cite{He2017SecureNetwork}.

-The use of authentication mechanisms~\cite{Shafique2021a,mo2010}.
 & -Repeated requests can flood the network and cause a possible DoS. \\ \hline

Network \newline Layer &
Blackhole~\cite{Tseng2011ANetworks}, Flooding~\cite{routing2007}, Sybil~\cite{Douceur2002TheAttack}, Wormhole~\cite{HuWormholeNetworks}, Sleep deprivation~\cite{Pirretti2006TheDefense}, Byzantine ~\cite{Awerbuch2004MitigatingNetworks}, and Forwarding~\cite{Hu2003RushingProtocols} attacks& -The use of secure-based routing protocols~\cite{Oubbati2019}. & -High computation overheads and delay. \newline -The security features are supported only by few routing protocols. \\ \hline

\rowcolor{grey}
Transport \newline Layer &
Attacks on communication protocols~\cite{Khan2020,Koubaa2019}& 
-Building a high-level architecture for resiliency and trustworthiness capable of repairing the flight mission despite the attack~\cite{Highnam2016AnArchitecture}.

-Embedding security services into hardware modules.

-The use of classical security approaches such as encryption techniques and IDS approaches.

-Exploiting the features of emerging technologies such as blockchain~\cite{Garcia-Magarino2019SecurityBlockchain}.
& -The introduction of trade-offs between performance and security. \\ \hline
\end{tabular}

\end{table*}

\textit{Countermeasures for the Transport Layer Attacks.} To prevent the adversary from disclosing sensitive information in the transport layer, it is important to implement security mechanisms enabling the confidentiality and integrity of the exchanged data (e.g., cryptographic protocols, secure key exchange, etc.). To mitigate MAVLink attacks, one approach proposes an architecture that consists of repairing and completing the mid-flight mission despite the cyber attack~\cite{Highnam2016AnArchitecture}. Other approaches also exist to secure the MAVLink communication protocol. In~\cite{Koubaa2019}, the researchers divided existing MAVLink security solutions into hardware-based solutions and software-based solutions. Hardware-based solutions rely on embedding security services into hardware modules, while software-based solutions include classical security approaches like encryption techniques and IDSs. Other solutions that aim to secure MAVLink communication protocol might benefit from the features of emerging technologies such as Blockchain and Software-Defined Networking (SDN)~\cite{Hassija2021}.

Table~\ref{SumComm} summarizes the UAV network and transport layer communication security issues, their existing countermeasures, and limitations. The communication-based attacks on UAVs at different layers enable the adversary to disrupt the communication link and jeopardize the flight mission. In the literature, specific countermeasures have been developed to guarantee the exchanged data's confidentiality, integrity, and availability. These countermeasures consist of building IDS solutions, adopting authenticated encryption to prevent eavesdropping attacks, enabling a multi-layer security framework, and using secure-based routing protocols. However, it is worth noting that the countermeasures mentioned above for UAV's communication-based attacks have some limitations and shortcomings. For example, building IDS solutions to prevent DoS attacks impact the performance of the UAV-2-GCS communication. Besides, latency issues occur when encrypting the communication control data. Moreover, the use of secure-based routing protocols significantly increases the computation overheads and introduces delays.

\section{Privacy Issues of UAVs}\label{sec:privacy-issues}

\begin{figure}[!t]
\begin{center}
\includegraphics[width=0.47\textwidth]{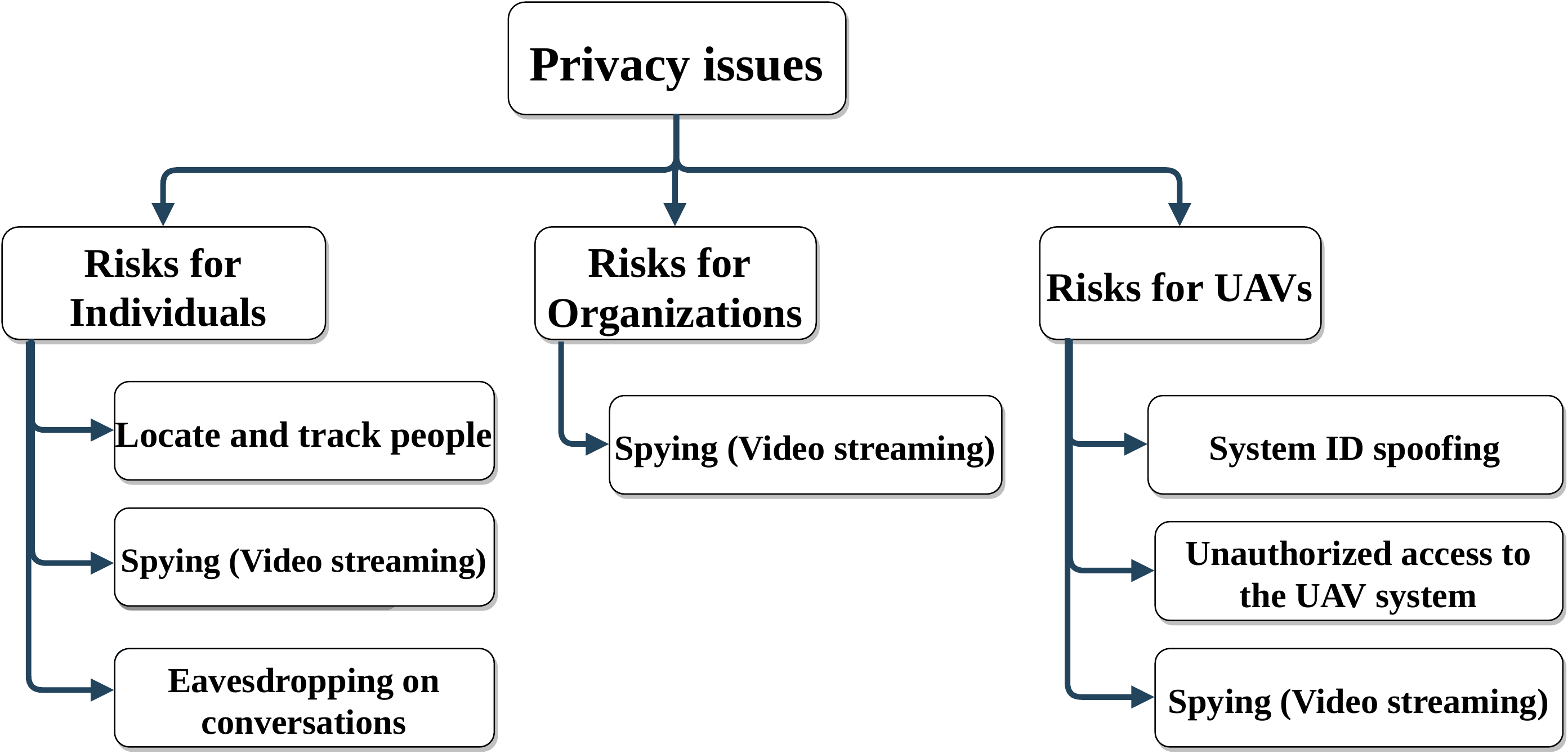}
\caption{Privacy issues of Unmanned Aerial Vehicles.}
\label{Privacy}
\end{center}
\end{figure}

The development of UAV technologies has raised a broad range of privacy issues at high risk. In this section, we categorize the privacy issues according to the entity being in risk and also the type of sensitive data that is leaked to unauthorized users. The categorization we apply and follow in this section is given in  Figure~\ref{Privacy}.

\subsection{Privacy Risks}
The privacy risks exposure of UAVs can be grouped into three categories: risks for individuals, risks for organizations, and risks for UAVs. The privacy risks for individuals are related to personal information obtained through a flying drone, while the risks for organizations are associated with organizational data that UAVs can collect. Risks for UAVs category concerns sensitive data disclosed to third parties. Compromising data privacy refers to compromising the secrecy of data that should not be revealed to third parties. 

\subsubsection{Risks For Individuals}

A recent study presented the privacy concerns posed by the use of UAVs in airborne photography~\cite{Jiang2020a}. In~\cite{Yaacoub2020}, the authors divided the privacy leakage into three classes: physical privacy, location privacy, and behavior privacy.

In \textit{physical privacy}, the attacker captures images and videos of people inside their houses for malicious purposes~\cite{ArthurHollandMichel2017DronesCases}. \textit{Spying} activity on people through UAVs is one of the significant physical privacy issues. Hence, the need to establish regulations governing the use of UAVs in civilian airspace. UAVs can also be equipped with directional microphones to eavesdrop on private conversations. 

\textit{Location privacy} targets people's physical locations and their movements without their knowledge of being under surveillance~\cite{Finn2013SevenPrivacy}. Third-parties could use it for business purposes (e.g., targeted advertising by location). Nowadays, the use of UAVs switched from aerial surveillance to tracking individuals~\cite{Mcneal2014DronesRobotics}. Indeed, one of the most challenging issues is to tell whether a flying drone is used for aerial surveillance or for tracking people~\cite{Nassi2019}. Detecting such privacy invasion attacks is still an open research problem.

In \textit{behavior privacy}, the attacker monitors the people's lifestyle and interests in public space~\cite{Clarke2014ThePrivacy}. Surveillance of individuals through systematic monitoring of their behaviors constitutes a major threat to behavioral privacy and may negatively impact people's psychological level~\cite{Clarke2014ThePrivacy}.      

\subsubsection{Risks For Organizations}

Information and resources that an adversary maliciously harvests from UAVs may not necessarily be personal~\cite{Lui2007IndividualAnalytics}. In particular, UAVs can \textit{spy} on organizations through video streaming such as industrial espionage, and the attacker can disclose the private information of government agencies and corporations to unauthorized parties. For example, in a farming business that uses a swarm of UAVs to optimize its operations and improve the corp production, an adversary can spy on this organization by using the same UAV model. In this case, we cannot distinguish between the friendly UAV and the malicious one. It is known as the identification problem and exploited to perform malicious activities such as terrorism and smuggling~\cite{Nassi2019}. 

\subsubsection{Risks For UAVs}

This category consists of leaking sensitive information monitored by the UAVs to unauthorized third parties, such as video footage, photos, and physical measurements. In addition, other types of sensitive data related to the flying UAVs, including the real-time GPS location, speed, height, and battery status, have to be preserved only for the operator. Preserving data privacy of the flying UAVs is a fundamental requirement for the safety of the flight mission~\cite{Lin2018}. In unencrypted communications, the adversary can perform a \textit{traffic analysis attack} by listening to the traffic and extracting sensitive flight mission information. This type of passive attack compromises the confidentiality and privacy of the UAV. Even in encrypted communications, forensics techniques, including data extraction and analysis, can recover digital data~\cite{Salamh2021AChallenges}. Another type of privacy attack targeting UAVs occurs when the adversary has an \textit{unauthorized access} to the critical components of the UAV system such as the sensors and the storage (e.g., hijacking attack, injecting hardware trojans, etc.). In this case, the adversary discloses flight data to the public and jeopardizes the flight mission.

\subsection{Defense Mechanisms Against Privacy Risks}

Several studies have suggested privacy-preserving mechanisms to prevent leaking secret information to unauthorized parties. These mechanisms include encryption techniques and the design of tamper-proof hardware so that even in scenarios where the drones are hijacked, they cannot reveal sensitive information. In~\cite{Birnbach2017Wi-FlyDrones}, the authors suggested an approach to detect privacy invasion attacks based on UAVs flight behavior. However, it fails to identify a UAV's purpose (whether it is legitimate or malicious). The researchers in~\cite{Chen2020ASystem} presented a privacy-preserving authentication scheme for UAV control systems. The proposed architecture has a mutual authentication to secure communication between entities and integrates cryptography mechanisms such as elliptic curve cryptography (ECC), digital signature, and hash functions.
Moreover, the suggested privacy protection protocol guarantees location privacy and proves its applicability in sensitive control areas. Similarly, A privacy-preserving authentication approach for UAVs was proposed in~\cite{Tian2019EfficientDrones}. It is a predictive authentication framework considering identity, location, and flying routes as sensitive information. Other solutions can overcome privacy issues, such as implementing access policies and lightweight cryptography approaches. Some manufacturers include a list of no-fly GPS coordinates covering sensitive areas in the firmware of their product. Moreover, regular users can register their home location in the NoFlyZone Database~\cite{Altawy2017}.
 
\section{Pitfalls and Future Research Directions}\label{sec:pitfalls}

As it appears and develops, the UAV technology brings certain advantages and benefits to our society. However, it can also create new potential threats and tools for malicious attacks for civilian users. Although the existing countermeasures aim to protect the operators from malicious activities, several open issues need to be addressed by the research community. In this section, we firstly present the lessons learned; then, identify open issues and discuss future research directions, which we believe will provide useful guidance for future UAV security research and practice. 

\subsection{Lessons Learned and Pitfalls}

The rise of UAV technology created a plethora of cyber attacks, such as intercepting unencrypted data links from UAVs or spoofing the UAV network. Protecting the flight mission requires a comprehensive defense-in-depth approach. 

\smallskip
\noindent \textbf{UAV Manufacturer Issues.} Our findings in this survey demonstrate that UAVs lack protection from various attacks at different levels. A possible reason for this shortcoming could be explained by manufacturers' interests in increasing the performance of their commercial products over security. Another reason is the additional cost needed from manufacturers to implement security mechanisms. Existing UAVs manufacturers should consider the security and privacy aspects while developing their products in all the supply chain phases.

\smallskip
\noindent \textbf{Sensor-level Issues.} At the Sensor-level, the diversity and complexity of onboard sensors (e.g., chemical, physical, mechanical, etc.) makes them targeted components for adversaries. Moreover, existing countermeasures against spoofing, sniffing, or jamming onboard sensors are limited due to the unique characteristics of UAVs. Although the existing security research covers sensor-based threats and attacks~\cite{Sikder2021AApplications}, in the context of UAVs, we need to consider additional parameters for UAVs such as the authenticity of sensor readings, the energy and computation costs when securing sensed data against malicious actors.

\smallskip
\noindent \textbf{Hardware-level Issues.} At the Hardware-level, despite the type and characteristics of different commercial UAVs such as the firmware and hardware type, UAV hardware could be targeted in the manufacturing process, or before or during the flight mission. These scenarios are possible due to the vulnerabilities that can occur in UAV firmware and also due to the lack of encryption in custom chipsets. Given the popularity and diversity of existing UAVs, it is important to build a unified hardware security strategy that consists of protecting UAVs from hardware-based attacks.    

\smallskip
\noindent \textbf{Software-level Issues.} At the Software-level, the adversaries can leverage the zero-day and existing software vulnerabilities in the flight stack as well as the GCS software to compromise the flight mission. The prevalence of software-based attacks demonstrates the need to develop robust  defense solutions for UAV software security. However, existing UAV manufacturers avoid integrating software security implementations in their products for performance reasons. Therefore, the adversaries can take advantage of this gap to build malicious software (e.g., Maldrone~\cite{PaganiniPierluigi2015AAffairs}, Snoopy~\cite{ThisPhone}, SkyJack~\cite{skyjack}).     

\smallskip
\noindent \textbf{Communication-level Issues.} At the Communication-level, designing a Multi-UAV network has to consider potential security issues according to the chosen network topology. Many UAV protocols are not properly secured and pose serious threats. Given that communication is a crucial part of the UAV system, we argue that standardized UAV protocols enabling reliable and secure communication have to be developed. Most of the existing communication protocols in UAVs are unencrypted or have limited cryptographic capabilities, thus enabling adversaries to compromise the communication channels. Moreover, existing security measures to protect civilian UAVs from malicious users are limited to single UAV systems~\cite{Shakhatreh2019}. Therefore, there is a need to develop countermeasures for multiple UAV scenarios.

\smallskip
\noindent \textbf{Security - Performance Tradeoff.} At any level of the UAV, when implementing security solutions, we need to assess the performance of the UAV system. In particular, the communication costs, the computation costs, the storage overheads, and the energy. However, adding an extra security layer for each level without considering the abovementioned parameters might significantly decrease the performance of the flight mission. Towards this point, we can derive possible tradeoffs between the performance and security considerations of UAVs.  

\smallskip
\noindent \textbf{Privacy Concerns.}
Besides security considerations, UAVs can also violate personal privacy, from spying on people's lifestyles to gathering sensitive data about organizations. The deployment of UAVs in the civilian airspace without specific regulations poses serious privacy concerns for individuals. Moreover, sensitive information collected by UAVs and transmitted to the GCS has to be protected from unauthorized parties. Therefore, privacy leakage has to be considered during the design of UAV systems. Two significant scientific gaps allow privacy invasion attacks: \textit{The purpose detection problem} and \textit{the identification problem}~\cite{Nassi2019}. The \textit{purpose detection problem} consists of distinguishing between a legitimate and a malicious nearby UAV that violates an individual's privacy. Existing approaches to solving the purpose detection problem are minimal since they cannot detect spying actions on a specific Point of Interest (POI)~\cite{Birnbach2017Wi-FlyDrones,Rozantsev2015FlyingCamera,CaseLow-costTracking}. A recent study demonstrated using a cryptanalysis approach, by applying a periodical physical stimulus (LED flicker) on the spying UAV cameras, it causes a watermark on the encrypted UAV-2-GCS communication traffic~\cite{Nassi2019a}. The detection of such a watermark determines the legitimate or illegitimate purpose of the drone. However, this approach is limited to the Wi-Fi First-Person-View (FPV) transmission in the UAV-2-GCS communication channel. In the \textit{identification problem}, given a multi-UAVs scenario, it is likely impossible to identify a foe UAV among legitimate ones. Although Identification Friend or Foe (IFF) methods~\cite{IdentificationWikipedia} exist, they fail to distinguish a foe UAV that is nearby to a legitimate one with the same altitude and GPS location (less than 4.9 m~\cite{GPS.gov:Accuracy}). Therefore, the malicious entities leverage the existing scientific gaps to violate individuals' privacy.    

\subsection{Future Research Directions}

In this subsection, we present promising security and privacy research directions of UAVs that could be investigated in future works. 


\smallskip
\noindent \textbf{UAV Forensics.} When security incidents occur during a flight mission, forensics analysts are required to analyze the compromised UAVs. However, it is likely impossible to gather evidence from the drones that do not implement logging capabilities~\cite{Altawy2017}. More specifically, important data such as flight trajectories and onboard-flight data are stored in the Flight Controller's volatile Random Access Memory (RAM), thus making the recovery process a challenging task. Therefore, building models is highly required to provide deep drone forensic analysis~\cite{Al-Dhaqm2021ResearchModels}. However, even with strong forensics models, the existing anti-forensics techniques could potentially thwart the digital investigation process~\cite{Atkinson2021DroneChallenges}. A possible mitigation strategy considers the adoption of a forensic-by-design approach, which integrates the forensics requirements into the design of the UAV system~\cite{Rahman2016Forensic-by-designSystems}. Forensics investigation of UAVs is an unexplored topic of research in UAV security. Existing digital forensics models lack proper unification and standardization to enclose a wider window of commercial UAVs. This is a major issue in UAV forensics, where an adversary could potentially compromise specific UAVs where their forensic models have not been covered yet. 

\smallskip
\noindent \textbf{UAV Intrusion Detection Systems.} Detecting intrusions against UAVs during a flight mission requires real-time analysis of the network traffic. To that end, implementing an IDS for UAVs enables the detection of different classes of intrusions such as signals modification, malware, routing attacks, and message forgery attacks~\cite{Choudhary2018IntrusionSurvey}. In addition, the development of anomaly detection frameworks to monitor malicious behaviors plays an important role in detecting attack patterns~\cite{Schumann2015R2U2:Systems}. Besides, the adoption of honeypot and honeynets along with the IDS can help to protect the flight mission from malicious entities~\cite{Franco2021ASystems}. Since UAV networks constitute a complex cyber-physical system that incorporates multiple components~\cite{Guo2020}, the intrusion detection approaches should consider different information gathering sources to increase the performance. However, more information sources can also increase the communication cost and also result in high computation overhead. Developing such solutions is challenging due to the existing tradeoffs between security and performance. Therefore, there is a need to implement lightweight IDSs to monitor UAV communications and detect attacks. In this respect, some solutions utilize the behavioral profiling of the flight to detect abnormal behavior and malicious intrusions~\cite{Birnbaum2015UnmannedProfiling}. However, cyber attacks that compromise UAVs while ensuring that the flight pattern is consistent cannot be detected by such approaches. 

\smallskip
\noindent \textbf{Secure UAV Communications.} The outcomes of our study at the UAV Communication-level demonstrates the need to develop proper UAV communication protocols and thus to provide reliable and secure communication between different components of the UAV system. However, securing UAV communication channels while achieving a maximum network throughput is still a challenge for the research community. Additionally, authentication of UAVs can secure the communication link and prevent impersonation and replay attacks~\cite{Rodrigues2019}. Developing access policies for UAVs such as authorization and authentication schemes is still a challenging research topic~\cite{Yaacoub2020}. Indeed, any unauthenticated UAV should not be part of the flight mission or gather exchanged data from other UAVs in the network. On the other hand, in multi-UAV scenarios, the use of specific networking models for UAVs such as FANETs~\cite{Bekmezci2013} enables the multi-UAV operations. However, FANETs are vulnerable to different attacks~\cite{Chriki2019FANET:Issues}, and establishing secure communication in Multi-UAVs network remains an open research topic. Although several FANETs routing protocols were proposed in the literature~\cite{Arafat2019a}, they cannot fully meet the security and privacy requirements, and further research in this category is needed~\cite{Maxa2017b}. 

\smallskip
\noindent \textbf{Realistic Implementations.} Practical development and deployment of UAVs requires an emphasis on the tradeoffs between security and performance. From a security point of view, we have to consider the security and privacy requirements of the UAV system. Moreover, we need to consider the energy, computation costs, and storage overheads from a performance perspective. For example, implementing authentication mechanisms or developing lightweight cryptographic protocols for energy-constrained UAVs incorporates the use of cryptographic primitives. However, such implementations might consume too much energy and increase the computational cost. Therefore, finding a strategic solution and balancing both sides is considered a major open research topic. Existing security countermeasures operate under specific hardware or software settings. Therefore, when proposing real-world implementations, we must consider the possible deployment challenges among different UAV systems. A possible solution consists of unifying a deployment interface for various types of UAV systems. Also, it should be emphasized that simulating cyber attack scenarios of UAVs in advance could demonstrate the resilience of existing security measures against cyber attacks before their deployments. Besides, existing simulation environments for UAV security analysis are limited~\cite{Javaid2013UAVSim:Analysis}, and this topic deserves further research efforts.

\smallskip
\noindent \textbf{Privacy Preservation.} The integration of UAVs in the national airspace has raised privacy preservation issues. These implications lead to the leakage of sensitive data collected by UAVs. The collected data might be uploaded to third-party organizations such as cloud servers for storage or processing purposes. In this context, there is a need to protect the privacy of outsourced data. Different privacy-preserving approaches have been proposed in the literature. Examples of mitigating privacy invasion attacks include using privacy-enhancing technologies to preserve consumers' data and guarantee privacy protection with third-party organizations. Namely, the secure computation or differential privacy mechanisms support the privacy of individual users and permit the data coordination between UAVs while guaranteeing privacy. Other examples include the use of homomorphic encryption to perform computational operations over encrypted data~\cite{Acar2018AImplementation} and the Zero Knowledge Proof (ZPF) to validate data without disclosing it.    

\smallskip
\noindent \textbf{Secure Data Aggregation.} The extensive use of UAVs in different application domains increased the amount of collected and shared data. The collected data is usually aggregated to use the resources efficiently. However, the data aggregation process needs to be protected against malicious actors. The deployment of aggregation schemes should consider encryption techniques to provide confidentiality, thus enabling a secure transfer of information between the GCS and UAVs. In addition, providing efficient and secure data aggregation approaches for UAVs will reduce the energy and communication costs while ensuring confidentiality. However, developing such schemes remains an ongoing challenge.

\smallskip
\noindent \textbf{Emerging Technologies.} Recently, there has been an extensive use of emerging technologies to secure UAVs: Artificial Intelligence, Blockchain technology, SDN, and Fog Computing~\cite{Hassija2021}~\cite{Syed2020a}. These technologies are applied in various civilian applications. The distributed architecture of Blockchain technology adds an extra layer of security at the communication level~\cite{Ghribi2020ANetworks}. With Smart Contracts, cryptographic hash functions to store data as a chain of blocks, and the consensus mechanisms, it becomes challenging for the adversary to tamper with UAVs communication. However, the major applications of Blockchain for UAV communication security suffer from real-time deployment for highly mobile UAVs~\cite{Kumari2020}.
Moreover, the real-world implementation of Blockchain technology to secure UAV networks is still an ongoing research topic. The evolution of Artificial Intelligence technology such as ML algorithms has demonstrated tremendous benefits for security-oriented applications, such as protecting UAV networks from attacks and privacy leakage. Different ML-based security frameworks have been proposed in the literature to address various security issues, including malicious drone detection and DoS attacks~\cite{Matson2019UAVModels}. Recently, federated learning techniques are reported to show better results compared to traditional ML algorithms. For example, the use of drone authentication models based on drone's Radio Frequency features in IoT networks~\cite{Yazdinejad2021FederatedAuthentication}.
Nonetheless, there is a lack of existing UAV datasets to train ML models (e.g., network traffic datasets, malware datasets, etc.). Furthermore, some ML models can fail to detect cyber attacks to UAVs~\cite{Hassija2021}. The use of SDN-based UAV networks enables the security of UAV communications. This technology offers dynamic flow control and a programmable network for different security functions. Hence, protecting the UAV network from potential cyber attacks. A major drawback of using  such a technology is the high end-to-end delay for non-delay tolerant UAV applications. Moreover, the link between the data plane and the control plane could be subject to attacks. In addition to these, the integration of UAVs in smart cities implies processing and storing of a large amount of data. To that end, the use of fog computing technology can help to process and store the data. Moreover, Fog Computing supports secure communication between the UAVs and the fog layer that is salable and that has low latency. However, the current fog architecture is not tailored for the UAV model, and adopting such an architecture might increase the data processing time, especially for multi-UAV networks. The next generations of UAVs will incorporate diverse emerging technologies~\cite{Nayyar2020TheDrones}. Therefore, there is a need from academia and industry for further research regarding the use of emerging technologies to secure UAVs in civilian applications.

\section{Conclusion}\label{sec:conclusion}
In this paper, we presented an exhaustive survey on security and privacy issues of Unmanned Aerial Vehicles. We thoroughly dissected UAV security issues at four levels: the \textit{Sensor-level}, the \textit{Hardware-level}, the \textit{Software-level}, and the \textit{Communication-level}. Furthermore, we discussed the privacy issues of UAVs, threats, and possible solutions. Next, we presented the lessons learned with the security and privacy aspects of UAVs, and also provided possible future research directions. With the increased number of commercial UAVs in civilian airspace, security and privacy issues became a highly urgent matter of national security. Therefore, industry, academia, and law enforcement need to collaborate and develop new security frameworks, standards, and regulations. Nowadays, existing drones manufacturers are deploying the next generation of commercial UAVs in the market, and security and privacy considerations are way behind. Our survey provides a valuable reference for the research community to learn more about building and designing secure UAV architectures.

\section*{ACKNOWLEDGMENTS}
This work is partially supported by the US National Science Foundation (Awards: NSF-CAREER-CNS-1453647,
NSF-1663051, and NSF-1718116). The views are those of the authors only.
\bibliographystyle{IEEEtran}
\bibliography{bibs/references}

\begin{thebibliography}{100}
\providecommand{\url}[1]{#1}
\csname url@samestyle\endcsname
\providecommand{\newblock}{\relax}
\providecommand{\bibinfo}[2]{#2}
\providecommand{\BIBentrySTDinterwordspacing}{\spaceskip=0pt\relax}
\providecommand{\BIBentryALTinterwordstretchfactor}{4}
\providecommand{\BIBentryALTinterwordspacing}{\spaceskip=\fontdimen2\font plus
\BIBentryALTinterwordstretchfactor\fontdimen3\font minus
  \fontdimen4\font\relax}
\providecommand{\BIBforeignlanguage}[2]{{%
\expandafter\ifx\csname l@#1\endcsname\relax
\typeout{** WARNING: IEEEtran.bst: No hyphenation pattern has been}%
\typeout{** loaded for the language `#1'. Using the pattern for}%
\typeout{** the default language instead.}%
\else
\language=\csname l@#1\endcsname
\fi
#2}}
\providecommand{\BIBdecl}{\relax}
\BIBdecl

\bibitem{Hayat2016}
S.~Hayat, E.~Yanmaz, and R.~Muzaffar, ``{Survey on Unmanned Aerial Vehicle
  Networks for Civil Applications: A Communications Viewpoint},'' \emph{IEEE
  Communications Surveys and Tutorials}, vol.~18, no.~4, pp. 2624--2661, 2016.

\bibitem{Gupta2016}
L.~Gupta, R.~Jain, and G.~Vaszkun, ``{Survey of Important Issues in UAV
  Communication Networks},'' \emph{IEEE Communications Surveys and Tutorials},
  vol.~18, no.~2, pp. 1123--1152, 2016.

\bibitem{HosseinMotlagh2016Low-AltitudePerspectives}
N.~Hossein~Motlagh, T.~Taleb, and O.~Arouk, ``{Low-Altitude Unmanned Aerial
  Vehicles-Based Internet of Things Services: Comprehensive Survey and Future
  Perspectives},'' \emph{IEEE Internet of Things Journal}, vol.~3, no.~6, pp.
  899--922, 2016.

\bibitem{Kellermann2020DronesReview}
R.~Kellermann, T.~Biehle, and L.~Fischer, ``{Drones for parcel and passenger
  transportation: A literature review},'' \emph{Transportation Research
  Interdisciplinary Perspectives}, vol.~4, p. 100088, 3 2020.

\bibitem{Market}
\BIBentryALTinterwordspacing
``{Commercial Drone Market Size}.'' [Online]. Available:
  \url{https://www.grandviewresearch.com/industry-analysis/global-commercial-drones-market}
\BIBentrySTDinterwordspacing

\bibitem{AviationAdministrationFAATables}
F.~Aviation~Administration, ``{FAA National Forecast FY 2019-2039 Full Forecast
  Document and Tables},'' Tech. Rep.

\bibitem{LiuRiseIssues}
Z.~Liu, Z.~Li, B.~Liu, X.~Fu, I.~Raptis, and K.~Ren, ``{Rise of Mini-Drones},''
  in \emph{Proceedings of the 2015 Workshop on Privacy-Aware Mobile Computing},
  2015, pp. 7--12.

\bibitem{DroneIncidents}
\BIBentryALTinterwordspacing
``{Map of World Wide Drone Incidents - Dedrone}.'' [Online]. Available:
  \url{https://www.dedrone.com/resources/incidents/all}
\BIBentrySTDinterwordspacing

\bibitem{Sedjelmaci2016}
H.~Sedjelmaci, S.~M. Senouci, and M.~A. Messous, ``{How to detect cyber-attacks
  in unmanned aerial vehicles network?}'' \emph{2016 IEEE Global Communications
  Conference, GLOBECOM 2016 - Proceedings}, 2016.

\bibitem{Guo2020}
R.~Guo, B.~Wang, and J.~Weng, ``{Vulnerabilities and Attacks of UAV Cyber
  Physical Systems},'' \emph{ACM International Conference Proceeding Series},
  pp. 8--12, 2020.

\bibitem{Yahuza2021a}
M.~Yahuza, M.~Y.~I. Idris, I.~B. Ahmedy, A.~W.~A. Wahab, T.~Nandy, N.~M. Noor,
  and A.~Bala, ``{Internet of Drones Security and Privacy Issues: Taxonomy and
  Open Challenges},'' \emph{IEEE Access}, vol.~9, pp. 57\,243--57\,270, 2021.

\bibitem{Yaacoub2020}
J.-P. Yaacoub, H.~Noura, O.~Salman, and A.~Chehab, ``{Security analysis of
  drones systems: Attacks, limitations, and recommendations},'' \emph{Internet
  of Things}, vol.~11, p. 100218, 2020.

\bibitem{Watkins2018}
L.~Watkins, J.~Ramos, G.~Snow, J.~Vallejo, W.~H. Robinson, A.~D. Rubin,
  J.~Ciocco, F.~Jedrzejewski, J.~Liu, and C.~Li, ``{Exploiting multi-vendor
  vulnerabilities as back-doors to counter the threat of rogue small unmanned
  aerial systems},'' \emph{Proceedings of the 1st ACM MobiHoc Workshop on
  Mobile IoT Sensing, Security, and Privacy, Mobile IoT SSP 2018}, 2018.

\bibitem{Li2019b}
Z.~Li, C.~Gao, Q.~Yue, and X.~Fu, ``{Toward Drone Privacy via Regulating
  Altitude and Payload},'' \emph{2019 International Conference on Computing,
  Networking and Communications}, pp. 562--566, 2019.

\bibitem{DronesSociety}
\BIBentryALTinterwordspacing
``{Drones as the New "Flying IoT" | IEEE Computer Society}.'' [Online].
  Available:
  \url{https://www.computer.org/publications/tech-news/research/flying-iot-toward-low-power-vision-sky}
\BIBentrySTDinterwordspacing

\bibitem{Iqtidar2021ASystems}
N.~Iqtidar, S.~Kumar, R.~Ashiqur, and U.~Selcuk, ``{A Survey on Security and
  Privacy Issues in Modern Healthcare Systems},'' \emph{ACM Transactions on
  Computing for Healthcare}, vol.~2, no.~3, pp. 1--44, 7 2021.

\bibitem{RondonSurveyPerspective2}
\BIBentryALTinterwordspacing
L.~P. Rondon, L.~Babun, A.~Aris, K.~Akkaya, and A.~S. Uluagac, ``{Survey on
  Enterprise Internet-of-Things Systems (E-IoT): A Security Perspective},''
  2021. [Online]. Available: \url{http://arxiv.org/abs/2102.10695}
\BIBentrySTDinterwordspacing

\bibitem{Altawy2017}
R.~Altawy and A.~M. Youssef, ``{Security, privacy, and safety aspects of
  civilian drones: A survey},'' \emph{ACM Transactions on Cyber-Physical
  Systems}, vol.~1, no.~2, pp. 1--25, 2017.

\bibitem{Krishna2017}
C.~G. Krishna and R.~R. Murphy, ``{A review on cybersecurity vulnerabilities
  for unmanned aerial vehicles},'' \emph{SSRR 2017 - 15th IEEE International
  Symposium on Safety, Security and Rescue Robotics, Conference}, pp. 194--199,
  2017.

\bibitem{Maxa2017b}
J.-A. Maxa, M.-S.~B. Mahmoud, and N.~Larrieu, ``{Survey on UAANET Routing
  Protocols and Network Security Challenges},'' \emph{Ad Hoc {\&} Sensor
  Wireless Networks}, 3 2017.

\bibitem{Choudhary2018a}
G.~Choudhary, V.~Sharma, T.~Gupta, J.~Kim, and I.~You, ``{Internet of drones
  (IoD): Threats, vulnerability, and security perspectives},'' in \emph{The 3rd
  International Symposium on Mobile Internet Security}, no.~37, 2018, pp.
  1--13.

\bibitem{Lin2018}
C.~Lin, D.~He, N.~Kumar, K.~K.~R. Choo, A.~Vinel, and X.~Huang, ``{Security and
  Privacy for the Internet of Drones: Challenges and Solutions},'' \emph{IEEE
  Communications Magazine}, vol.~56, no.~1, pp. 64--69, 2018.

\bibitem{Shakhatreh2019}
H.~Shakhatreh, A.~H. Sawalmeh, A.~Al-Fuqaha, Z.~Dou, E.~Almaita, I.~Khalil,
  N.~S. Othman, A.~Khreishah, and M.~Guizani, ``{Unmanned Aerial Vehicles
  (UAVs): A Survey on Civil Applications and Key Research Challenges},''
  \emph{IEEE Access}, vol.~7, pp. 48\,572--48\,634, 2019.

\bibitem{Nassi2019}
\BIBentryALTinterwordspacing
B.~Nassi, A.~Shabtai, R.~Masuoka, and Y.~Elovici, ``{SoK - Security and privacy
  in the age of drones: Threats, challenges, solution mechanisms, and
  scientific gaps},'' \emph{arXiv}, pp. 1--17, 2019. [Online]. Available:
  \url{http://arxiv.org/abs/1903.05155}
\BIBentrySTDinterwordspacing

\bibitem{Fotouhi2019a}
A.~Fotouhi, H.~Qiang, M.~Ding, M.~Hassan, L.~G. Giordano, A.~Garcia-Rodriguez,
  and J.~Yuan, ``{Survey on UAV cellular communications: Practical aspects,
  standardization advancements, regulation, and security challenges},''
  \emph{IEEE Communications Surveys and Tutorials}, vol.~21, no.~4, pp.
  3417--3442, 2019.

\bibitem{Chriki2019FANET:Issues}
A.~Chriki, H.~Touati, H.~Snoussi, and F.~Kamoun, ``{FANET: Communication,
  mobility models and security issues},'' \emph{Computer Networks}, vol. 163,
  p. 106877, 2019.

\bibitem{Boccadoro2020}
\BIBentryALTinterwordspacing
P.~Boccadoro, D.~Striccoli, and L.~A. Grieco, ``{An Extensive Survey on the
  Internet of Drones},'' Tech. Rep., 2020. [Online]. Available:
  \url{http://arxiv.org/abs/2007.12611}
\BIBentrySTDinterwordspacing

\bibitem{Wang2020}
H.~Wang, H.~Zhao, J.~Zhang, D.~Ma, J.~Li, and J.~Wei, ``{Survey on Unmanned
  Aerial Vehicle Networks: A Cyber Physical System Perspective},'' \emph{IEEE
  Communications Surveys and Tutorials}, vol.~22, no.~2, pp. 1027--1070, 2020.

\bibitem{Hentati2020}
A.~I. Hentati and L.~C. Fourati, ``{Comprehensive survey of UAVs communication
  networks},'' \emph{Computer Standards and Interfaces}, vol.~72, no. September
  2019, p. 103451, 2020.

\bibitem{Zhi2020}
Y.~Zhi, Z.~Fu, X.~Sun, and J.~Yu, ``{Security and Privacy Issues of UAV: A
  Survey},'' \emph{Mobile Networks and Applications}, vol.~25, no.~1, pp.
  95--101, 2020.

\bibitem{Sharma2020}
A.~Sharma, P.~Vanjani, N.~Paliwal, C.~M. Basnayaka, D.~N.~K. Jayakody, H.~C.
  Wang, and P.~Muthuchidambaranathan, ``{Communication and networking
  technologies for UAVs: A survey},'' \emph{Journal of Network and Computer
  Applications}, vol. 168, no. June, p. 102739, 2020.

\bibitem{Noor2020a}
F.~Noor, M.~A. Khan, A.~Al-Zahrani, I.~Ullah, and K.~A. Al-Dhlan, ``{A review
  on communications perspective of flying AD-HOC networks: Key enabling
  wireless technologies, applications, challenges and open research topics},''
  \emph{Drones}, vol.~4, no.~4, pp. 1--14, 2020.

\bibitem{Mishra2020}
D.~Mishra and E.~Natalizio, ``{A survey on cellular-connected UAVs: Design
  challenges, enabling 5G/B5G innovations, and experimental advancements},''
  \emph{Computer Networks}, vol. 182, no. August, p. 107451, 2020.

\bibitem{Syed2020a}
F.~Syed, S.~K. Gupta, S.~Hamood~Alsamhi, M.~Rashid, and X.~Liu, ``{A survey on
  recent optimal techniques for securing unmanned aerial vehicles
  applications},'' \emph{Transactions on Emerging Telecommunications
  Technologies}, vol.~32, no.~7, 2021.

\bibitem{Yahuza2021}
M.~Yahuza, M.~Y.~I. Idris, I.~B. Ahmedy, A.~W.~A. Wahab, T.~Nandy, N.~M. Noor,
  and A.~Bala, ``{Internet of Drones Security and Privacy Issues: Taxonomy and
  Open Challenges},'' \emph{IEEE Access}, vol.~9, pp. 57\,243--57\,270, 2021.

\bibitem{Nassi2021SoK:Drones}
B.~Nassi, R.~Bitton, R.~Masuoka, A.~Shabtai, and Y.~Elovici, ``{SoK: Security
  and Privacy in the Age of Commercial Drones},'' \emph{2021 2021 IEEE
  Symposium on Security and Privacy (SP)}, no. Section IV, pp. 73--90, 2021.

\bibitem{Shafique2021a}
A.~Shafique, A.~Mehmood, and M.~Elhadef, ``{Survey of Security Protocols and
  Vulnerabilities in Unmanned Aerial Vehicles},'' \emph{IEEE Access}, vol.~9,
  pp. 46\,927--46\,948, 2021.

\bibitem{Hassija2021}
V.~Hassija, V.~Chamola, A.~Agrawal, A.~Goyal, N.~C. Luong, D.~Niyato, F.~R. Yu,
  and M.~Guizani, ``{Fast, Reliable, and Secure Drone Communication: A
  Comprehensive Survey},'' \emph{IEEE Communications Surveys {\&} Tutorials},
  vol.~PP, no.~c, p.~1, 2021.

\bibitem{TeachPi}
\BIBentryALTinterwordspacing
``{Teach, Learn, and Make with Raspberry Pi}.'' [Online]. Available:
  \url{https://www.raspberrypi.org/}
\BIBentrySTDinterwordspacing

\bibitem{BeagleBoard.orgMaking}
\BIBentryALTinterwordspacing
``{BeagleBoard.org - community supported open hardware computers for making}.''
  [Online]. Available: \url{https://beagleboard.org/}
\BIBentrySTDinterwordspacing

\bibitem{patent}
U.~States, ``{(12) Patent Application Publication (10) Pub. No.: US
  2007/0244608 A1},'' vol.~1, no.~19, 2007.

\bibitem{CopterDocumentation}
\BIBentryALTinterwordspacing
``{Copter Home — Copter documentation}.'' [Online]. Available:
  \url{https://ardupilot.org/copter/}
\BIBentrySTDinterwordspacing

\bibitem{crazy}
\BIBentryALTinterwordspacing
B.~Ab, ``{Crazyflie 2.1},'' pp. 7--9. [Online]. Available:
  \url{https://www.seeedstudio.com/crazyflie-V2-1-p-2894.html}
\BIBentrySTDinterwordspacing

\bibitem{KKMulticopterlazyzero.de}
\BIBentryALTinterwordspacing
``{KKMulticopter Flashtool [lazyzero.de]}.'' [Online]. Available:
  \url{https://lazyzero.de/en/modellbau/kkmulticopterflashtool/start}
\BIBentrySTDinterwordspacing

\bibitem{UASOverview}
\BIBentryALTinterwordspacing
``{UAS Remote Identification Overview}.'' [Online]. Available:
  \url{https://www.faa.gov/uas/getting_started/remote_id/}
\BIBentrySTDinterwordspacing

\bibitem{Petricca2011}
L.~Petricca, P.~Ohlckers, and C.~Grinde, ``{Micro- and nano-air vehicles: State
  of the art},'' \emph{International Journal of Aerospace Engineering}, vol.
  2011, 2011.

\bibitem{Andre2014}
T.~Andre, K.~A. Hummel, A.~P. Schoellig, E.~Yanmaz, M.~Asadpour,
  C.~Bettstetter, P.~Grippa, H.~Hellwagner, S.~Sand, and S.~Zhang,
  ``{Application-driven design of aerial communication networks},'' \emph{IEEE
  Communications Magazine}, vol.~52, no.~5, pp. 129--137, 2014.

\bibitem{Hooper2016SecuringAttacks}
M.~Hooper, Y.~Tian, R.~Zhou, B.~Cao, A.~P. Lauf, L.~Watkins, W.~H. Robinson,
  and W.~Alexis, ``{Securing commercial WiFi-based UAVs from common security
  attacks},'' in \emph{Proceedings - IEEE Military Communications Conference
  MILCOM}, 2016, pp. 1213--1218.

\bibitem{Valavanis2015}
K.~P. Valavanis and G.~J. Vachtsevanos, \emph{{Handbook of unmanned aerial
  vehicles}}.\hskip 1em plus 0.5em minus 0.4em\relax Springer Netherlands, 1
  2015.

\bibitem{Mozaffari2019}
M.~Mozaffari, W.~Saad, M.~Bennis, Y.~H. Nam, and M.~M. Debbah, ``{A tutorial on
  UAVs for wireless networks: Applications, challenges, and open problems},''
  \emph{IEEE Communications Surveys and Tutorials}, vol.~21, no.~3, pp.
  2334--2360, 2019.

\bibitem{Bekmezci2013}
I.~Bekmezci, O.~K. Sahingoz, and S.~Temel, ``{Flying Ad-Hoc Networks (FANETs):
  A survey},'' \emph{Ad Hoc Networks}, vol.~11, no.~3, pp. 1254--1270, 2013.

\bibitem{Li2013}
J.~Li, Y.~Zhou, and L.~Lamont, ``{Communication architectures and protocols for
  networking unmanned aerial vehicles},'' \emph{2013 IEEE Globecom Workshops,
  GC Wkshps 2013}, pp. 1415--1420, 2013.

\bibitem{Chriki2019UAV-GCSApplications}
A.~Chriki, H.~Touati, H.~Snoussi, and F.~Kamoun, ``{UAV-GCS centralized
  data-oriented communication architecture for crowd surveillance
  applications},'' in \emph{2019 15th International Wireless Communications and
  Mobile Computing Conference, IWCMC 2019}, 2019, pp. 2064--2069.

\bibitem{Arafat2019a}
M.~Y. Arafat and S.~Moh, ``{Routing protocols for unmanned aerial vehicle
  networks: A survey},'' \emph{IEEE Access}, vol.~7, pp. 99\,694--99\,720,
  2019.

\bibitem{Oubbati2019}
O.~S. Oubbati, M.~Atiquzzaman, P.~Lorenz, M.~H. Tareque, and M.~S. Hossain,
  ``{Routing in flying Ad Hoc networks: Survey, constraints, and future
  challenge perspectives},'' \emph{IEEE Access}, vol.~7, pp. 81\,057--81\,105,
  2019.

\bibitem{ShumeyeLakew2020}
D.~Shumeye~Lakew, U.~Sa'Ad, N.~N. Dao, W.~Na, and S.~Cho, ``{Routing in Flying
  Ad Hoc Networks: A Comprehensive Survey},'' \emph{IEEE Communications Surveys
  and Tutorials}, vol.~22, no.~2, pp. 1071--1120, 4 2020.

\bibitem{Arafat2019}
M.~Y. Arafat and S.~Moh, ``{A Survey on Cluster-Based Routing Protocols for
  Unmanned Aerial Vehicle Networks},'' \emph{IEEE Access}, vol.~7, pp.
  498--516, 2019.

\bibitem{Koubaa2019}
A.~Koubaa, A.~Allouch, M.~Alajlan, Y.~Javed, A.~Belghith, and M.~Khalgui,
  ``{Micro Air Vehicle Link (MAVlink) in a Nutshell: A Survey},'' \emph{IEEE
  Access}, vol.~7, pp. 87\,658--87\,680, 2019.

\bibitem{Kwon2018}
Y.~M. Kwon, J.~Yu, B.~M. Cho, Y.~Eun, and K.~J. Park, ``{Empirical Analysis of
  MAVLink Protocol Vulnerability for Attacking Unmanned Aerial Vehicles},''
  \emph{IEEE Access}, vol.~6, pp. 43\,203--43\,212, 2018.

\bibitem{Kriz2015}
V.~Kriz and P.~Gabrlik, ``{UranusLink-Communication protocol for UAV with small
  overhead and encryption ability},'' \emph{IFAC-PapersOnLine}, vol.~28, no.~4,
  pp. 474--479, 2015.

\bibitem{Khan2020}
N.~A. Khan, N.~Z. Jhanjhi, S.~N. Brohi, and A.~Nayyar, ``{Emerging use of
  UAV’s: secure communication protocol issues and challenges},'' in
  \emph{Drones in Smart-Cities}.\hskip 1em plus 0.5em minus 0.4em\relax
  Elsevier Inc., 2020, pp. 37--55.

\bibitem{Choudhary2018}
G.~Choudhary, V.~Sharma, I.~You, K.~Yim, I.~R. Chen, and J.~H. Cho,
  ``{Intrusion Detection Systems for Networked Unmanned Aerial Vehicles: A
  Survey},'' \emph{2018 14th International Wireless Communications and Mobile
  Computing Conference, IWCMC 2018}, pp. 560--565, 2018.

\bibitem{Uluagac2014SensoryCall}
A.~S. Uluagac, V.~Subramanian, and R.~Beyah, ``{Sensory channel threats to
  cyber physical systems: A wake-up call},'' \emph{2014 IEEE Conference on
  Communications and Network Security}, pp. 301--309, 12 2014.

\bibitem{Sikder2021AApplications}
A.~K. Sikder, G.~Petracca, H.~Aksu, T.~Jaeger, and A.~S. Uluagac, ``{A Survey
  on Sensor-Based Threats and Attacks to Smart Devices and Applications},''
  \emph{IEEE Communications Surveys and Tutorials}, vol.~23, no.~2, pp.
  1125--1159, 4 2021.

\bibitem{AruSaputro2020ImplementationTest}
J.~Aru~Saputro, E.~Egistian~Hartadi, and M.~Syahral, ``{Implementation of GPS
  Attacks on DJI Phantom 3 Standard Drone as a Security Vulnerability Test},''
  \emph{Proceeding - 1st International Conference on Information Technology,
  Advanced Mechanical and Electrical Engineering, ICITAMEE 2020}, pp. 95--100,
  10 2020.

\bibitem{mo2010}
E.~Deligne, ``{ARDrone corruption},'' \emph{Journal in Computer Virology},
  vol.~8, no. 1-2, pp. 15--27, 2012.

\bibitem{Wesson2013HackingDrones}
K.~Wesson and T.~Humphreys, ``{Hacking drones},'' \emph{Scientific American},
  vol. 309, no.~5, pp. 54--59, 2013.

\bibitem{Kerns2014UnmannedSpoofing}
A.~J. Kerns, D.~P. Shepard, J.~A. Bhatti, and T.~E. Humphreys, ``{Unmanned
  aircraft capture and control via GPS spoofing},'' \emph{Journal of Field
  Robotics}, vol.~31, no.~4, pp. 617--636, 2014.

\bibitem{Seo2015EffectSignal}
S.-H. Seo, B.-H. Lee, S.-H. Im, and G.-I. Jee, ``{Effect of Spoofing on
  Unmanned Aerial Vehicle using Counterfeited GPS Signal},'' \emph{Journal of
  Positioning, Navigation, and Timing}, vol.~4, no.~2, pp. 57--65, 2015.

\bibitem{Roth2009SimulationGyroscopes}
\BIBentryALTinterwordspacing
G.~Roth, ``{Simulation of the Effects of Acoustic Noise on MEMS Gyroscopes},''
  2009. [Online]. Available: \url{https://etd.auburn.edu//handle/10415/1773}
\BIBentrySTDinterwordspacing

\bibitem{Fischler1981RandomCartography}
M.~A. Fischler and R.~C. Bolles, ``{Random sample consensus: A Paradigm for
  Model Fitting with Applications to Image Analysis and Automated
  Cartography},'' \emph{Communications of the ACM}, vol.~24, no.~6, pp.
  381--395, 1981.

\bibitem{Feng2020}
Z.~Feng, N.~Guan, M.~Lv, W.~Liu, Q.~Deng, X.~Liu, and W.~Yi, ``{Efficient drone
  hijacking detection using two-step GA-XGBoost},'' \emph{Journal of Systems
  Architecture}, vol. 103, pp. 1414--1419, 2020.

\bibitem{Vaeshosaz2019}
M.~Varshosaz, A.~Afary, B.~Mojaradi, M.~Saadatseresht, and E.~G. Parmehr,
  ``{Spoofing detection of civilian UAVs using visual odometry},'' \emph{ISPRS
  International Journal of Geo-Information}, vol.~9, no.~1, 2019.

\bibitem{Feng2019AnUnit}
Z.~Feng, N.~Guan, M.~Lv, W.~Liu, Q.~Deng, X.~Liu, and W.~Yi, ``{An efficient
  UAV hijacking detection method using onboard inertial measurement unit},''
  \emph{ACM Transactions on Embedded Computing Systems}, vol.~17, no.~6, pp.
  1--19, 2019.

\bibitem{Abera2019DIAT:Systems}
T.~Abera, R.~Bahmani, F.~Brasser, A.~Ibrahim, A.-R. Sadeghi, and M.~Schunter,
  ``{DIAT: Data Integrity Attestation for Resilient Collaboration of Autonomous
  Systems}.''\hskip 1em plus 0.5em minus 0.4em\relax NDSS, 2019.

\bibitem{Wu2013AutonomousSensors}
A.~D. Wu, E.~N. Johnson, M.~Kaess, F.~Dellaert, and G.~Chowdhary, ``{Autonomous
  flight in gps-denied environments using monocular vision and inertial
  sensors},'' in \emph{Journal of Aerospace Information Systems}, vol.~10,
  no.~4, 2013, pp. 172--186.

\bibitem{Wu2013AutonomousSensorsb}
------, ``{Autonomous flight in gps-denied environments using monocular vision
  and inertial sensors},'' \emph{Journal of Aerospace Information Systems},
  vol.~10, no.~4, pp. 172--186, 2013.

\bibitem{Whelan2020}
J.~Whelan, T.~Sangarapillai, O.~Minawi, A.~Almehmadi, and K.~El-Khatib,
  ``{Novelty-based Intrusion Detection of Sensor Attacks on Unmanned Aerial
  Vehicles},'' \emph{Q2SWinet 2020 - Proceedings of the 16th ACM Symposium on
  QoS and Security for Wireless and Mobile Networks}, pp. 23--28, 2020.

\bibitem{Arthur2019DetectingIDS}
M.~P. Arthur, ``{Detecting signal spoofing and jamming attacks in UAV networks
  using a lightweight IDS},'' in \emph{CITS 2019 - Proceeding of the 2019
  International Conference on Computer, Information and Telecommunication
  Systems}.\hskip 1em plus 0.5em minus 0.4em\relax Institute of Electrical and
  Electronics Engineers Inc., 8 2019.

\bibitem{Choi2018DetectingApproach}
H.~Choi, W.~C. Lee, Y.~Aafer, F.~Fei, Z.~Tu, X.~Zhang, D.~Xu, and X.~Deng,
  ``{Detecting attacks against robotic vehicles: A control invariant
  approach},'' \emph{Proceedings of the ACM Conference on Computer and
  Communications Security}, pp. 801--816, 2018.

\bibitem{Quinonez2020SAVIOR:Invariants}
R.~Quinonez, J.~Giraldo, L.~Salazar, E.~Bauman, A.~Cardenas, and Z.~Lin,
  ``{SAVIOR: Securing autonomous vehicles with robust physical invariants},''
  in \emph{Proceedings of the 29th USENIX Security Symposium}, 2020, pp.
  895--912.

\bibitem{Son2015}
Y.~Son, H.~Shin, D.~Kim, Y.~Park, J.~Noh, K.~Choi, J.~Choi, and Y.~Kim,
  ``{Rocking drones with intentional sound noise on gyroscopic sensors},''
  \emph{Proceedings of the 24th USENIX Security Symposium}, pp. 881--896, 2015.

\bibitem{Davidson2016}
D.~Davidson, H.~Wu, R.~Jellinek, T.~Ristenpart, and V.~Singh, ``{Controlling
  UAVs with sensor input spoofing attacks},'' \emph{10th USENIX Workshop on
  Offensive Technologies, WOOT 2016}, 2016.

\bibitem{Vosatka2017IntroductionTrojans}
J.~Vosatka, ``{Introduction to hardware Trojans},'' in \emph{The Hardware
  Trojan War: Attacks, Myths, and Defenses}.\hskip 1em plus 0.5em minus
  0.4em\relax Springer International Publishing, 11 2017, pp. 15--51.

\bibitem{Rahman2020IntrusionVehicles}
M.~A. Rahman, M.~T. Rahman, M.~Kisacikoglu, and K.~Akkaya, ``{Intrusion
  Detection Systems-Enabled Power Electronics for Unmanned Aerial Vehicles},''
  in \emph{2020 IEEE CyberPELS}, 2020, pp. 1--5.

\bibitem{GilCasals2013GenericDetection}
S.~Gil~Casals, P.~Owezarski, and G.~Descargues, ``{Generic and autonomous
  system for airborne networks cyber-threat detection},'' \emph{AIAA/IEEE
  Digital Avionics Systems Conference - Proceedings}, 2013.

\bibitem{Collision}
J.~N. Yasin, S.~A.~S. Mohamed, M.~H. Haghbayan, J.~Heikkonen, H.~Tenhunen, and
  J.~Plosila, ``{Unmanned Aerial Vehicles (UAVs): Collision Avoidance Systems
  and Approaches},'' \emph{IEEE Access}, vol.~8, pp. 105\,139--105\,155, 2020.

\bibitem{Hannah2020}
J.~Hannah, R.~Mills, and R.~Dill, ``{Traffic Collision Avoidance System: Threat
  Actor Model and Attack Taxonomy},'' in \emph{Proceedings of the 22nd
  International Conference on New Trends in Civil Aviation 2020, NTCA
  2020}.\hskip 1em plus 0.5em minus 0.4em\relax Institute of Electrical and
  Electronics Engineers Inc., 11 2020, pp. 17--26.

\bibitem{Alwateer2019DroneProcessing}
M.~Alwateer, S.~W. Loke, and A.~M. Zuchowicz, ``{Drone services: issues in
  drones for location-based services from human-drone interaction to
  information processing},'' \emph{Journal of Location Based Services},
  vol.~13, no.~2, pp. 94--127, 4 2019.

\bibitem{Lee2020}
J.~Lee, S.~Ryu, and H.~J. Kim, ``{Stable Flight of a Flapping-Wing Micro Air
  Vehicle under Wind Disturbance},'' \emph{IEEE Robotics and Automation
  Letters}, vol.~5, no.~4, p.~1, 2020.

\bibitem{hodgkins}
\BIBentryALTinterwordspacing
K.~Hodgkins, ``{Anti-drone shoulder rifle lets police take control of UAVs with
  radio pulses.(2015)},'' 2015. [Online]. Available:
  \url{https://www.digitaltrends.com/cool-tech/battle-innovations-anti-drone-gun/}
\BIBentrySTDinterwordspacing

\bibitem{BelikovetskyDr0wned-Cyber-PhysicalManufacturing}
S.~Belikovetsky, M.~Yampolskiy, J.~Toh, and J.~Gatlin,
  ``{dr0wned-Cyber-Physical Attack with Additive Manufacturing},'' in
  \emph{11th USENIX Workshop on Offensive Technologies (WOOT 17)}.

\bibitem{Desnitsky2021}
V.~Desnitsky and I.~Kotenko, ``{Simulation and assessment of battery depletion
  attacks on unmanned aerial vehicles for crisis management infrastructures},''
  \emph{Simulation Modelling Practice and Theory}, vol. 107, p. 102244, 2 2021.

\bibitem{lopez2017}
A.~B. Lopez, K.~Vatanparvar, A.~P. Deb~Nath, S.~Yang, S.~Bhunia, and M.~A.
  Al~Faruque, ``{A Security Perspective on Battery Systems of the Internet of
  Things},'' \emph{Journal of Hardware and Systems Security}, vol.~1, no.~2,
  pp. 188--199, 2017.

\bibitem{Rodday2016HackingDrone}
\BIBentryALTinterwordspacing
N.~Rodday, ``{Hacking a Professional Drone},'' \emph{RSAConference2016}, 2016.
  [Online]. Available:
  \url{https://www.rsaconference.com/writable/presentations/file_upload/ht-w03-hacking_a_professional_police_drone.pdf}
\BIBentrySTDinterwordspacing

\bibitem{Nigh2021AdaTrust:Construction}
C.~Nigh and A.~Orailoglu, ``{AdaTrust: Combinational Hardware Trojan Detection
  through Adaptive Test Pattern Construction},'' \emph{IEEE Transactions on
  Very Large Scale Integration (VLSI) Systems}, vol.~29, no.~3, pp. 544--557, 3
  2021.

\bibitem{McNeely2016DetectionStatistics}
J.~McNeely, M.~Hatfield, A.~Hasan, and N.~Jahan, ``{Detection of UAV hijacking
  and malfunctions via variations in flight data statistics},''
  \emph{Proceedings - International Carnahan Conference on Security
  Technology}, vol.~0, 7 2016.

\bibitem{Williams2008SupplyAgenda}
Z.~Williams, J.~E. Lueg, and S.~A. Lemay, ``{Supply chain security: An overview
  and research agenda},'' \emph{The International Journal of Logistics
  Management}, vol.~19, no.~2, pp. 254--281, 2008.

\bibitem{Podhradsky2017ImprovingLink}
M.~Podhradsky, C.~Coopmans, and N.~Hoffer, ``{Improving communication security
  of open source UAVs: Encrypting radio control link},'' \emph{2017
  International Conference on Unmanned Aircraft Systems, ICUAS 2017}, pp.
  1153--1159, 7 2017.

\bibitem{GitHubSMACCM}
\BIBentryALTinterwordspacing
``{GitHub - GaloisInc/gec: Embedded-friendly crypto a la SMACCM}.'' [Online].
  Available: \url{https://github.com/GaloisInc/gec/}
\BIBentrySTDinterwordspacing

\bibitem{He2021AnAttack}
D.~He, G.~Yang, H.~Li, S.~Chan, Y.~Cheng, and N.~Guizani, ``{An Effective
  Countermeasure against UAV Swarm Attack},'' \emph{IEEE Network}, vol.~35,
  no.~1, pp. 380--385, 3 2021.

\bibitem{Pu2020}
C.~Pu and Y.~Li, ``{Lightweight Authentication Protocol for Unmanned Aerial
  Vehicles Using Physical Unclonable Function and Chaotic System},'' \emph{IEEE
  Workshop on Local and Metropolitan Area Networks}, vol. 2020-July, 2020.

\bibitem{Paul2008TamperDevices}
P.~Paul, S.~Moore, and S.~Tam, ``{Tamper protection for security devices},''
  \emph{Proceedings BLISS 2008 - 2008 ECSIS Symposium on Bio-inspired,
  Learning, and Intelligent Systems for Security}, pp. 92--96, 2008.

\bibitem{PaganiniPierluigi2015AAffairs}
\BIBentryALTinterwordspacing
{Paganini Pierluigi}, ``{A hacker developed Maldrone, the first malware for
  drones Security Affairs},'' \emph{2015}, 2015. [Online]. Available:
  \url{https://securityaffairs.co/wordpress/32767/hacking/maldrone-malware-for-drones.html}
\BIBentrySTDinterwordspacing

\bibitem{skyjack}
J.~Crook, ``{Infamous Hacker Creates SkyJack To Hunt, Hack, And Control Other
  Drones},'' Tech. Rep., 2013.

\bibitem{ThisPhone}
\BIBentryALTinterwordspacing
``{This drone can steal what's on your phone}.'' [Online]. Available:
  \url{https://money.cnn.com/2014/03/20/technology/security/drone-phone/index.html}
\BIBentrySTDinterwordspacing

\bibitem{Oz2021ASolutions}
\BIBentryALTinterwordspacing
H.~Oz, A.~Aris, A.~Levi, and A.~S. Uluagac, ``{A Survey on Ransomware:
  Evolution, Taxonomy, and Defense Solutions},'' 2 2021. [Online]. Available:
  \url{https://arxiv.org/abs/2102.06249v1}
\BIBentrySTDinterwordspacing

\bibitem{amazon}
{Amazon.com Inc.}, ``{Revising the Airspace Model for the Safe Integration of
  Small Unmanned Aircraft Systems},'' \emph{NASA UTM 2015: The Next Era of
  Aviation}, no.~1, pp. 2--5, 2015.

\bibitem{Seshadri2004SWATT:Devices}
A.~Seshadri, A.~Perrig, L.~Van~Doom, and P.~Khosla, ``{SWATT: SoftWare-based
  ATTestation for embedded devices},'' \emph{Proceedings - IEEE Symposium on
  Security and Privacy}, vol. 2004, pp. 272--282, 2004.

\bibitem{Dushku2020SARA:Systems}
E.~Dushku, M.~M. Rabbani, M.~Conti, L.~V. Mancini, and S.~Ranise, ``{SARA:
  Secure Asynchronous Remote Attestation for IoT Systems},'' \emph{IEEE
  Transactions on Information Forensics and Security}, vol.~15, pp. 3123--3136,
  2020.

\bibitem{He2017}
D.~He, S.~Chan, and M.~Guizani, ``{Drone-Assisted Public Safety Networks: The
  Security Aspect},'' \emph{IEEE Communications Magazine}, vol.~55, no.~8, pp.
  218--224, 2017.

\bibitem{Wu2007ANetworks}
B.~Wu, J.~Chen, J.~Wu, and M.~Cardei, ``{A Survey of Attacks and
  Countermeasures in Mobile Ad Hoc Networks},'' \emph{Wireless Network
  Security}, pp. 103--135, 12 2007.

\bibitem{Pleban2014Hacking}
\BIBentryALTinterwordspacing
J.-S. Pleban, R.~Band, and R.~Creutzburg, ``{Hacking and securing the AR.Drone
  2.0 quadcopter: investigations for improving the security of a toy},''
  \emph{SPIE}, vol. 9030, p. 90300L, 2 2014. [Online]. Available:
  \url{https://ui.adsabs.harvard.edu/abs/2014SPIE.9030E..0LP/abstract}
\BIBentrySTDinterwordspacing

\bibitem{Kolias2016IntrusionDataset}
C.~Kolias, G.~Kambourakis, A.~Stavrou, and S.~Gritzalis, ``{Intrusion detection
  in 802.11 networks: Empirical evaluation of threats and a public dataset},''
  \emph{IEEE Communications Surveys and Tutorials}, vol.~18, no.~1, pp.
  184--208, 1 2016.

\bibitem{Shin2016SecurityController}
H.~Shin, K.~Choi, Y.~Park, J.~Choi, and Y.~Kim, ``{Security analysis of
  FHSS-type drone controller},'' in \emph{International Workshop on Information
  Security Applications}, vol. 9503.\hskip 1em plus 0.5em minus 0.4em\relax
  Springer Verlag, 2015, pp. 240--253.

\bibitem{Garbelini2021BRAKTOOTH:Manager}
M.~E. Garbelini, S.~Chattopadhyay, V.~Bedi, S.~Sun, and E.~Kurniawan,
  ``{BRAKTOOTH: Causing Havoc on Bluetooth Link Manager},'' 2021.

\bibitem{Olawumi2003ThreeLearned}
O.~Olawumi, K.~Haataja, M.~Asikainen, N.~Vidgren, and P.~Toivanen, ``{Three
  practical attacks against ZigBee security: Attack scenario definitions,
  practical experiments, countermeasures, and lessons learned},'' in \emph{2014
  14th International Conference on Hybrid Intelligent Systems, HIS 2014}, 2003,
  pp. 199--206.

\bibitem{WrightKillerBee:World}
J.~Wright, ``{KillerBee: Practical ZigBee Exploitation Framework or "Wireless
  Hacking and the Kinetic World"},'' in \emph{11th ToorCon conference, San
  Diego}, 2009.

\bibitem{Gemalto2017LoRaWANPROVIDERS}
A.~Gemalto and S.~And, ``{LoRaWAN ™ Security a white paper prepared for the
  LoRa ALLIANCE™ Full End-To-End encryption for IoT application providers},''
  Tech. Rep., 2017.

\bibitem{You2018AnSystem}
I.~You, S.~Kwon, G.~Choudhary, V.~Sharma, and J.~Seo, ``{An Enhanced LoRaWAN
  Security Protocol for Privacy Preservation in IoT with a Case Study on a
  Smart Factory-Enabled Parking System},'' \emph{Sensors}, vol.~18, no.~6, p.
  1888, 2018.

\bibitem{Chacko2018SecurityLPWAN}
S.~Chacko and M.~D. Job, ``{Security mechanisms and Vulnerabilities in
  LPWAN},'' in \emph{IOP Conference Series: Materials Science and Engineering},
  vol. 396, no.~1, 2018, p. 12027.

\bibitem{Neji2019CommunicationAgriculture}
N.~Neji and T.~Mostfa, ``{Communication technology for unmanned aerial
  vehicles: A qualitative assessment and application to Precision
  Agriculture},'' in \emph{2019 International Conference on Unmanned Aircraft
  Systems, ICUAS 2019}, 2019, pp. 848--855.

\bibitem{Kumar2020NB-IoTSurvey}
V.~Kumar, R.~K. Jha, and S.~Jain, ``{NB-IoT Security: A Survey},''
  \emph{Wireless Personal Communications}, vol. 113, no.~4, pp. 2661--2708,
  2020.

\bibitem{Saranya2017ANetworks}
K.~Saranya and M.~A. Dorairangaswamy, ``{A Study on Evaluation of DoS Attacks
  in WiMAX Networks},'' \emph{International Research Journal of Engineering and
  Technology}, 2017.

\bibitem{Han2009PotentialNetworks}
J.~H.~K. Han, M.~Y. Alias, and G.~B. Min, ``{Potential denial of service
  attacks in IEEE802.16e-2005 networks},'' \emph{2009 9th International
  Symposium on Communications and Information Technology, ISCIT 2009}, pp.
  1207--1212, 2009.

\bibitem{Vasconcelos2016}
G.~Vasconcelos, G.~Carrijo, R.~Miani, J.~Souza, and V.~Guizilini, ``{The Impact
  of DoS Attacks on the AR.Drone 2.0},'' in \emph{Proceedings - 13th Latin
  American Robotics Symposium and 4th Brazilian Symposium on Robotics, LARS/SBR
  2016}, 2016, pp. 127--132.

\bibitem{Muzzi2015}
F.~A.~G. Muzzi, P.~R. d.~M. Cardoso, D.~F. Pigatto, and K.~R. L. J.~C. Branco,
  ``{Using Botnets to provide security for safety critical embedded systems - A
  case study focused on UAVs},'' in \emph{Journal of Physics: Conference
  Series}, vol. 633, no.~1, 2015, p. 012053.

\bibitem{Conti2016AAttacks}
M.~Conti, N.~Dragoni, and V.~Lesyk, ``{A Survey of Man in the Middle
  Attacks},'' \emph{IEEE Communications Surveys and Tutorials}, vol.~18, no.~3,
  pp. 2027--2051, 7 2016.

\bibitem{Rodday2016ExploringVehicles}
N.~M. Rodday, R.~O. De~Schmidt, and A.~Pras, ``{Exploring security
  vulnerabilities of unmanned aerial vehicles},'' in \emph{Proceedings of the
  NOMS 2016 - 2016 IEEE/IFIP Network Operations and Management Symposium},
  2016, pp. 993--994.

\bibitem{VideoJak:Calls}
\BIBentryALTinterwordspacing
``{VideoJak: hijacking ip video calls},'' 2011. [Online]. Available:
  \url{http://videojak.sourceforge.net/}
\BIBentrySTDinterwordspacing

\bibitem{Maxa2015}
J.-a. Maxa, M.-s.~B. Mahmoud, N.~Larrieu, J.-a. Maxa, M.-s.~B. Mahmoud,
  N.~Larrieu, S.~Routing, and P.~Design, ``{Secure Routing Protocol Design for
  UAV Ad Hoc Networks},'' in \emph{2015 IEEE/AIAA 34th Digital Avionics Systems
  Conference (DASC)}, 2015, pp. 4A5--1.

\bibitem{Tseng2011ANetworks}
F.~H. Tseng, L.~D. Chou, and H.~C. Chao, ``{A survey of black hole attacks in
  wireless mobile ad hoc networks},'' \emph{Human-centric Computing and
  Information Sciences}, vol.~1, no.~1, pp. 1--16, 12 2011.

\bibitem{Pirretti2006TheDefense}
M.~Pirretti, S.~Zhu, N.~Vijaykrishnan, P.~McDaniel, M.~Kandemir, and R.~Brooks,
  ``{The sleep deprivation attack in sensor networks: Analysis and methods of
  defense},'' \emph{International Journal of Distributed Sensor Networks},
  vol.~2, no.~03, pp. 0--0, 9 2006.

\bibitem{Douceur2002TheAttack}
J.~R. Douceur, ``{The sybil attack},'' in \emph{International workshop on
  peer-to-peer systems}, vol. 2429.\hskip 1em plus 0.5em minus 0.4em\relax
  Springer Verlag, 2002, pp. 251--260.

\bibitem{HuWormholeNetworks}
Y.~C. Hu and A.~Perrig, ``{Wormhole attacks in wireless networks},'' \emph{IEEE
  Journal on Selected Areas in Communications}, vol.~24, no.~2, pp. 370--379,
  2006.

\bibitem{routing2007}
B.~Kannhavong, H.~Nakayama, Y.~Nemoto, N.~Kato, and A.~Jamalipour, ``{A survey
  of routing attacks in mobile ad hoc networks},'' \emph{IEEE Wireless
  Communications}, vol.~14, no.~5, pp. 85--91, 10 2007.

\bibitem{Awerbuch2004MitigatingNetworks}
B.~Awerbuch, R.~Curtmola, D.~Holmer, C.~Nita-Rotaru, and H.~Rubens,
  ``{Mitigating byzantine attacks in ad hoc wireless networks},''
  \emph{Department of Computer Science, Johns Hopkins University, Tech. Rep.
  Version}, vol.~1, no. March, p.~16, 2004.

\bibitem{Hu2003RushingProtocols}
Y.-C. Hu, A.~Perrig, and D.~B. Johnson, ``{Rushing attacks and defense in
  wireless ad hoc network routing protocols},'' in \emph{Proceedings of the 2nd
  ACM workshop on Wireless security}, 2003, pp. 30--40.

\bibitem{Highnam2016AnArchitecture}
K.~Highnam, K.~Angstadt, K.~Leach, W.~Weimer, A.~Paulos, and P.~Hurley, ``{An
  Uncrewed Aerial Vehicle Attack Scenario and Trustworthy Repair
  Architecture},'' in \emph{Proceedings - 46th Annual IEEE/IFIP International
  Conference on Dependable Systems and Networks, DSN-W 2016}, 2016, pp.
  222--225.

\bibitem{Zhang2017SecuringOverview}
J.~Zhang, T.~Q. Duong, R.~Woods, and A.~Marshall, ``{Securing Wireless
  Communications of the Internet of Things from the Physical Layer, An
  Overview},'' \emph{Entropy 2017, Vol. 19, Page 420}, vol.~19, no.~8, p. 420,
  8 2017.

\bibitem{Bellare2000AuthenticatedParadigm}
M.~Bellare and C.~Namprempre, ``{Authenticated encryption: Relations among
  notions and analysis of the generic composition paradigm},'' in
  \emph{International Conference on the Theory and Application of Cryptology
  and Information Security}, vol. 1976.\hskip 1em plus 0.5em minus 0.4em\relax
  Springer Verlag, 2000, pp. 517--530.

\bibitem{Zhang2017SecuringOptimization}
G.~Zhang, Q.~Wu, M.~Cui, and R.~Zhang, ``{Securing UAV Communications via
  Trajectory Optimization},'' in \emph{GLOBECOM 2017-2017 IEEE Global
  Communications Conference}, vol. 2018-Janua, 2017, pp. 1--6.

\bibitem{Barros2006SecrecyChannels}
J.~Barros and M.~R. Rodrigues, ``{Secrecy capacity of wireless channels},''
  \emph{IEEE International Symposium on Information Theory - Proceedings}, pp.
  356--360, 2006.

\bibitem{Alladi2020SecAuthUAV:Communication}
T.~Alladi, {Naren}, G.~Bansal, V.~Chamola, and M.~Guizani, ``{SecAuthUAV: A
  Novel Authentication Scheme for UAV-Ground Station and UAV-UAV
  Communication},'' \emph{IEEE Transactions on Vehicular Technology}, vol.~69,
  no.~12, pp. 15\,068--15\,077, 12 2020.

\bibitem{He2017SecureNetwork}
S.~He, Q.~Wu, J.~Liu, W.~Hu, B.~Qin, and Y.~N. Li, ``{Secure communications in
  unmanned aerial vehicle network},'' in \emph{International Conference on
  Information Security Practice and Experience}, vol. 10701 LNCS.\hskip 1em
  plus 0.5em minus 0.4em\relax Springer Verlag, 2017, pp. 601--620.

\bibitem{WonAObjects}
J.~Won, S.-H. Seo, and E.~Bertino, ``{A Secure Communication Protocol for
  Drones and Smart Objects},'' in \emph{Proceedings of the 10th ACM Symposium
  on Information, Computer and Communications Security}, 2015, pp. 249--260.

\bibitem{Shoufan2015}
A.~Shoufan, H.~Alnoon, and J.~Baek, ``{Secure communication in civil drones},''
  \emph{Communications in Computer and Information Science}, vol. 576, pp.
  177--195, 2015.

\bibitem{Choudhary2018IntrusionSurvey}
G.~Choudhary, V.~Sharma, I.~You, K.~Yim, I.~R. Chen, and J.~H. Cho,
  ``{Intrusion Detection Systems for Networked Unmanned Aerial Vehicles: A
  Survey},'' in \emph{2018 14th International Wireless Communications and
  Mobile Computing Conference, IWCMC 2018}, 2018, pp. 560--565.

\bibitem{ADS2016}
T.~Kacem, D.~Wijesekera, P.~Costa, and A.~Barreto, ``{An ADS-B Intrusion
  Detection System},'' in \emph{IEEE Trustcom/BigDataSE/ISPA}, 2016.

\bibitem{Condomines2019NetworkValidation}
J.~P. Condomines, R.~Zhang, and N.~Larrieu, ``{Network intrusion detection
  system for UAV ad-hoc communication: From methodology design to real test
  validation},'' \emph{Ad Hoc Networks}, vol.~90, p. 101759, 2019.

\bibitem{Maxa2015a}
J.~A. Maxa, M.~S. Ben~Mahmoud, and N.~Larrieu, ``{Secure routing protocol
  design for UAV ad hoc networks},'' in \emph{AIAA/IEEE Digital Avionics
  Systems Conference - Proceedings}, 2015.

\bibitem{Sbeiti2016}
M.~Sbeiti, N.~Goddemeier, D.~Behnke, and C.~Wietfeld, ``{PASER: Secure and
  Efficient Routing Approach for Airborne Mesh Networks},'' \emph{IEEE
  Transactions on Wireless Communications}, vol.~15, no.~3, pp. 1950--1964,
  2016.

\bibitem{Aggarwal2012}
A.~Aggarwal, ``{AODVSEC: A Novel Approach to Secure Ad Hoc on-Demand Distance
  Vector (AODV) Routing Protocol from Insider Attacks in MANETs},''
  \emph{International journal of Computer Networks {\&} Communications},
  vol.~4, no.~4, pp. 191--210, 2012.

\bibitem{Maxa2016}
J.~A. Maxa, M.~S. Ben~Mahmoud, and N.~Larrieu, ``{Joint Model-Driven design and
  real experiment-based validation for a secure UAV Ad hoc Network routing
  protocol},'' in \emph{ICNS 2016: Securing an Integrated CNS System to Meet
  Future Challenges}.\hskip 1em plus 0.5em minus 0.4em\relax AIAA/IEEE, 2016,
  pp. 1--2.

\bibitem{Garcia-Magarino2019SecurityBlockchain}
I.~Garc{\'{i}}a-Magari{\~{n}}o, R.~Lacuesta, M.~Rajarajan, and J.~Lloret,
  ``{Security in networks of unmanned aerial vehicles for surveillance with an
  agent-based approach inspired by the principles of blockchain},'' \emph{Ad
  Hoc Networks}, vol.~86, pp. 72--82, 2019.

\bibitem{Jiang2020a}
B.~Jiang, J.~Yang, and H.~Song, ``{Protecting privacy from aerial photography:
  State of the art, opportunities, and challenges},'' \emph{IEEE INFOCOM 2020 -
  IEEE Conference on Computer Communications Workshops, INFOCOM WKSHPS 2020},
  pp. 799--804, 2020.

\bibitem{ArthurHollandMichel2017DronesCases}
A.~H. Michel and D.~Gettinger, ``{Drone incidents: A survey of legal cases},''
  Bard College Center for Study of the Drone: Annandale-on-the-Hudson, NY, USA,
  Tech. Rep., 2017.

\bibitem{Finn2013SevenPrivacy}
R.~L. Finn, D.~Wright, and M.~Friedewald, ``{Seven types of privacy},'' in
  \emph{European Data Protection: Coming of Age}.\hskip 1em plus 0.5em minus
  0.4em\relax Springer Netherlands, 2013, pp. 3--32.

\bibitem{Mcneal2014DronesRobotics}
G.~Mcneal, ``{Drones and Aerial Surveillance: Considerations For Legislators
  The Robots Are Coming: The Project On Civilian Robotics},'' Tech. Rep., 2014.

\bibitem{Clarke2014ThePrivacy}
R.~Clarke, ``{The regulation of civilian drones' impacts on behavioural
  privacy},'' \emph{Computer Law and Security Review}, vol.~30, no.~3, pp.
  286--305, 2014.

\bibitem{Lui2007IndividualAnalytics}
S.~M. Lui and L.~Qiu, ``{Individual privacy and organizational privacy in
  business analytics},'' in \emph{Proceedings of the Annual Hawaii
  International Conference on System Sciences}, 2007.

\bibitem{Salamh2021AChallenges}
F.~E. Salamh, U.~Karabiyik, M.~K. Rogers, and E.~T. Matson, ``{A Comparative
  UAV Forensic Analysis: Static and Live Digital Evidence Traceability
  Challenges},'' \emph{Drones}, vol.~5, no.~2, p.~42, 2021.

\bibitem{Birnbach2017Wi-FlyDrones}
S.~Birnbach, R.~Baker, and I.~Martinovic, ``{Wi-Fly?: Detecting Privacy
  Invasion Attacks by Consumer Drones},'' \emph{NDSS}, 2017.

\bibitem{Chen2020ASystem}
C.-L. Chen, Y.-Y. Deng, W.~Weng, C.-H. Chen, Y.-J. Chiu, and C.-M. Wu, ``{A
  Traceable and Privacy-Preserving Authentication for UAV Communication Control
  System},'' \emph{Electronics}, vol.~9, no.~1, p.~62, 2020.

\bibitem{Tian2019EfficientDrones}
Y.~Tian, J.~Yuan, and H.~Song, ``{Efficient privacy-preserving authentication
  framework for edge-assisted Internet of Drones},'' \emph{Journal of
  Information Security and Applications}, vol.~48, p. 102354, 2019.

\bibitem{Rozantsev2015FlyingCamera}
A.~Rozantsev, V.~Lepetit, and P.~Fua, ``{Flying objects detection from a single
  moving camera},'' in \emph{Proceedings of the IEEE Conference on Computer
  Vision and Pattern Recognition}, 2015, pp. 4128--4136.

\bibitem{CaseLow-costTracking}
E.~E. Case, A.~M. Zelnio, and B.~D. Rigling, ``{Low-cost acoustic array for
  small UAV detection and tracking},'' in \emph{IEEE National Aerospace and
  Electronics Conference}, 2008, pp. 110--113.

\bibitem{Nassi2019a}
B.~Nassi, R.~Ben-Netanel, A.~Shamir, and Y.~Elovici, ``{Drones' cryptanalysis -
  Smashing cryptography with a flicker},'' \emph{Proceedings - IEEE Symposium
  on Security and Privacy}, vol. 2019-May, pp. 1397--1414, 2019.

\bibitem{IdentificationWikipedia}
\BIBentryALTinterwordspacing
``{Identification friend or foe - Wikipedia}.'' [Online]. Available:
  \url{https://en.wikipedia.org/wiki/Identification_friend_or_foe}
\BIBentrySTDinterwordspacing

\bibitem{GPS.gov:Accuracy}
\BIBentryALTinterwordspacing
``{GPS.gov: GPS Accuracy}.'' [Online]. Available:
  \url{https://www.gps.gov/systems/gps/performance/accuracy/}
\BIBentrySTDinterwordspacing

\bibitem{Al-Dhaqm2021ResearchModels}
A.~Al-Dhaqm, R.~A. Ikuesan, V.~R. Kebande, S.~Razak, and F.~M. Ghabban,
  ``{Research challenges and opportunities in drone forensics models},'' p.
  1519, 6 2021.

\bibitem{Atkinson2021DroneChallenges}
S.~Atkinson, G.~Carr, C.~Shaw, and S.~Zargari, ``{Drone Forensics: The Impact
  and Challenges},'' \emph{Advanced Sciences and Technologies for Security
  Applications}, pp. 65--124, 2021.

\bibitem{Rahman2016Forensic-by-designSystems}
N.~H. Ab~Rahman, W.~B. Glisson, Y.~Yang, and K.~K.~R. Choo,
  ``{Forensic-by-Design Framework for Cyber-Physical Cloud Systems},''
  \emph{IEEE Cloud Computing}, vol.~3, no.~1, pp. 50--59, 2016.

\bibitem{Schumann2015R2U2:Systems}
J.~Schumann, P.~Moosbrugger, and K.~Y. Rozier, ``{R2U2: Monitoring and
  Diagnosis of Security Threats for Unmanned Aerial Systems},'' in
  \emph{Runtime Verification}.\hskip 1em plus 0.5em minus 0.4em\relax Springer,
  Cham, 2015, vol. 9333, pp. 233--249.

\bibitem{Franco2021ASystems}
J.~Franco, A.~Aris, B.~Canberk, and A.~S. Uluagac, ``{A Survey of Honeypots and
  Honeynets for Internet of Things, Industrial Internet of Things, and
  Cyber-Physical Systems},'' \emph{IEEE Communications Surveys {\&} Tutorials},
  pp. 1--1, 8 2021.

\bibitem{Birnbaum2015UnmannedProfiling}
Z.~Birnbaum, A.~Dolgikh, V.~Skormin, E.~O'Brien, D.~Muller, and
  C.~Stracquodaine, ``{Unmanned Aerial Vehicle security using behavioral
  profiling},'' \emph{2015 International Conference on Unmanned Aircraft
  Systems, ICUAS 2015}, pp. 1310--1319, 7 2015.

\bibitem{Rodrigues2019}
M.~Rodrigues, J.~Amaro, F.~S. Osorio, and B.~R. Kalinka, ``{Authentication
  Methods for UAV Communication},'' \emph{Proceedings - International Symposium
  on Computers and Communications}, vol. 2019-June, pp. 1210--1215, 2019.

\bibitem{Javaid2013UAVSim:Analysis}
A.~Y. Javaid, W.~Sun, and M.~Alam, ``{UAVSim: A simulation testbed for unmanned
  aerial vehicle network cyber security analysis},'' \emph{2013 IEEE Globecom
  Workshops, GC Wkshps 2013}, pp. 1432--1436, 2013.

\bibitem{Acar2018AImplementation}
A.~Acar, H.~Aksu, A.~S. Uluagac, and M.~Conti, ``{A survey on homomorphic
  encryption schemes: Theory and implementation},'' \emph{ACM Computing
  Surveys}, vol.~51, no.~4, 7 2018.

\bibitem{Ghribi2020ANetworks}
E.~Ghribi, T.~T. Khoei, H.~T. Gorji, P.~Ranganathan, and N.~Kaabouch, ``{A
  Secure Blockchain-based Communication Approach for UAV Networks},''
  \emph{IEEE International Conference on Electro Information Technology}, vol.
  2020-July, pp. 411--415, 7 2020.

\bibitem{Kumari2020}
A.~Kumari, R.~Gupta, S.~Tanwar, and N.~Kumar, ``{A taxonomy of
  blockchain-enabled softwarization for secure UAV network},'' \emph{Computer
  Communications}, vol. 161, no. August, pp. 304--323, 2020.

\bibitem{Matson2019UAVModels}
E.~Matson, B.~Yang, A.~Smith, E.~Dietz, and J.~Gallagher, ``{UAV Detection
  System with Multiple Acoustic Nodes Using Machine Learning Models},''
  \emph{Proceedings - 3rd IEEE International Conference on Robotic Computing,
  IRC 2019}, pp. 493--498, 3 2019.

\bibitem{Yazdinejad2021FederatedAuthentication}
A.~Yazdinejad, R.~M. Parizi, A.~Dehghantanha, and H.~Karimipour, ``{Federated
  learning for drone authentication},'' \emph{Ad Hoc Networks}, vol. 120, p.
  102574, 9 2021.

\bibitem{Nayyar2020TheDrones}
A.~Nayyar, B.~L. Nguyen, and N.~G. Nguyen, ``{The internet of drone things
  (Iodt): Future envision of smart drones},'' in \emph{First international
  conference on sustainable technologies for computational intelligence}, vol.
  1045.\hskip 1em plus 0.5em minus 0.4em\relax Springer, 2020, pp. 563--580.

\end{thebibliography}

\end{document}